\documentclass[useAMS,usenatbib]{mn2e}
\usepackage{graphicx,fleqn,natbib,amssymb}
\usepackage{epstopdf}
\usepackage{xcolor}
\usepackage{longtable}
\usepackage{appendix}

\DeclareGraphicsExtensions{.eps,.cps,.ps,.eps.gz,.ps.gz,.eps.Z}
\def \sourcename{J221951 }
\def \sourcenamefull{{\it Swift} J221951-484240 }
\def \srcname{J221951}

\usepackage{dcolumn}
\usepackage{pdflscape}
\usepackage{dblfloatfix}
\usepackage{xcolor}

\newcommand\nh{$N_{HI}$}

\title[J221951-484240]{\it{Swift}/UVOT discovery of {\it Swift} J221951-484240: a UV luminous ambiguous nuclear transient}

\author[Oates et al.]{S. R. Oates$^{1}$\thanks{E-mail: s.r.oates@bham.ac.uk}, N. P. M. Kuin$^{2}$, M. Nicholl$^{3,1}$, F. Marshall$^{4}$, E. Ridley$^{1}$, K.~Boutsia$^{5}$,
\newauthor A. A. Breeveld$^{2}$, D. A. H.~Buckley$^{6,7,8}$, S. B. Cenko$^{4,9}$, M. De Pasquale$^{10}$, 
\newauthor P. G. Edwards$^{11}$, M. Gromadzki$^{12}$, R. Gupta$^{13,14}$, S. Laha$^{4,15,16}$, N.~Morrell$^{5}$, 
\newauthor M.~ Orio$^{17,18}$, S.~B.~Pandey$^{13}$, M. J. Page$^{2}$, K.L. Page$^{19}$, T. Parsotan$^{4,15,16}$, 
\newauthor A. Rau$^{20}$, P. Schady$^{21}$, J. Stevens$^{22}$, P. J. Brown$^{23}$, P.A. Evans$^{19}$, C. Gronwall$^{24,25}$,  
\newauthor J.A. Kennea$^{24}$, N.J. Klingler$^{4,15,16}$, M. H. Siegel$^{24}$, A. Tohuvavohu$^{26}$, E. Ambrosi$^{27}$, 
\newauthor S.D. Barthelmy$^{4}$, A.P. Beardmore$^{19}$, M.G. Bernardini$^{28}$, C. Bonnerot$^{1}$, S. Campana$^{28}$, 
\newauthor R. Caputo$^{4}$, S. Ciroi$^{30}$, G. Cusumano$^{27}$, A. D'A{\` i}$^{27}$, P. D'Avanzo$^{28}$, V. D'Elia$^{29,31}$,  
\newauthor P. Giommi$^{31}$, D.H. Hartmann$^{32}$, H.A. Krimm$^{33}$,  D. B. Malesani$^{34,35,36}$, A. Melandri$^{29}$, 
\newauthor J. A. Nousek$^{24}$, P.T.  O'Brien$^{19}$, J.P. Osborne$^{19}$, C. Pagani$^{19}$, D.M. Palmer$^{37}$,  
\newauthor M. Perri$^{31,29}$, J. L. Racusin$^{4}$, T. Sakamoto$^{38}$, B. Sbarufatti$^{28,24}$, J. E. Schlieder$^{4}$, 
\newauthor G. Tagliaferri$^{28}$, E. Troja$^{39,40}$ \& D. Xu$^{40,41}$
  }

\begin{document}

\date{Accepted...Received...}

\maketitle

\label{firstpage}

\begin{abstract} 
We report the discovery of \sourcenamefull (hereafter: \srcname), a luminous slow-evolving blue transient that was detected by the Neil Gehrels Swift Observatory Ultra-violet/Optical Telescope (\textit{Swift}/UVOT) during the follow-up of Gravitational Wave alert S190930t, to which it is unrelated. {\it Swift}/UVOT photometry shows the UV spectral energy distribution of the transient to be well modelled by a slowly shrinking black body with an approximately constant temperature of $T\sim2.5\times10^4$\,K. At a redshift $z=0.5205$, \sourcename had a peak absolute magnitude of $M_{u,AB} = -23$ mag, peak bolometric luminosity $L_{max}=1.1\times10^{45}~{\rm erg\,s}^{-1}$ and a total radiated energy of $E>2.6\times10^{52}$ erg. The archival {\it WISE} IR photometry shows a slow rise prior to a peak near the discovery date. Spectroscopic UV observations display broad absorption lines in N\,V and O\,VI, pointing toward an outflow at coronal temperatures. The lack of emission in the higher H~Lyman lines, N\,I and other neutral lines is consistent with a viewing angle close to the plane of the accretion or debris disc. The origin of \sourcename can not be determined with certainty but has properties consistent with a tidal disruption event and the turn-on of an active galactic nucleus. 
\end{abstract}


\section{Introduction}
\label{intro}
The Advanced Laser Interferometer Gravitational Wave Observatory \citep[LIGO; LIGO Scientific Collaboration; ][]{aa15} and the Advanced Virgo detector \citep[Virgo; the Virgo Scientific Collaboration;][]{ace15} began the third observing run (“O3”) in search of Gravitational Wave (GW) events on 2019 April 1 \citep{GCN:24045}. The Neil Gehrels Swift Observatory \citep[henceforth {\it Swift};][]{geh04} participated in the search for the electromagnetic (EM) counterpart of GW sources. In total {\it Swift} observed, with varying degrees of coverage, 18 of the GW candidate alerts released by the LIGO-Virgo Collaboration (LVC). One effect of scanning very large areas of the sky for the EM counterpart is the discovery of a multitude of transient phenomena that are not necessarily related to the GW itself. During the O3 run the {\it Swift} Ultra-violet/Optical Telescope \citep[UVOT; 1600–8000 \AA;][]{roming} serendipitously found 27 optical transients that changed in magnitude at 3$\sigma$ level compared with archival $u$ or $g$-band catalogued values \citep{oat21}. Determining the nature of all the optical/UV transients that reside in GW error regions is important to confirm or rule out their possible association with the GW trigger and these serendipitous UV sources detected may also be of interest in their own right. Indeed this is the case for \sourcenamefull \citep[henceforth \srcname;][]{GCN:25901,GCN:25939}, which we investigate further in this paper. This source was observed by the {\it Swift}/UVOT telescope as part of the follow-up campaign to identify the EM counterpart to the GW trigger S190930t, which was initially classified as a neutron star - black hole (NS-BH) merger \citep{GCN:25876}. 

Since detection, {\it Swift} has continued to monitor this source and we have obtained additional photometric and spectral observations with {\it HST} ACS+COS, SALT, Magellan/IMACS, VLT/X-shooter, ATCA, {\it AstroSat} and GROND. With the {\it HST} COS spectrum, we identify \sourcename at a redshift of $z=0.5205\pm 0.0003$ (see \S \ref{optical_spectra}), which is outside the distance range of the GW source reported on GraceDB, ruling out its association with S190930t. At this redshift it had a peak absolute magnitude of $M_{u,AB} = -23$ mag and total energy release in the optical/UV over the 2.5 years of observations of $>2.6\times 10^{52}$ erg, making it one of the most luminous transients ever recorded. In the following, we report on these observations and investigate the nature of this extremely luminous UV transient, ultimately comparing it to tidal disruption events (TDEs\footnote{A TDE is a bright flare that arises as a consequence of a star being torn apart as it passes too close to the centre of a supermassive black hole \citep{hil75,ree88,loe97}.}) and active galactic nuclei (AGN).

This paper is organized as follows. We provide the data analysis in \S~\ref{data_reduction}, results in \S~\ref{results} and discussion and conclusions follow in \S~\ref{discussion} and \S~\ref{conclusions}. All uncertainties throughout this paper are quoted at 1$\sigma$ unless otherwise stated. Throughout, we assume the Hubble parameter $H_0 = 70\;{\rm km\,s}^{-1}\;{\rm Mpc}^{-1}$ and density parameters $\Omega_{\Lambda}= 0.7$ and $\Omega_m= 0.3$. All magnitudes are given in the AB system, except for {\it WISE} photometry which is provided in the Vega system.

\section{Observations}
\label{data_reduction}
S190930t triggered LIGO/Virgo at 14:34:08 UT on the 2019 September 30, $T_{0,GW}$ \citep{GCN:25876}. It was reported to be at a distance of $108\pm38$ Mpc, on GraceDB\footnote{ https://gracedb.ligo.org/} using the BAYESTAR skymap. S190930t had a high false alarm probability of 2.05 $\rm {yr}^{-1}$. At the time of the announcement, this trigger met the {\it Swift} follow-up criteria and {\it Swift}/UVOT observed 50.1 deg$^2$, equating to 2 per cent of the total localization probability; determined from a convolution of the LIGO-Virgo probability map and the 2MASS Photometric Redshift catalogue \citep{bil14,eva16}. As this GW event was only detected by a single detector, this event did not meet the criteria to be included in the Gravitational-Wave Transient Catalog (GWTC-2) of compact binary coalescences observed by Advanced LIGO and Advanced Virgo \citep{abb20d}. 

\sourcename was identified in the UVOT GW pipeline as a Q0 source\footnote{For a source to be given a Q0 identification it implies that in the detection image it must be brighter than 19.9 mag and is either a new source or a known source that is two magnitudes brighter than a catalogued value \citep[see][for further details on the {\it Swift}/UVOT GW pipeline and quality flags]{oat21}.} and given the initial identification Q0\_src93 \citep{oat21}. \sourcename was detected at a $u$-band magnitude of $19.48\pm0.20$ mag at $T_{0,GW}+14.37\,{\rm ks}$ \citep{GCN:25901}. The UVOT position is RA, Dec(J2000)= 334.96599, -48.71116 deg with an estimated uncertainty of 0.7 arcsec (radius, 90 per cent confidence). Examining archival images, no source was detected at the location in GALEX NUV or FUV images \citep{bia14,bia17}. However, a faint source consistent with this position was identified in the DSS archival image and in the catalogues of the Dark Energy Survey \citep[DES; ][]{abb18,abb21}, VISTA \citep{mcm13} and {\it WISE} \citep{cut12,cut14}. The source appears nuclear when comparing our ACS imagery (see below) to DES images going back to 2014.

Initially, it was suggested that this source was a flare of a red dwarf star \citep{GCN:25901}. However, follow-up observations, performed by UVOT at $T_{0,GW}+2.1$ days \citep{GCN:25939}, showed the source was blue and at a magnitude consistent with the initial detection ($u=19.67\pm0.22$). At the time of detection, follow-up observations were also performed by J-GEM \citep{GCN:25941}, Chilescope observatory \citep{GCN:25963} and spectrally with SALT \citep{GCN:25962}. However, no spectral features could be identified and therefore the redshift of this source could not be constrained.

\begin{figure*}
\includegraphics[angle=0,scale=0.6]{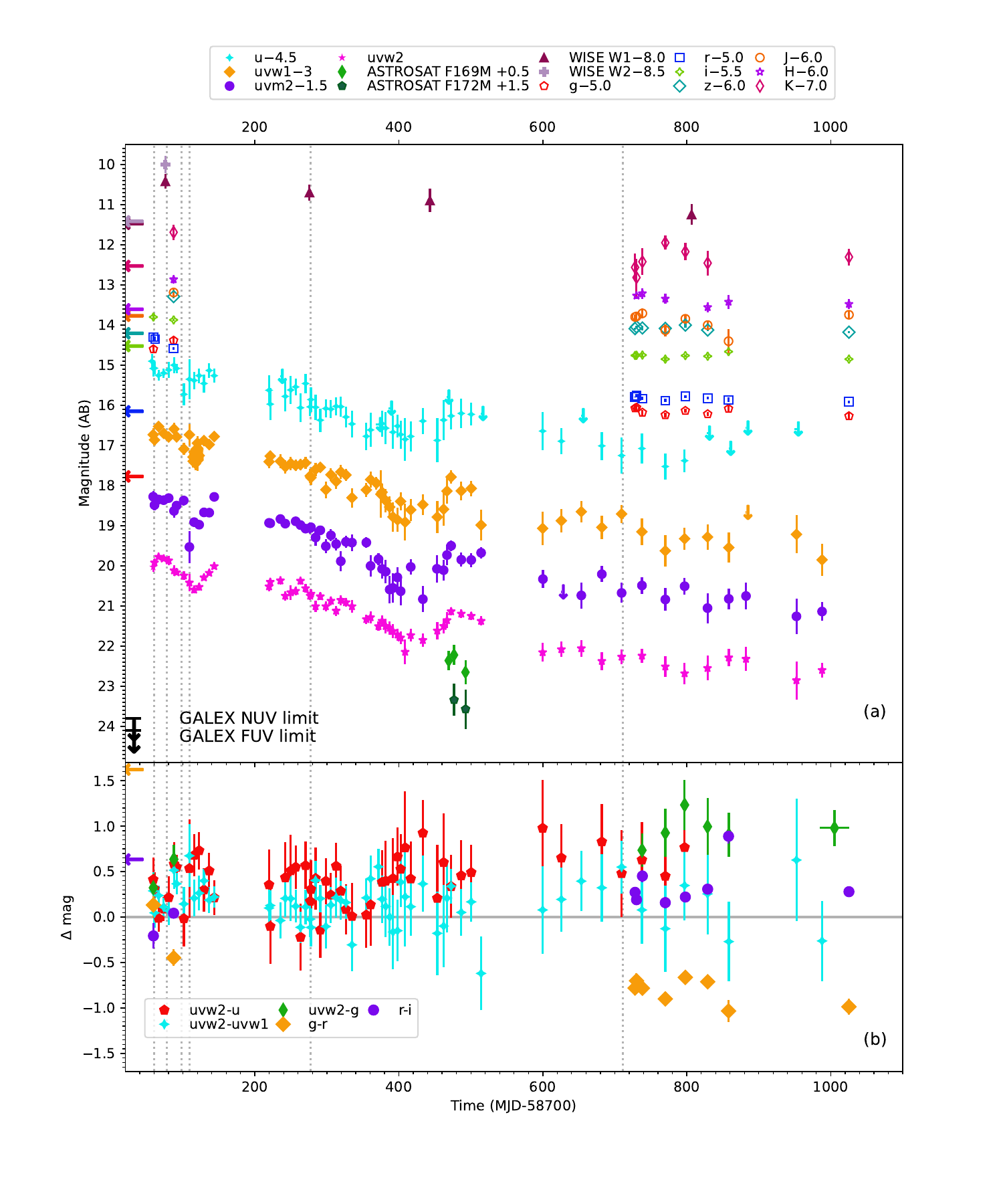}
\caption{Panel (a) displays the 4 UVOT UV filter light curves of \sourcename together with data from {\it AstroSat}, GROND, GALEX, {\it WISE}, J-GEM and Chilescope observatory; data from the latter two telescopes are from GCNs \protect\citep{GCN:25941, GCN:25963}. The different filter light curves have been scaled, with the scaling given in the legend. No correction for Galactic extinction, corresponding to a reddening of $E(B-V) = 0.012$ \citep{sch11} has been applied. Upper limits for the GALEX FUV and NUV observations are given by black down-pointing arrows. Included as left pointing arrows are the archival detections for the different filters, obtained from DES, VISTA and {\it WISE}. No correction has been made to the photometry of \sourcename for this archival source. The dotted vertical lines indicate the times at which spectra were taken. The light curve shows a gradual decrease in brightness over the course of observations. The light curve brightens three times ($\sim$58766, 58843 and 59172 MJD), which appear to reset the brightness level. At peak \sourcename is brighter than archival values at all UV/optical/IR wavelengths, by more than 1-3 mag. Superimposed on the decay, most apparent in the $uvw2$ light curve, are three rebrightenings which show changes in magnitude of $\sim 0.5$ mag, after which, the light curve continues to decay but at an elevated brightness compared to that pre-brightening. Panel (b) displays the change over time of 5 different colours, which are given in the legend. The colour curves have been corrected for Galactic extinction and host subtracted. The left-pointing arrows indicate archival colours. Compared to historic values \sourcename is much bluer, changing in $g-r$ by $-0.68$~mag by the first observation. Within errors, $uvw2-uvw1$ is constant in colour, while $g-r$ and $r-i$ become bluer and $uvw2-g$ becomes redder with time.} 
\label{J221951_lc}
\end{figure*}

Below we summarize the archival data and the photometry and spectroscopy obtained for \srcname. The photometry is provided in Fig. \ref{J221951_lc} and Supplementary Table S.1. A log of the spectroscopic observations is given in Table \ref{tab:spectobs}.

\subsection{{\it Swift} BAT Observations}
\label{batanalysis}
We analyzed all of the publicly available (on HEASARC\footnote{https://heasarc.gsfc.nasa.gov/docs/archive.html}) BAT ``survey" mode data, from 29th Oct 2019 to 21st April 2022, which are also known as detector plane histograms or DPHs. BAT survey data are accumulated in histograms onboard the spacecraft, with typical integration times of between 300 seconds and around $2000$ seconds. An 80-channel binned spectrum is recorded for each of the active detectors, which are saved in the DPH files. For a detailed explanation of the reprocessing and analysis of the BAT survey mode data see \cite{sib22}. We do not detect a signal at the $3\sigma$ level above the background in any of the BAT exposures. Integrating from the time of detection until the last observation we find the average $5 \sigma$ upper limit on the $14-195 \, \rm keV$ flux is $6\times 10^{-9}\rm erg\, cm^{-2}\, s^{-1}$.

\subsection{{\it Swift}/XRT Observations}
\label{xrtanalysis}
We processed the XRT data using the online analysis tools provided by the UK {\it Swift} Science Data Centre \citep{eva07,eva09}. \sourcename is not detected in single visits or in a stacked image created from the $279\,{\rm ks}$ of observations taken over 2.5 years, from 2019 Sep 30 to 2022 Apr 21. Individual visits are typically of a few ks duration, with limiting count rates of $(2-3)\times10^{-3}\,{\rm counts\,s^{-1}}$ (0.3-10\,keV). This is equivalent to an unabsorbed flux density of $< 1\times 10^{-13}\,{\rm erg}\, {\rm cm}^{-2}\,{\rm s}^{-1}$ and luminosity $<10^{44}\,{\rm erg \,s^{-1}}$ in the 0.3-10 keV energy range, assuming a photon index of $\Gamma =$\,1.7, a Galactic absorbing column $N_H= 9.8\times10^{19}\,{\rm cm}^{-2}$ and an intrinsic absorbing column of $N_H = 3\times10^{20}\,{\rm cm}^{-2}$. Stacking all the XRT images, we find a deeper limiting count rate of $1.4\times10^{-4}\,{\rm counts\,s^{-1}}$ (0.3-10\,keV), in $279\,{\rm ks}$, equivalent to an unabsorbed flux density of $<5.5\times 10^{-15}\,{\rm erg}\, {\rm cm}^{-2}\,{\rm s}^{-1}$ and a luminosity of $L_X<6\times10^{42}\,{\rm erg s^{-1}}$. Assuming this limit over the duration of \textit{Swift} observations this places an upper limit on the total energy released in X-rays (0.3-10\,keV) as $E_X<5\times10^{50}\,{\rm erg}$.

\subsection{{\it Swift}/UVOT Observations}
\label{uvotanalysis}
After the detection of \sourcename in the $u$ band on 2019 September 30, UVOT continued to observe \sourcename in all six optical/UV filters until 2021 Aug 11, after which observations were performed in the $u$ and UV filters. Observations were taken in image mode only. To begin with, observations were performed every few days to a week cadence, which decreased to approximately a monthly cadence as the source faded. On the 3rd of December 2020, we observed with a higher cadence to investigate variability. All images were downloaded from the {\it Swift} data archive\footnote{https://www.swift.ac.uk/archive/index.php}. 
The source counts were extracted from single or summed exposures using a source region of 5 arcsec radius. Background counts were extracted 
using an annular region with an inner radius of 15 arcsec and an outer radius of 35 arcsec. The count rates were obtained from the images using the 
{\it Swift} tool {\sc uvotsource}. Finally, the count rates were converted to AB magnitudes using the UVOT photometric zero-points \citep{poole,bre11}. The analysis pipeline used UVOT calibration 20201215. The UVOT detector is less sensitive in a few small patches\footnote{https://heasarc.gsfc.nasa.gov/docs/heasarc/caldb/swift/\\docs/uvot/uvotcaldb\_sss\_01b.pdf} for which a correction has not yet been determined. Therefore, we have checked to see if any of the sources of interest fall on any of these patches in any of our images and exclude 15 individual UV exposures for this reason.

\subsection{Dark Energy Survey}
DES \citep{abb18} observed the field of \sourcename over several occasions in the $g$, $r$, $i$, $z$ and $Y$ filters, starting November 2013. Images from 2013 until 2018 are available for \sourcename in NOIRLab Astro Data Lab\footnote{https://datalab.noirlab.edu/}. 

We used a custom wrapper for \textsc{photutils} to perform both aperture and point spread function (PSF) fitting photometry at the location of \sourcename in these DECam images, finding consistent results between the two approaches. We calibrated the zeropoint of each image using local stars in the Pan-STARRS DR2 catalogue \citep{fle20}. At the location of \srcname, a red point-like source is well detected in all epochs taken prior to the detection of \sourcename by UVOT. The source shows no significant temporal variability.

\subsection{HST COS \& ACS}
\label{cosanalysis}
\begin{figure}
    \centering
    \includegraphics[angle=0, scale=0.27]{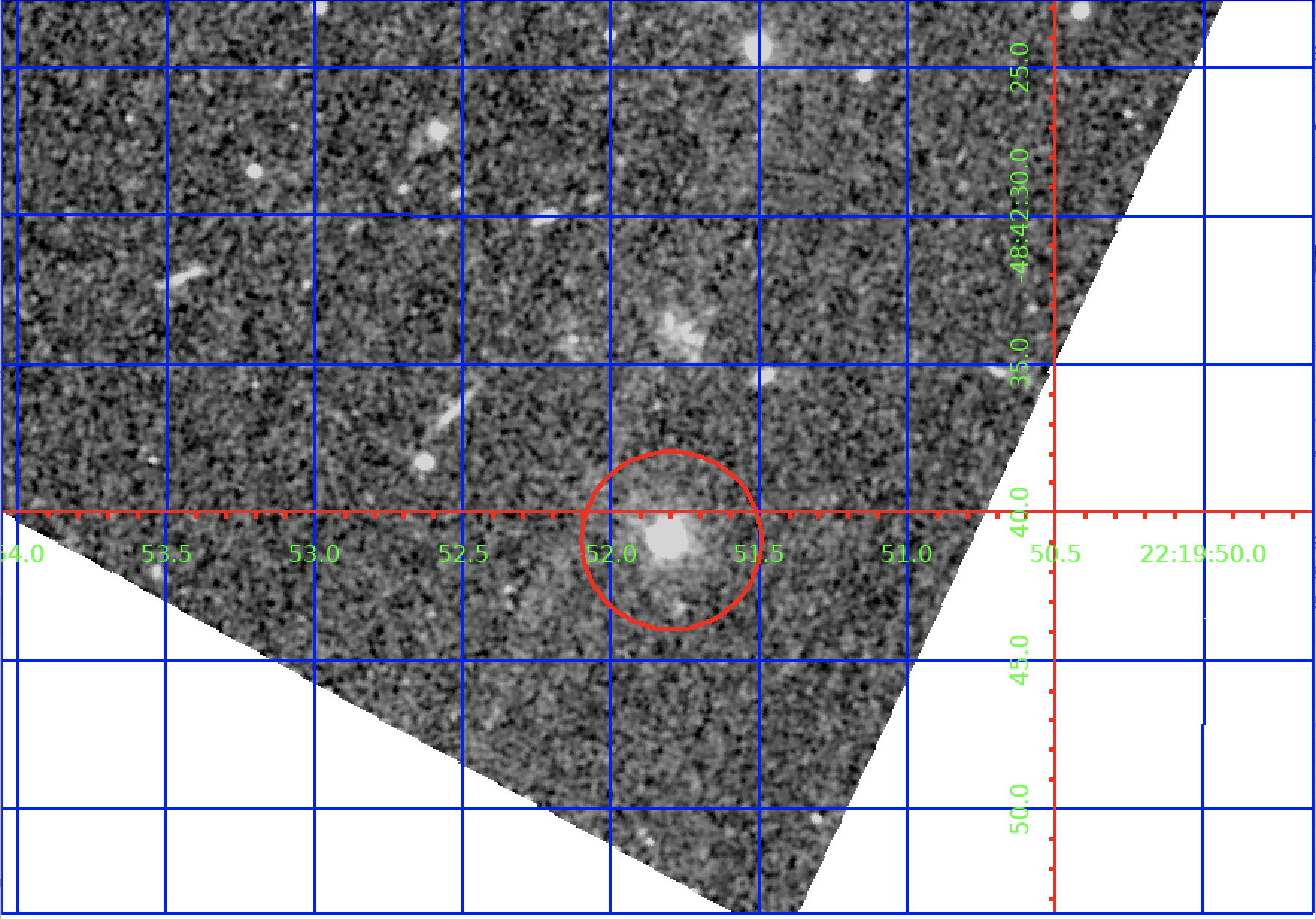}
    \caption{{\it HST} ACS image of \srcname. \srcname, indicated by a red circle, is consistent with being a point source.}
    \label{ACS_position}
\end{figure}

We observed \sourcename  with the Cosmic Origins Spectrograph \citep{gre2003,dix2010} on the Hubble Space Telescope \citep{HST1986} on the 8th May 2020. In two orbits we obtained medium resolution spectra with the G130M and G140L gratings (program ID 16076, P.I. S. Oates). 
The spectra  were processed with the standard pipeline (OPUS\_VER=HSTDP 2020\_5; calibration software system version caldp\_20201012; and CALCOS code version 3.3.10) on MJD 59139.38. The spectra were spliced together using the IRAF `splice' task. 
We also obtained an image, split in 4 exposures, with the Advanced Camera for Surveys \citep{acs2000} (total exposure time = 2256\,s) during one orbit using the F475W filter (4746 \AA; width 420 \AA) which led to an improved position for \sourcename of RA=22:19:51.80, Dec=-48:42:40.90 (J2000), Fig.~\ref{ACS_position}. The source magnitude was $M(F475W)=20.255\pm0.002\,{\rm mag}$. Standard observation and processing were used.
We investigated if the source was nuclear by comparing to DES DECam ``Resampled" images in $g, r, z$ and $Y$ bands from 2014 onward by overlaying the ACS. We find that \sourcename is centred on the nucleus in DECam to within 0.045 arcsec (3-sigma) which is equivalent to an angular separation of $<0.3$kpc.

\subsection{GROND}
\label{grondanalysis}
On 2019 October 31, $g'r'i'z'JHK$ observations of \sourcename were taken with the Gamma-Ray Optical/Near-infrared Detector \citep[GROND;][]{gbc+08} mounted on the MPG 2.2\,m telescope at the ESO La Silla observatory, Chile. The source was re-observed with GROND on a further seven occasions between August and December of 2021, and again in May 2022. The data were reduced and analysed with the standard tools and methods described in \citet{kkg+08}. The optical and NIR magnitudes  were obtained using aperture photometry and absolute calibration was performed using field stars within the GROND field of view covered by the Sky Mapper Southern Sky Survey \citep{ksb+07} and the Two Micron Sky Survey \citep[2MASS;][]{scs+06} for the $g'r'i'z'$ and the $JHK$ bands respectively.

\subsection{AstroSat}
\label{astrosatanalysis}
The Ultra-violet Imaging Telescope (UVIT) onboard {\it AstroSat} \citep{pat03} observed \sourcename thrice on the 16th November 2020, 23rd November 2020, and a few days later on the 9th December 2020 (proposal ID A10\_024, P.I. S. B. Pandey). Observations were taken with the F169M Sapphire (central wavelength of $\lambda=1608{\rm \AA}$, and width $\Delta \lambda =290{\rm \AA}$) and F172M Silica (central wavelength of $\lambda=1717{\rm \AA}$, and width $\Delta \lambda =125{\rm \AA}$) filters, respectively. We aligned the images by comparing field stars against those also found in GALEX FUV images. We performed the aperture photometry on Level 2 UVIT images to extract the source brightness. We used the standard zero points provided in \cite{tan17b}. \sourcename was detected at all three epochs (see Fig. \ref{J221951_lc}).

\subsection{SALT}
\label{saltanalysis}
The South African Large Telescope (SALT; \citealt{Buckley2006}) using the Robert Stobie Spectrograph (RSS, \citealt{Burgh2003}) obtained several spectra with the PG0300 grating starting 2019 October 4 (1800 s), then 2019 October 22 ($2\times1200$ s), and on 2020 May 11 (1500 s); the latter on the same day of the {\it HST} COS observation though with full moon. The mean resolving power of $R\sim420$ (14.8 \AA~resolution). To reduce the spectra we used the PyRAF-based PySALT package (\citealt{Crawford2010})\footnote{https://astronomers.salt.ac.za/software/}, which includes corrections for gain and cross-talk, and performs bias subtraction. We extracted the science spectrum using standard IRAF\footnote{https://iraf-community.github.io/} tasks, including wavelength calibration (Argon calibration lamp exposures were taken, one immediately before and one immediately after the science spectra), background subtraction, and 1D spectra extraction. Due to the SALT design, absolute flux calibration is not possible\footnote{The pupil (i.e. the view of the mirror from the tracker) moves during all SALT observations, causing the effective area of the telescope to change during exposures. Therefore, absolute flux calibration cannot be done. See Buckley et al. (2006) and Crawford et al. (2010) for details.}. However, by observing spectrophotometric standards during twilight, we were able to obtain relative flux calibration, i.e. allowing recovery of the correct spectral shape and relative line strengths.

\subsection{Las Campanas Observatory}
\label{LCOananalysis}
On 2019 November 11 a low-resolution spectrum was taken, consisting of $2\times1200$\,s exposures in the range 3700-9250 \AA, using the IMACS instrument on the Baade Telescope of Las Campanas Observatory. The S/N at 1.16\AA/pix resolution is low, 5-6, and there is a gap between 6430 \AA~and 6524 \AA~due to the location of the spectrum on the detector, which lies across two of the eight CCDs. The spectrum is blue and fairly featureless. Calibration was done using a standard star spectrum. 

\subsection{ATCA}
\label{radioanalysis}

On 2020 Jan 11, we observed the source position with the Australia Telescope Compact Array (ATCA), which
consists of six 22 m diameter dishes \citep{atca}.
Observations, made under project code C1730, were made when the telescope was in its 6A array configuration, with a maximum baseline of 5.9 km. The observing bands were centred at 5.5 and 9.0\,GHz, with a 2\,GHz bandwidth in both bands. Four 10-minute scans of \sourcename  were made, with each scan bookended by 2.5 minute scans on the phase calibrator PKS 2204-540. PKS 1934-638 was used as the primary flux density calibrator. Data reduction was carried out following the standard procedures in miriad \citep{miriad1,miriad2}.
No source was detected at the position of \sourcename with 3\,$\sigma$ upper limits of 117\,$\mu$Jy at 5.5 GHz and 90\,$\mu$Jy at 9 GHz. Assuming a flat spectrum, the 5.5 GHz flux is equivalent to a luminosity of $\sim 2\times10^{39}\, {\rm erg\,s^{-1}}$, across a bandwidth of 2 GHz.

\subsection{GALEX}
{\it GALEX} observed the field of \sourcename  6 times in the NUV and 4 times in the FUV between 2004 and 2008 for a total of 2.5 ks and 2.6 ks. No source is detected in either band at the location of \srcname. Using the Galex Merged catalog of sources \citep[MCAT;][]{mor07} we derive $3\sigma$ magnitude upper limits in the NUV and FUV of $23.8$ and $24.1$, respectively. For comparison, the {\it GALEX} NUV filter spans 1771-2831 \AA, which is broader than the individual UVOT UV filters. The NUV filter covers a similar wavelength range as the UVOT's $uvw1$ and $uvm2$ filters. The FUV filter spans 1344-1786 \AA, which is bluer than UVOT $uvw2$.

\subsection{WISE}
\begin{figure*}
\includegraphics[angle=0,scale=0.8, trim={0.5cm 0.cm 0.5cm 0.cm},clip]{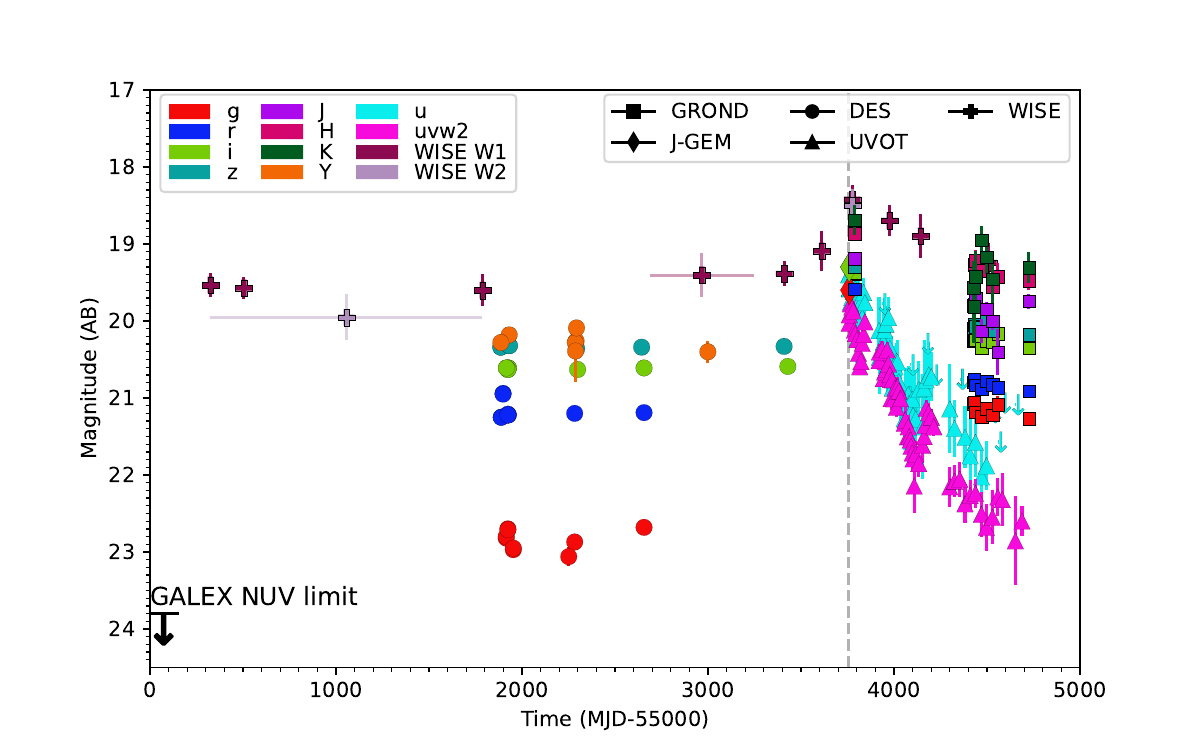}
\caption{Long term light curve of \sourcename containing data from DES, {\it WISE}, GROND, UVOT and J-GEM. The grey dotted line represents $T_{0,GW}$, the trigger time of the gravitational wave event S190930t. The {\it GALEX} observations took place between 53248 MJD and 54679 MJD (August 2004 and July 2008).} 
\label{J221951_archival}
\end{figure*}

In the ALLWISE catalogue \citep{cut14} a source is detected within 1.6 arcsec in the W1 and W2 filters only, with non-detections in the W3 and W4 filters (W1 $=16.90$, W2 $=16.78$, W3 $>12.47$, W4 $>9.12$). {\it WISE/NEOWISE} \citep{wri10,mai11} has been observing the field of \sourcename biannually since May 2014. At the time of writing, the most recent data release provides observations until October 2021. We obtained photometry for the individual exposures from the IRSA/IPAC infrared data science archive. The individual images during a single visit are taken within a 1-2 day period. A weak source is detected at the location of \sourcename in most of the W1 images. In W2, a weak source is detected in approximately half of the exposures. In the single exposure photometry, there appears to be a slight increase in flux in observations taken between 2019 October 18 and 2019 October 21 close to the time \sourcename was detected by {\it Swift}/UVOT, however, the data are noisy. We, therefore, used the coadder tool\footnote{https://irsa.ipac.caltech.edu/applications/ICORE/} to produce stacked images. We used {\sc sextractor} \citep{ber96} to obtain the photometry. We display all the {\it WISE} visits in Fig. \ref{J221951_archival}. We created a stack of the W1 and W2 images taken from May 2010 until May 2014 and measure a magnitude of $16.81\pm0.10$ and $16.61\pm0.30$, in W1 and W2, respectively, consistent with that reported in the ALLWISE catalogue \citep{cut14}. In a stack of the October 2019 observations, the first {\it WISE} visit after \sourcename was detected by UVOT, the source is 1.1, and 1.4 magnitudes brighter in the W1 and W2 filters, respectively, compared to the stacks of the data taken prior to May 2014. \sourcename is not detected in later W2 per visit stacks. In W1 it fades in the first 6 months by $\sim 0.3$ magnitudes and by October 2021 is consistent with pre-2014 level. 

\subsection{X-shooter}
We observed \sourcename  with the X-shooter echelle spectrograph \citep{vernet2011}, mounted on the European Southern Observatory (ESO) Very Large Telescope, on 16th July 2021 (PI Oates, program ID 107.22RT). X-shooter provides continuous coverage from $\approx3000-25000$\,\AA\ in the observer frame. Data were obtained in on-slit nodding mode and reduced using the ESO \textsc{reflex} pipeline. The pipeline applies de-biasing, flat-fielding, geometric transformations of the echelle orders, wavelength calibration, cosmic ray removal, and extraction to a one-dimensional spectrum. Flux calibration is achieved using standard star observations in the same setup.

\begin{table*} 
\centering 
\caption{Log of the spectroscopic observations\label{tab:spectobs}}
\begin{tabular}{lcccccc} 
\hline 
UT Start & Time (MJD) &  Telescope  & Instrument & Grating & Exposure Time (s) \\
\hline
2019-10-04 & 58760 & SALT & RSS & PG0300 & 1800  \\
2019-10-22 & 58778 & SALT & RSS & PG0300 & $2\times1200$ & \\
2019-11-11 & 58798 & Magellan-Baade & IMACS & Spectroscopic2/Gri-300-17.5& $2\times 1200$\\ 
2019-11-22 & 58809 & SALT & RSS & PG0300& 1600  \\
2020-05-08 & 58977 & SALT & RSS &PG0300 & 1500  \\
2020-05-08 & 58977 & {\it HST}  & COS/FUV & G140L & 300     \\
2020-05-08 & 58977 & {\it HST}  & COS/FUV & G130M & 2275     \\
2021-07-15 & 59411 & VLT & X-shooter & UVB,VIS,NIR & 2500,2400,2600 \\
\hline 

\hline 
\end{tabular} 
\end{table*} 

\section{Results}
\label{results}

\begin{figure*}
\includegraphics[angle=0,scale=0.45, trim={0cm 0.cm 1.5cm 0.cm},clip]{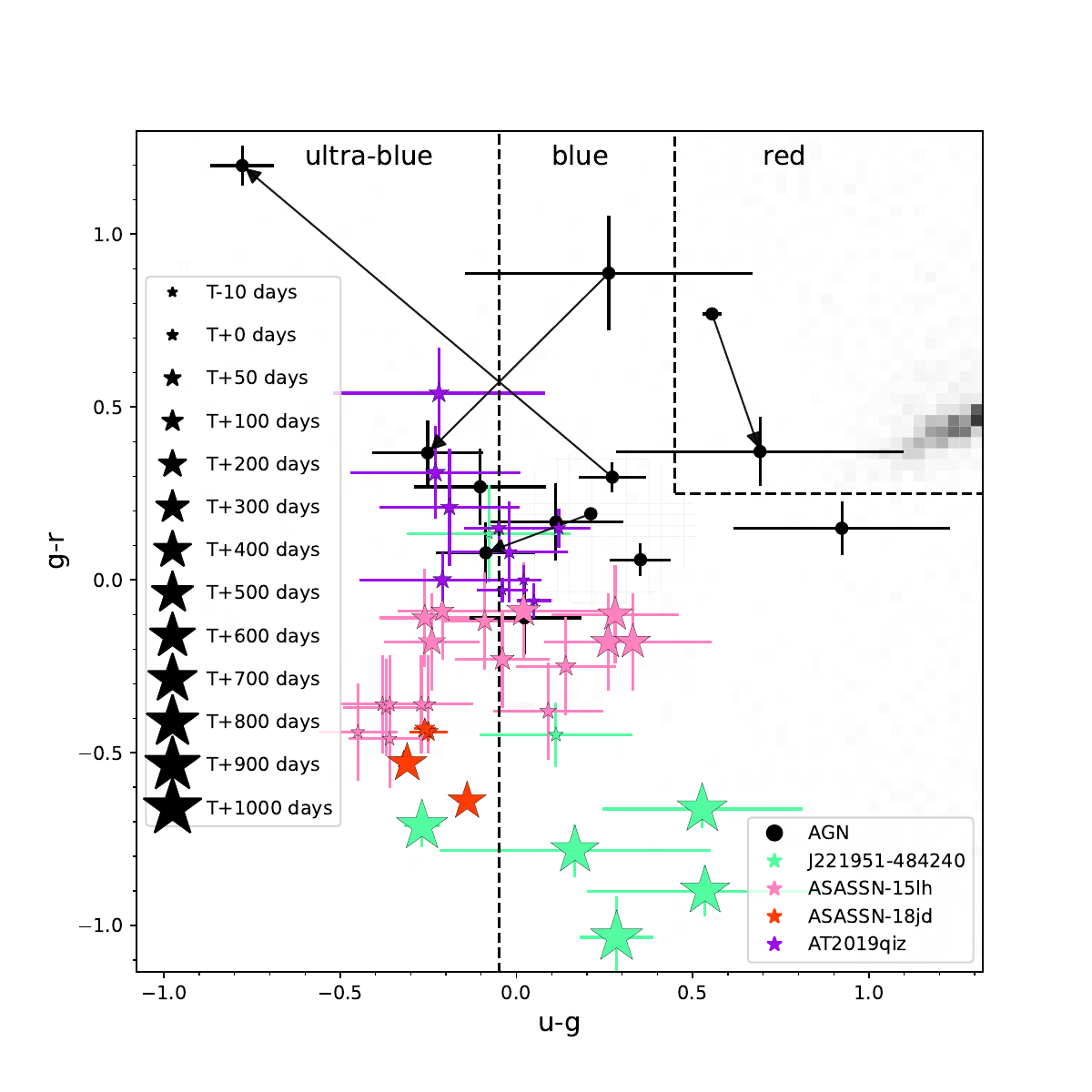}
\includegraphics[angle=0,scale=0.45, trim={0cm 0.cm 1.5cm 0.cm},clip]{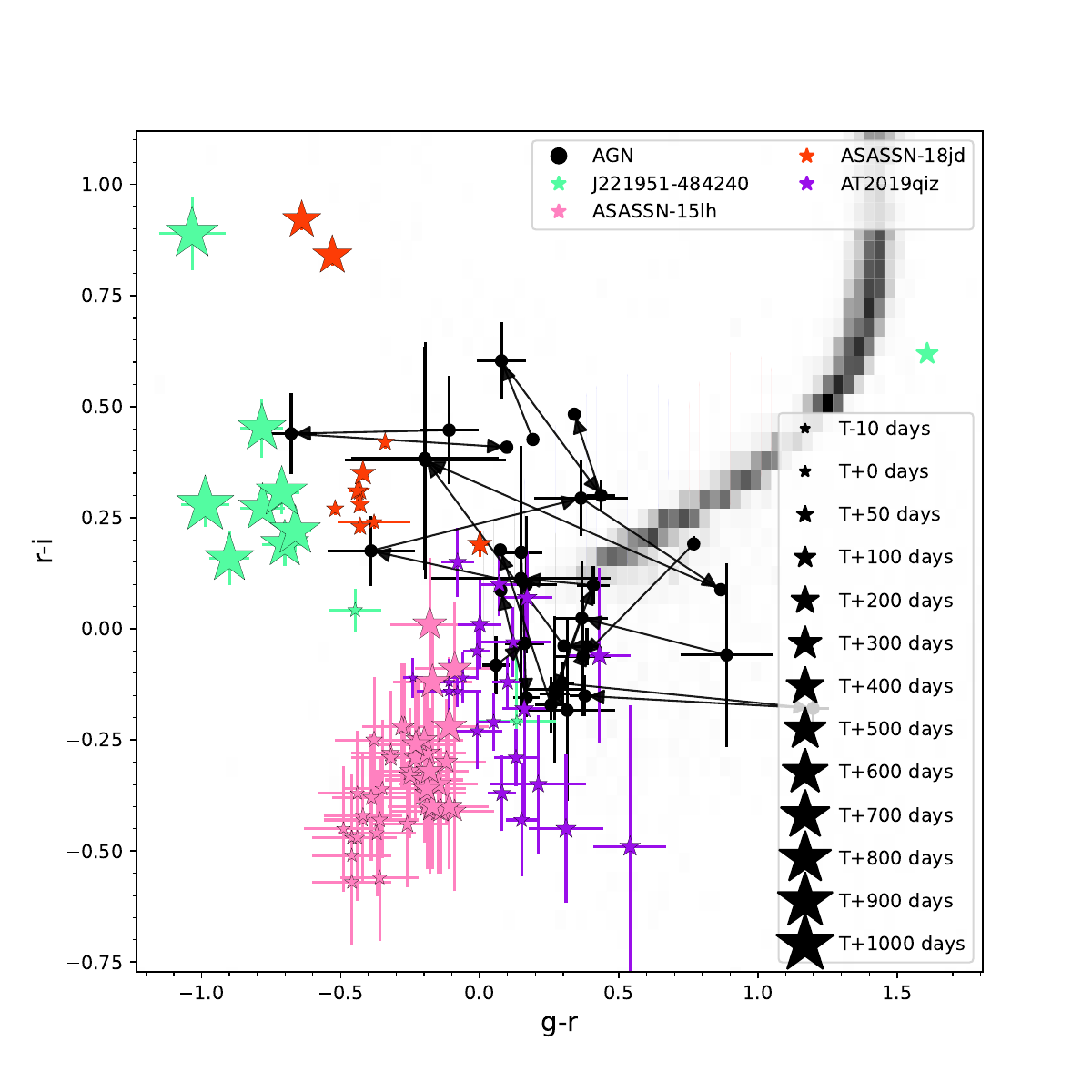}
\caption{Colour-colour diagram of the photometry of \srcname, together with ambiguous nuclear transients (ANTs) ASASSN-15lh \citep{lel16} and ASASSN-18jd \citep{neu20}, a classic tidal disruption event AT2019qiz \citep{nic20b} and the UV transient sources followed-up by {\it Swift}/UVOT during the O3 follow-up and identified as candidate AGN, see \protect\cite{oat21}. Left: $u-g$ versus $g-r$. Right: $g-r$ versus $r-i$. In both panels, the size of the markers for the black-edged data points indicates the time since peak or time since discovery in days, the key is given in the legend. For \srcname, we use the time since UVOT discovery, MJD 58756. For the AGN arrows connect points of the same source in chronological order. In the left panel, the dotted lines divide the figure into regions identifying objects as red, blue and ultra-blue, adapted from \protect\cite{law16}. In both panels, the grey region indicates the colour location of 90 per cent of SDSS spectroscopic quasars \protect\citep[see][for details]{law16}, and we display a 2-D histogram, given in grey, of the SDSS colours of 10,000 stars from the $Gaia$ DR2 catalogue, selected from a region at high Galactic latitude. In the right panel, the blue region represents the location of the blue cloud galaxies, and the red region represents the red sequence galaxies, both out to $z = 0.22$ \protect\citep[adapted from][]{law16}. The green star, without a black edge, in the right-hand panel (on the right-hand side of the figure) is the pre-outburst colour of \sourcename and represents the colour of the host galaxy. All the other $gri$ values have been host corrected. All values have been corrected for Galactic extinction. We have not corrected for host extinction. Correction for host extinction would move points down and to the left in both panels.}
\label{Optical_colours}
\end{figure*}

\subsection{Temporal Evolution}
\label{lightcurve}
In the top panel of Fig. \ref{J221951_lc}, we display the IR/optical/UV photometry of \sourcename obtained by UVOT (only the 3 UV and $u$ filters are displayed), GROND, {\it AstroSat}, {\it WISE}, J-GEM
and the Chilescope observatory \citep[the latter two from GCNs;][]{GCN:25941, GCN:25963} along with archival values obtained with {\it GALEX}, DES and {\it WISE}. Overall, the light curve shows a gradual decrease in brightness with time. In addition to the decaying nature of the light curve, there are three rebrightenings ($\sim$58766, 58843 and 59172 MJD) which show changes in magnitude of $\sim 0.5$ mag and which appear to reset the brightness level. The light curve continues to decay from the peak of the flare, rather than returning to the value expected from the extrapolation of the power-law observed pre-flare. These flaring episodes are most clearly observed in the $uvw2$ filter. The long-term light curve of \srcname, including photometry taken pre-and post-UVOT detection, is displayed in Fig.\, \ref{J221951_archival}. At peak \sourcename is brighter than archival values at all UV/optical/IR wavelengths, by more than 1-3 mags, with the largest change observed in the blue filters. There is some marginal evidence to suggest the rise of the optical/UV emission lags the IR emission, though this cannot be claimed with certainty due to the lack of optical/UV emission between 58429 and 58759 MJD. The most recent observations, around 1000 days after the initial detection, indicate \sourcename continues to be brighter than archival values in the UV through to the $r$-band, while in redder filters, \sourcename is comparable in brightness with historic values. 

The start time of \sourcename is uncertain, but constraints can be placed using individual DES, J-GEM $i$ band measurements. The last visit by DES taken in November 2018 is at a brightness consistent with historic values, suggesting \sourcename began no earlier than 10 months prior to September 2019 (see Fig. \ref{J221951_archival}). This implies the start time $T_0$ is within a $\sim330$ day window between 58429 and 58759 MJD. For comparison with other types of objects, we will take the mid-time of this range, 58594 MJD, as $T_0$.

Using $T_0$ as the start time, we fit the $uvw2$ count rate (CR) light curve, with a series of functions of increasing complexity. We initially fit a power-law ($CR=Nt^{\alpha}$, where $N$ is the normalisation, and $\alpha$ is the temporal decay index), which gives $\alpha =-1.32\pm 0.06$, although the fit is poor with $\chi^2/d.o.f=286/62$. With a broken power-law (two power-laws connected by a break at $t_{break})$, the fit traces the general underlying behaviour. The fit has improved, but the $\chi^2/d.o.f$ is still poor with $\chi^2/d.o.f=229/60$. An F-test suggests the break is required with a confidence of $3\sigma$. The broken power-law fit has parameters $\alpha_1=-0.84^{+0.17}_{-0.09}$, the break time $t_{break}=58935^{+30}_{-18}$ MJD and $\alpha_2=-1.82\pm0.15$. The poor $\chi^2/d.o.f$ is likely due to the rebrightenings, which appear to reset the brightness level. The uncertainty on the $T_0$ will also affect our estimate of the decay indices of the fits. For instance setting $T_0$ as $T_{0,GW}$ and fitting a power-law results in an $\alpha =-0.39\pm 0.01$, but with a much worse $\chi^2/d.o.f=847/62$.

In the bottom panel of Fig. \ref{J221951_lc} we display the colour evolution. The $uvw2-uvw1$ and $uvw2-u$ colours do not show strong evolution with time although there is some variation, which appears to correspond to the peaks and troughs of the light curve behaviour in the panel above. However, the error bars are large compared to the colour curves built with the ground-based data. For the $uvw2-g$, $g-r$, $r-i$ we notice there is a strong change in colour between the first two data points, becoming redder for $uvw2-g$ and $r-i$, but bluer for $g-r$. There is a gap in these colour curves, until 59420 MJD, after which the colour curves remain approximately at the same level until the end of observations, however, this level is slightly redder for $uvw2-g$ and $r-i$ and slightly bluer for $g-r$, compared to the last data point before the gap. Also shown in Fig. \ref{J221951_lc} is the archival $g-r$ colour (indicated by an arrow). Compared to historic values \sourcename has become bluer, changing in $g-r$ by -0.68 mag by the first observation post-detection. 

The {\it WISE} $W1-W2$ colour changes from $0.20\pm0.32$ pre-outburst to $0.56\pm0.28$ near the peak. The $W1-W2$ has been used to provide a rough assessment of the AGN or stellar dominance of a galaxy \citep[$W1–W2>0.8$ for AGN-like or $W1–W2<0.5$ for galaxy-like;][]{ste12,yan13}. A value of 0.20 mag is consistent with being galaxy-like and not dominated by an AGN. The change to 0.56 at peak, suggests that the surrounding dust is being heated, but $W1-W2$ is still less than that observed in AGN. 

In Fig. \ref{Optical_colours}, we compare the $u-g$, $g-r$ and $r-i$ colours determined from archival photometry and the photometry taken during the evolution of \srcname. The $u-g$ vs $g-r$ colour evolves downwards with time, while the $g-r$ vs $r-i$ colour moves towards the top left corner with time. In both instances, the strongest colour evolution is observed in the first few days after detection. Initially, the colour of the transient is similar to that of the candidate AGN from \cite{oat21} and quasars (QSOs) and it changes in colour away from these objects as it evolves with time. Note there is a substantial gap in GROND observations and so we only have colour information in these panels from observations at very early and very late times.

\subsection{Spectral Energy Distributions}
\label{SED}
We also constructed spectral energy distributions (SEDs) for 8 epochs where we have quasi-simultaneous data from facilities in addition to {\it Swift}. To each data point of each SED, which includes UVOT and ground observations we add a $5\%$ systematic error. We fit the SEDs with {\textsc XSPEC} version 12.12.0 \citep{arn96}. In these SEDs, we use host subtracted photometry, with the host values taken as the archival DES, VISTA and {\it WISE} values. The UVOT data are not host subtracted; this is reasonable given that \sourcename is $>1$ magnitude brighter than the underlying host galaxy in the redder g-band at all epochs and similarly \sourcename is brighter than the NUV GALEX limit of 23.7 by $>1$ mag at all epochs. Two of the SEDs were built using values from GCNs, $gri$ for 58758 MJD \citep{GCN:25941} and $r$ for 58760 MJD \citep{GCN:25963}. For the $gri$ filters in the SED at 58758 MJD, we assume an error of 0.1 mag since photometric errors are not provided in \cite{GCN:25941}. The remaining 6 SEDs were built using the optical/UV UVOT filters and GROND filters. For the MJD 58787 SED, we also include host subtracted {\it WISE} W1 and W2 photometry. We fit each SED with a power-law and then with a single blackbody. In both cases, we include a dust component with Galactic reddening of $E(B-V) = 0.012$ \citep{sch11}. We tested whether a host extinction component improved the fits. The host $E(B-V)$ was consistent with zero, similar to that derived from the X-shooter spectrum. The X-shooter spectrum does not display a dip at 2175 \AA~ in the rest-frame, implying that either the object has a Small Magellanic Cloud type extinction law without extra absorption at 2175 \AA\, or a very low $E(B-V)$ of the host. We, therefore, do not use a second dust component in our SED modelling. 

\begin{figure}
\includegraphics[angle=0, scale=0.48,trim={0.7cm 0 0 0.7cm},clip]{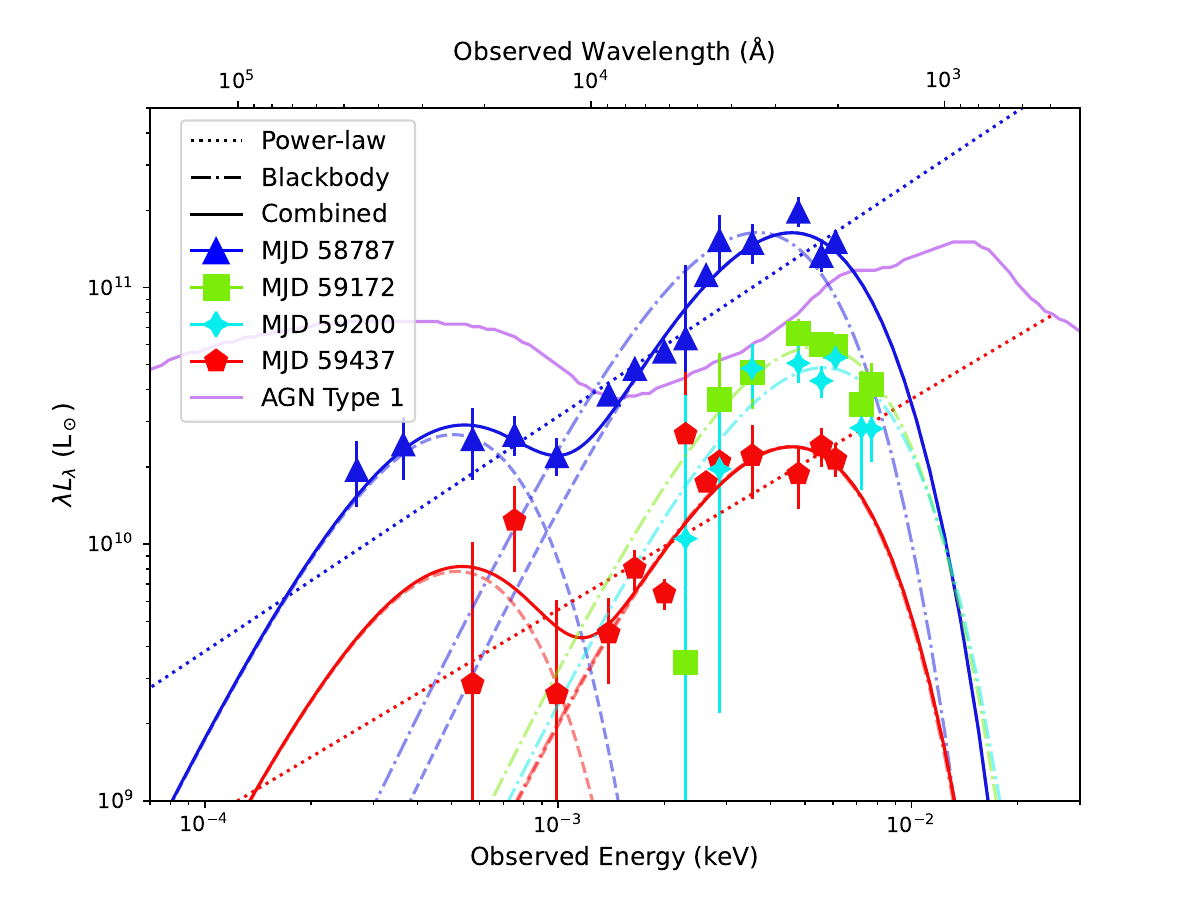}
\caption{Spectral energy distributions of \sourcename at MJD 58787 (blue), MJD 59172 (green), MJD 59200 (cyan) and MJD 59437 (red). For the MJD 58787 and MJD 59437 we overlay three models: power-law (dotted), single blackbody (dot-dashed) and two blackbodies (dashed and solid lines). For the two blackbody model we display the two single blackbodies (dashed) and the combination of these components (solid).  A two blackbody model is the best fit for the MJD 58787 SED. Visually a two blackbody appears to be the best fit for the MJD 59437 SED, although is not statistically required. The MJD 59172, MJD 59200 SEDs include {\it AstroSat} data and show the turnover of the higher temperature black body. We only overlay the best-fit blackbody for these two SEDs. The typical AGN Type 1 spectrum \citep{ric06}, divided by a factor of 10, is given in purple. Overall, the SEDs of \sourcename do not resemble the typical AGN Type 1 SED.}
\label{fig:SEDs}
\end{figure}

For the first 3 SEDs, a power-law is preferred over a single blackbody, while for the latter 5 epochs, a single blackbody is preferred. However, the $\chi^2/d.o.f$ for both the power-law and single blackbody fits for the SEDs from MJD 58787 (the third SED) onwards were poor. We therefore also tried fitting a model consisting of two blackbodies. For the MJD 58787 SED, this model is marginally preferred over the single power-law at $2.5\sigma$. For the subsequent SEDs, the two blackbody model does not provide a better fit.  We note that by these late epochs, the brightness of \sourcename in the reddest filters is comparable to the host making it difficult to constrain the second blackbody component. In Fig. \ref{fig:SEDs}, we display the SEDs of MJD 58787 and MJD 59437, together with the three models. The necessity of two blackbody components for the MJD 58787 SED is apparent. For the MJD 59437 SED, the H-band data point is well above the extrapolation of the single component blackbody model, suggesting that the second thermal component is required. The temperature of the second component decreases with time. The spectral fits are provided in Table \ref{tab:sed}, with the convention for flux density, $F\propto\nu^{-\beta}$.

We also constructed SEDs using UVOT data only for each observation performed by {\it Swift}/UVOT for which at least three filters were obtained. We fit two simple models to each UVOT SED: a power-law and a blackbody, again including in both instances a dust component with Galactic reddening. From 64 SEDs, we find a weighted average index of $\beta=0.49\pm0.04$, which is $4 \sigma$ shallower than the $\beta =2/3$ predicted for a standard thin accretion disc at UV/optical wavelengths \citep{sha73}. For the blackbody fits the average temperature is $23000\pm410\,{\rm K}$. The mean and standard deviation of the $\chi^2$/d.o.f of the power-law fits and the blackbody fits is $0.80\pm0.65$ and $0.89\pm0.62$, respectively.

\begin{table*} 
\centering 
\caption{The best-fit parameters to eight SEDs with ground-based photometry in addition to 6 filter UVOT photometry. The parameters given are: spectral index $\beta$, intrinsic blackbody temperature $T_{BB}$ and the $\chi^2/d.o.f$. 
\label{tab:sed}}
\begin{tabular}{lccccccc} 
\hline 
Time (MJD) &  Model   & $\beta$ & $T_{BB,1}$ (K) & $T_{BB,2}$ (K) & $\chi^2/d.o.f$ & Null Hypothesis\\
\hline
\hline 
58758 & Pow         & $0.19\pm0.10$ & - & -  & 5/7 &$6.1\times10^{-1}$\\
58758 & Bbody       & - & $21300\pm1000$    & -  & 7/7 & $4.3\times10^{-1}$\\
\hline
58760 & Pow         & $0.25\pm0.10$ & - & -   & 2/5 &$8.1\times10^{-1}$\\
58760 & Bbody       & - & $21100\pm1000$    & -  & 7/5 & $1.9\times10^{-1}$\\
\hline
58787 & Pow         & $0.03\pm0.06$ & - & -   & 35/11 &$2.2\times10^{-4}$\\
58787 & Bbody       & - & $20200\pm700$    & -  & 38/11 & $6.8\times10^{-5}$\\
58787 & Bbody Bbody &  -  & $20800\pm700$ & $2800\pm400$ & 13/9 & $1.6\times10^{-1}$\\
\hline
59437 & Pow         &  $0.12\pm0.12$ & - & - & 29/11 & $2.2\times10^{-3}$ \\ 
59437 & Bbody       & - & $20500\pm1100$    & -  & 18/11 & $8.7\times10^{-2}$\\
59437 & Bbody Bbody &  -  & $20700\pm1200$ & $2300\pm500$& 15/9 & $1.0\times10^{-1}$\\
\hline
59470 & Pow         &  $0.15\pm0.14$ & - & - & 51/9 & $6.4\times10^{-8}$ \\ 
59470 & Bbody       & - & $19400\pm1300$    & -  & 36/9 & $4.4\times10^{-5}$\\
59470 & Bbody Bbody &  -  & $19500\pm1300$ & $1200\pm100$ & 28/7 & $2.2\times10^{-4}$\\
\hline
59497 & Pow         &  $0.33\pm0.11$ & - & - & 46/8 & $1.9\times10^{-7}$ \\ 
59497 & Bbody       & - & $18300\pm1000$    & -  & 22/8 & $5.2\times10^{-3}$\\
59497 & Bbody Bbody &  -  & $18300\pm1000$ & $580\pm 50$ & 20/6 & $2.8\times10^{-3}$\\
\hline
59528 & Pow         &  $0.12\pm0.13$ & - & - & 46/9 & $6.9\times10^{-7}$ \\ 
59528 & Bbody       & - & $18900\pm1200$    & -  & 21/9 & $1.1\times10^{-2}$\\
\hline
59558 & Pow         &  $0.12\pm0.13$ & - & - & 46/9 & $6.9\times10^{-7}$ \\ 
59558 & Bbody       & - & $18900\pm1200$    & -  & 21/9 & $1.1\times10^{-2}$\\
\hline 
\end{tabular} 
\end{table*}

\subsection{Optical to X-ray flux ratio, $\alpha_{OX}$}
For both TDEs and AGN, the UV to X-ray spectral slope, $\alpha_{OX}$ \citep{tan79,wev20}, can be measured as: 
\vfil
\begin{equation}
        \alpha_{OX} = -\frac{\log(f_{\nu,X}/f_{\nu,O})}{\log(\nu_X/\nu_O)}
\end{equation}
where $f_{\nu,X}$ and $f_{\nu,O}$ are the X-ray and optical flux densities at rest-frame $2\,{\rm keV}$ and $2500$ \AA, respectively, and $\nu_X$ and $\nu_O$ are the X-ray and optical rest-frame frequencies at $2\,{\rm keV}$ and $2500$ \AA, respectively. Using the extinction-corrected observed $u$-band flux ($\lambda_{central} =3501$ \AA), at peak brightness, as a proxy for $f_{\nu,O}$ at rest-frame $2500$ \AA~and the X-ray unabsorbed flux limit from an individual XRT visit, scaled to rest-frame 2\,keV, assuming a photon index of $\Gamma =$\,1.7, we obtain a value of $\alpha_{OX}\gtrsim 1.6$ at 58788 MJD.

\subsection{Bolometric Light curve}
\label{bol}
\begin{figure}
    \centering
    \includegraphics[angle=0, scale=0.43]{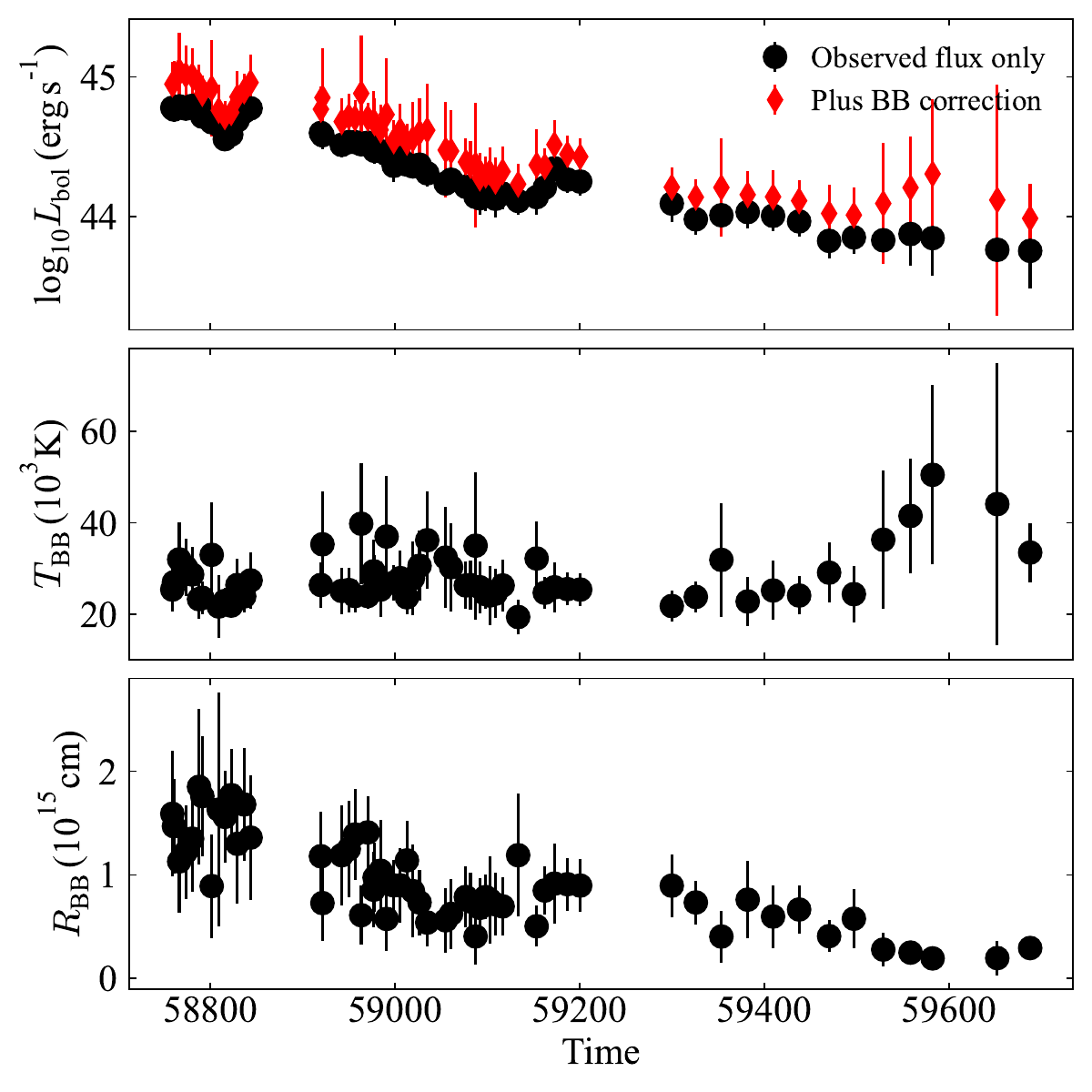}
    \caption{Top: the bolometric light curve of \sourcename derived from the UVOT photometry.
Middle: temperature evolution. Bottom: evolution of the blackbody radius. The luminosity and blackbody radius evolve similarly, while the temperature is approximately constant. Time is given in MJD.}
    \label{bolometric}
\end{figure}

We construct the bolometric light curve of \sourcename from the {\it Swift}/UVOT data using {\sc superbol} \citep{nic18}. The method allows us to integrate under the SED inferred from the multi-colour data at each epoch, and fit a blackbody function to estimate the temperature, radius, and missing energy outside of the observed wavelength range. For TDEs, a blackbody is an excellent approximation of the near-UV and optical emission \citep[e.g.][]{van21}. However, we note that the radius is computed under the assumption of spherical symmetry, which may not reflect the potentially complex geometry in TDEs. We include Galactic extinction but do not correct for the host extinction, which for this source is likely to be negligible (see \S\ref{SED}). The bolometric light curve is plotted in Fig. \ref{bolometric}. In the middle and bottom panels of Fig. \ref{bolometric}, we display the effective temperature and radius of \srcname. 
The maximum luminosity is $L_{max} = (1.1\pm0.7)\times10^{45}\,{\rm erg}\,{\rm s}^{-1} = (2.9\pm1.1)\times10^{11} L_\odot$. The bumps seen in the photometry are also seen in the luminosity evolution, although the increase in the size of the errors makes these features less apparent. The effective temperature of \sourcename has remained roughly constant throughout with an average temperature of $T=2.8\times 10^4\,{\rm K}$, with a typical error for each inferred temperature of $\sim7000\,{\rm K}$. This is consistent with the values determined in \S \ref{SED}. The blackbody radius evolves in a similar fashion as the luminosity of \srcname. Trapezoidal integration of the blackbody luminosity over the span of \textit{Swift} observations in rest frame days gives a total emitted energy of $E=2.6\times10^{52}\,{\rm erg}$, a lower limit since we miss the peak of the light curve. This corresponds to a lower limit on the accreted mass of $M_{acc}=0.14 M_\odot$, for an accretion efficiency of $\eta = 0.1$.

\begin{figure*}
    \centering
    \includegraphics[angle=0, scale=0.7, trim={1.5cm 0.0cm 1.cm 1.8cm},clip]{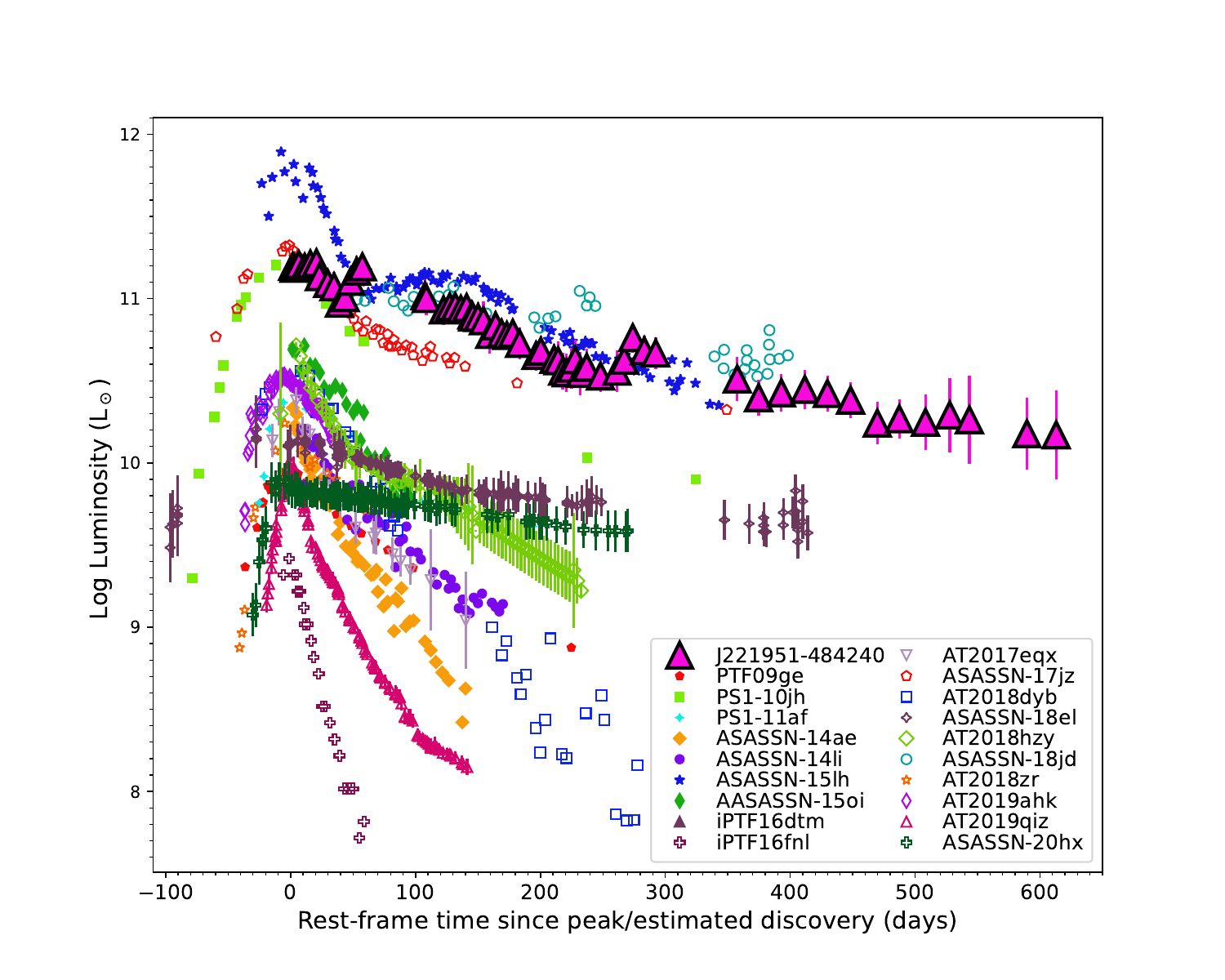}
    \caption{The bolometric light curve of \srcname, with peak time taken as the UVOT discovery date, MJD 58756, together with a sample of tidal disruption events: PTF09ge \citep{arc14}; PS1-10jh \citep{gez12}; PS1-11af \citep{cho14}; ASASSN-14ae \citep{hol14}; ASASSN-14li \citep{hol16b}; ASASSN-15oi \citep{hol16}; iPTF16fnl \citep{bla17b}; AT2017eqx \citep{nic19}; AT2018dyb \citep{lel19}; AT2018hyz \citep{sho20,gom20b}; AT2018zr \citep{van19b}; AT2019ahk \citep{hol19b}; AT2019qiz \citep{nic20b} and ambiguous nuclear transients: ASASSN-15lh \citep{don16,lel16,god17,bro14}, ASASSN-18jd \citep{neu20}, ASASSN-17jz \citep{hol21}, ASASSN-18el \citep{tra19b}, PS16dtm \citep{bla17}, ASASSN-20hx \citep{hin22b}. \sourcename is comparable in luminosity to ASASSN-15lh and ASASSN-18jd.}
    \label{luminosiy_comparison}
\end{figure*}

In Fig. \ref{luminosiy_comparison}, we compare the luminosity of \sourcename with a sample of TDEs. \sourcename is more luminous than the bulk population of TDEs and is on par with the luminosity of ASASSN-15lh, ASASSN-17jz and ASASSN-18jd \citep{don16, hol21, neu20}, which are luminous members of the population of ambiguous nuclear transients \cite[ANTs;][]{hol21,mar17}. ANTs are transients for which it is not clear if they are TDEs or related to AGN activity, with properties consistent with both classes. The temporal evolution of \sourcename most closely resembles ASASSN-15lh after its initial peak. Other objects showing similar slowly declining light curve behaviour are the ANTs ASASSN-18el \citep{tra19b,ric20,lah22,hin22a} and ASASSN-20hx \citep{hin22b}, though they are overall an order of magnitude fainter. \sourcename continues to be detected and, whether it is a TDE or an ANT, it is one of the longest observed to date \citep[see also][]{van19}.

\subsection{TDE model fit}
\label{TDE_model}
If we assume J221951 is a TDE, we can derive physical parameters from our multiband light curves using the Modular Open Source Fitter for Transients \citep[MOSFIT;][]{gui18} with the TDE model from \citet{moc19}. This model assumes a mass fallback rate derived from simulated disruptions of polytropic stars by a super massive black hole (SMBH) of $10^6 M_\odot$ \citep{gui14}, and uses scaling relations and interpolations for a range of black hole masses, star masses, and impact parameters. The free parameters of the model, as defined by \cite{moc19}, are the masses of the black hole, $M_{BH}$, and star, $M_\ast$; the scaled impact parameter $b$; the efficiency $\eta$ of converting accreted mass to energy; the normalisation and power-law index, $R_{ph,0}$ and $l_{ph}$, connecting the radius to the instantaneous luminosity; the viscous delay time $T_\nu$ (the time taken for matter to circularise and/or move through the accretion disc) which acts approximately as a low pass filter on the light curve; the time of first fallback, $t_0$, which is equivalent to our $T_0$, assuming the TDE model is correct; the extinction, proportional to the hydrogen column density $N_H$ in the host galaxy; and a white noise parameter, $\sigma$. The priors follow those used by \cite{nic22}, and reflect the range of SMBH masses where optically-bright TDEs are expected \citep[e.g.][]{van18}, the range of impact parameters covering both full and partial disruptions, accretion efficiencies for non-rotating to maximally-rotating black holes, and a broad range of possible photospheres and viscous timescales \citep[see][or details]{moc19}. In addition, we use the time of the last DES i-band archival observation as a lower limit on the $T_0$ prior.

\begin{figure}
    \centering
    \includegraphics[angle=0, scale=0.42]{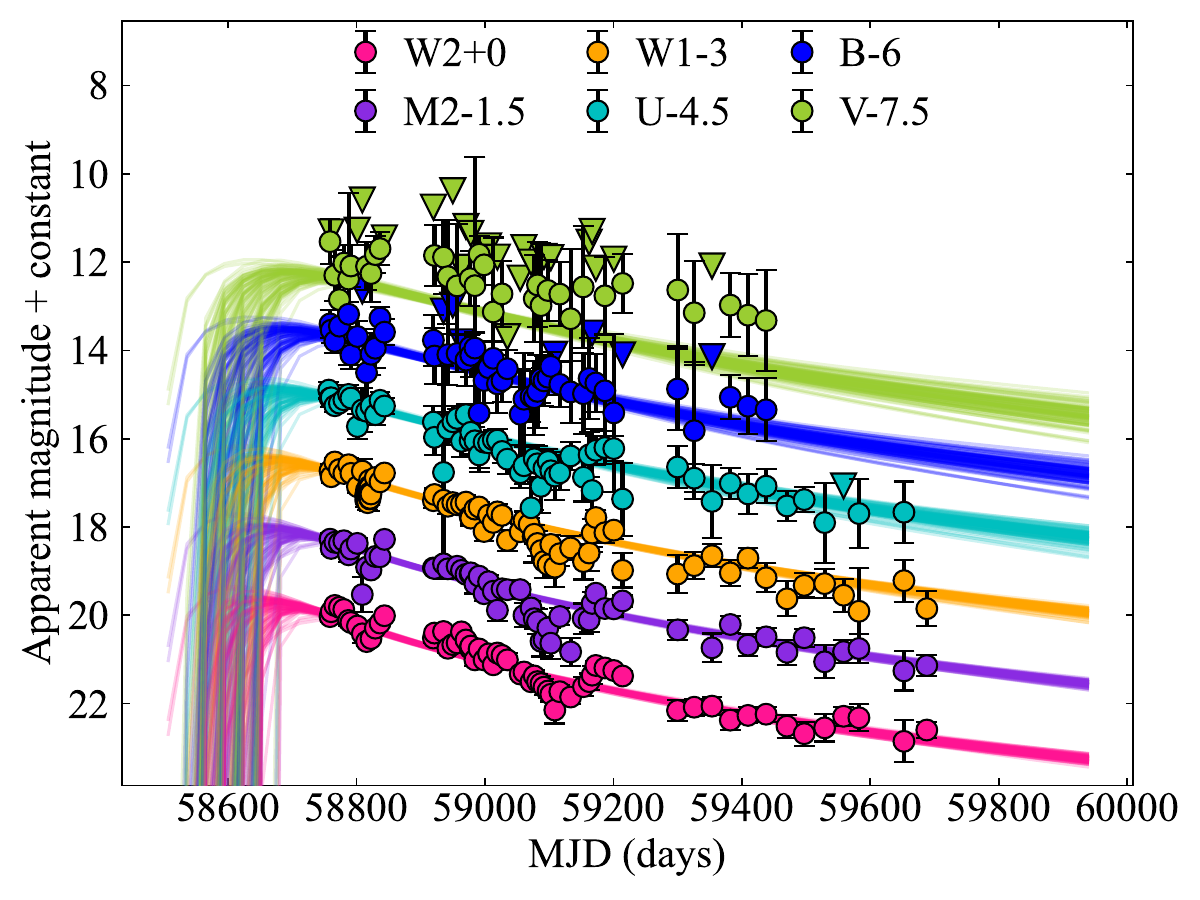}
    \caption{Fits to the multicolour light curve using the TDE model in MOSFIT \citep{gui18,moc19}. We plot the median and 16th-84th percentiles of the light curve posterior distribution from 100 realisations of the Markov Chain.}
    \label{TDE_fit}
\end{figure}

The fits are applied using the python package, {\sc Dynasty} \citep{spe20}, which implements dynamic nested sampling methods to evaluate the posterior distributions of the model parameters. We plot the median and 16th-84th percentiles of the light curve posterior distribution from 100 realisations of the Markov Chain in Fig. \ref{TDE_fit}. The model provides a good fit to the optical/UV bands but is unable to reproduce the undulations that are most clearly observed in the UV bands. \cite{moc19} also found for ASSASN-14li and ASASSN-15oi, that the observed photometry deviates from the decline of the model light curve. They suggest that for these TDEs additional late-time components may contribute, which are not modelled by MOSFIT. From the fit to \sourcename we derive the posterior probability distributions of the parameters, listed in Table \ref{tab:mosfit}, with two-dimensional posteriors plotted in Supplementary Figure S.1. With the light curve observed to decay from the start of observations, the start time, $T_0$, inferred with MOSFIT is MJD $58579.85^{+33.70}_{-31.41}$. The physical parameters point to the disruption of a $\sim 0.6{\rm M_\odot}$ star by a black hole of mass $\log (M_{BH}/M_\odot)= \,7.12$. The scaled impact parameter, $b = 0.63^{+0.08}_{-0.05}$, corresponds to a median physical impact parameter $\beta = R_t /R_p = 1.04$, where $R_t$ is the tidal radius and $R_p$ is the orbital pericentre. For the inferred SMBH mass, $R_t = 3.4R_{S}$, where $R_{S}$ is the Schwarzschild radius, which is equivalent to $4.3\times 10^{-6}\,{\rm pc}$. Using the remnant mass versus $\beta$ curve from \citet[][their Fig. 4]{ryu20} for a 0.5-0.7 $M_\odot$ star, up to $\sim 25$ per cent of the star could have survived this encounter. 

Comparing these parameters to those derived in the same fashion for a sample of 32 TDEs \citep{nic22}, we find the black hole mass at the high end, star mass in the normal range and that the impact parameter is more consistent with the TDE-H spectroscopic class than the TDE-H+He\footnote{TDEs may be divided into sub-classes spectroscopically \citep{van21}. These classes are: TDE-H: TDEs with H I lines, TDE-He: TDEs with He II lines only, and TDE-H+He: TDEs with a mixture of H I, He II and N III lines.}. Compared to other TDEs from \citet{moc19} and \citet{nic22}, the $T_\nu$ posterior is flat up to relatively high values, $\sim10\,{\rm days}$, though there is similar posterior support for lower values $\lesssim$ days. This broad distribution is likely due to the lack of constraints on the rise time of this source, together with the slow decay rate of the light curve. Viscous delays would broaden the light curve (relative to the fallback rate) around the peak, and the wide range of possible rise times allows for a wider range of viscous reprocessing than in faster-evolving TDEs with early data.

\begin{table} 
\centering 
\caption{Priors and marginalized posteriors for the MOSFIT TDE model. Priors are flat within the stated ranges, except for $M_\ast$, which uses a Kroupa initial mass function. The quoted results are the median of each distribution, and the error bars are the 16th and 84th percentiles. These errors are purely statistical; \protect\cite{moc19} provide estimates of the systematic uncertainty. $t_0$ is observer frame days before the first detection.
\label{tab:mosfit}}
\begin{tabular}{lccc} 
\hline 
Parameter   & Prior         &   Posterior           &  Units \\
\hline
\hline 
$\log (M_{BH}/M_\odot)$    &    [5,8]        &  $7.12^{+0.12}_{-0.09}$  &  \\
$M\ast$         &    [0.001,100]  &  $0.62^{+0.41}_{-0.16}$  &  $M_\odot$ \\
$b$               &    [0,2]        &  $0.63^{+0.08}_{-0.05}$&            \\
log $\epsilon$  &    [-2.3, -0.4] &  $-0.60^{+0.13}_{-0.22}$ &            \\
log $R_{ph,0}$  &    [-4,4]       &  $0.43\pm0.07 $          &            \\
$l_{ph}$        &    [0,4]        &  $0.63^{+0.09}_{-0.08}$  &            \\
log $T_\nu$     &    [-3,3]       &  $-0.20^{+1.01}_{-1.30}$ &  d         \\
$t_0$           &    [-500,0]      &  $-177^{+34}_{-31}$    &  d         \\
log $N_{H,host}$&    [19,23]      &  $18.79^{+1.01}_{-1.42}$ & cm$^{-2}$  \\
log $\sigma$    &    [-4,2]       &  $-0.76^{+0.03}_{-0.04}$ &            \\
\hline 
\hline 
\end{tabular} 
\end{table}

\subsection{Spectroscopic Analysis}
\label{optical_spectra}
\begin{figure*}
    \centering
    \includegraphics[angle=0, scale=0.6]{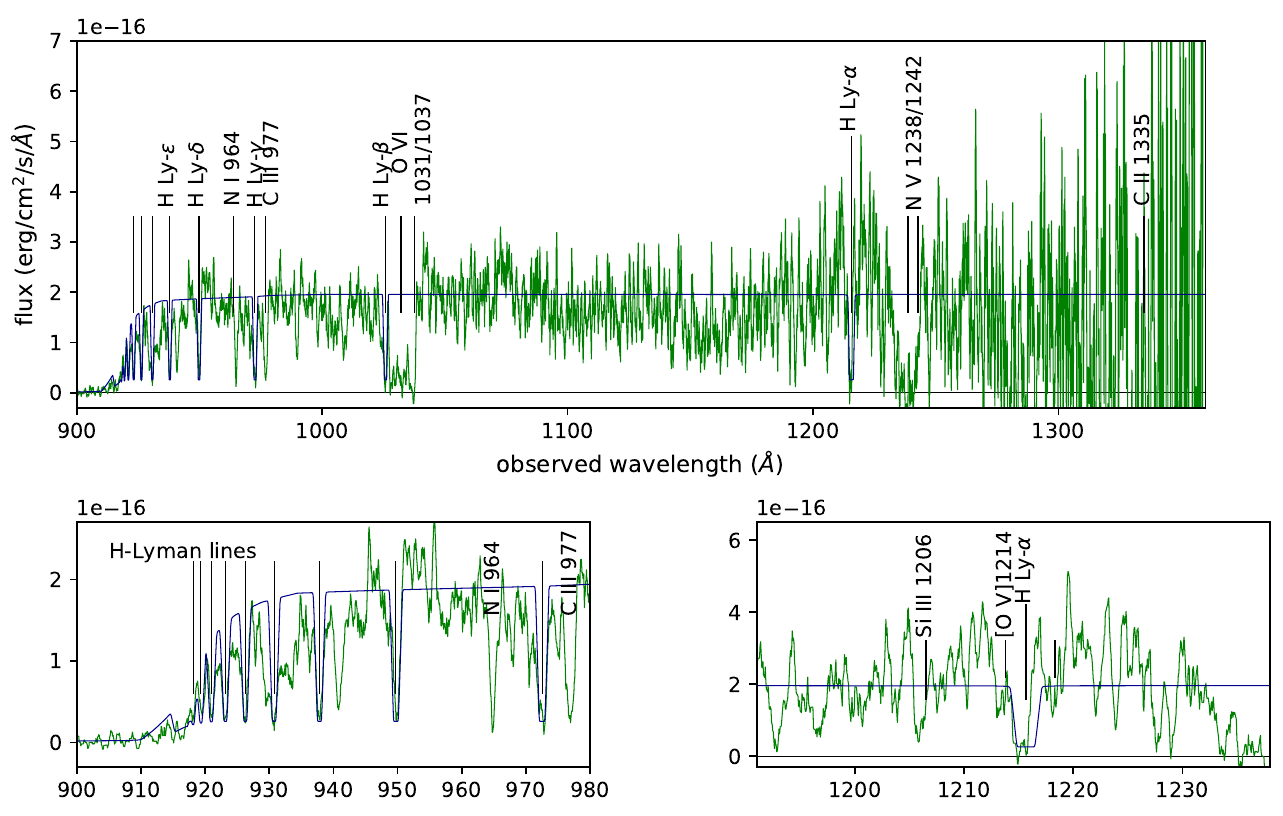}
    \caption{The {\it HST} COS spectrum between 900-1360\AA. The blue line is a Voigt profile fit to all the Lyman lines for \nh=$1 \times 10^{18}$ cm$^{-2}$ and a line-broadening velocity of 90~km s$^{-1}$ The bottom figures show details. 
    An underlying free-free emission component has been adopted. 
    Left: the Lyman series can be seen up to quite high numbers.   Right: The Ly-$\alpha$ absorption is all confined to the blue wing, the red wing shows weak emission. A blue shifted absorption from Ly-$\alpha$ at higher velocity of 650~km\,s$^{-1}$ seems to be present blue-ward of the [O V] 1213.8 \AA~emission line. There are similar high-velocity features in Ly-$\beta$ and Ly-$\gamma$ }
    \label{cosspectrum}
\end{figure*}

From the {\it HST} UV spectrum, given in Fig. \ref{cosspectrum}, we are able to determine the redshift of \srcname. The UV spectrum shows the cutoff caused by the Lyman limit absorption and, exceptionally, also the higher level Lyman lines, up to Ly$-11$, giving a redshift of $z=0.5205\pm0.0003$. We use the higher level Lyman lines for the redshift determination since they are cleaner (not saturated or blended with other lines) than Ly-$\alpha$ and Ly-$\beta$. A P-Cygni profile is observed in the Ly-$\alpha$ line core, indicative of an outflow. Ly-$\beta$ is blended with the absorption edge of the nearby O~VI resonance doublet. 
A fit with Voigt profiles of the Lyman lines from Ly-$\alpha$ up to Ly$-11$ shows that the Lyman lines are broadened by velocities of $\approx90~{\rm km\,s^{-1}}$ with natural line broadening from interstellar hydrogen in the host galaxy \nh$\approx 1\times 10^{18}\,{\rm cm^{-2}}$. For the fit an additional emission at a level of  ${\rm 2.5 10^{-17} erg\,cm^{-2}\,s^{-1}\,\AA^{-1}}$ has been adopted and velocity broadening for the Lyman continuum of {\rm $1800\,{\rm km\,s^{-1}}$}. 

\begin{figure}
    \centering
    \includegraphics[angle=0, scale=0.65,trim={1.0cm 0 0 0.7cm},clip]{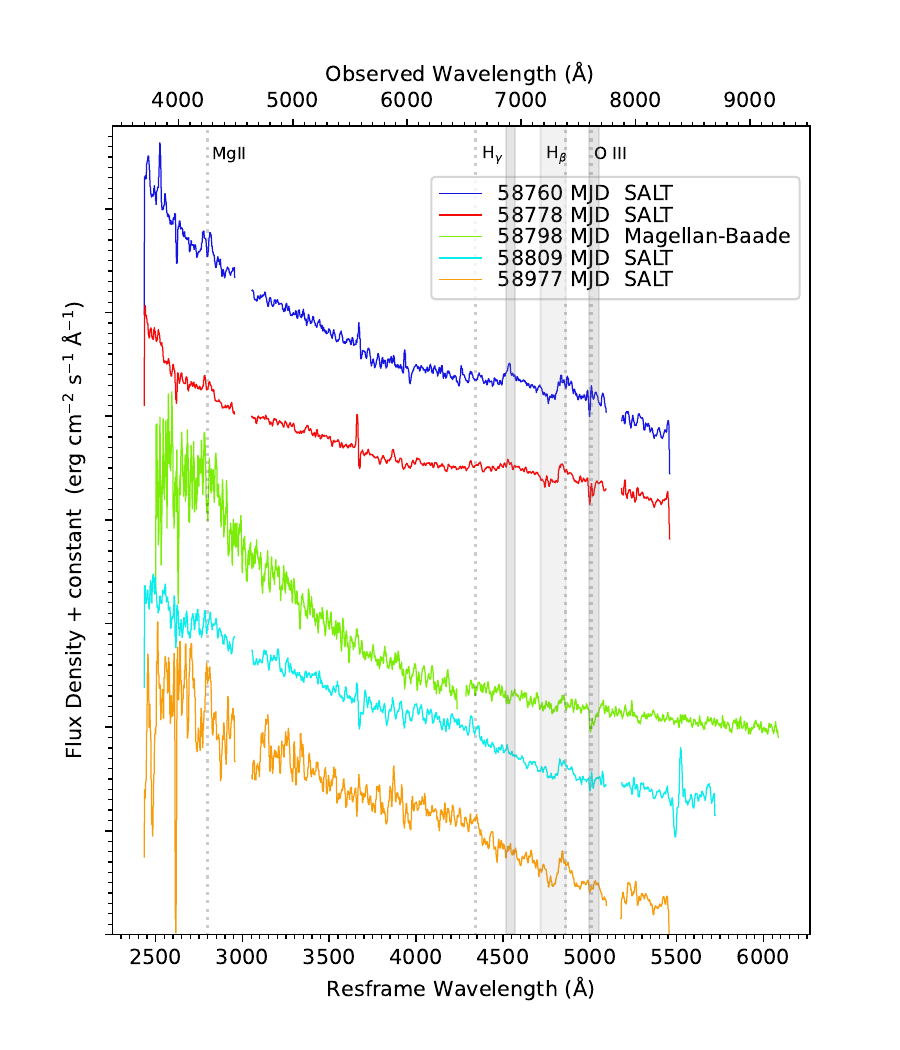}
    \caption{SALT and LCO-Magellan-Baade optical spectra of \srcname. The spectra have been smoothed using a 1D Box filter kernel, with a kernel width of 5 for the SALT spectra, and a kernel width of 10 for the LCO-Magellan-Baade spectrum. The LCO-Magellan-Baade spectrum is markedly different in shape compared to the SALT spectra. The $B$ band flux from the acquisition image matches well, suggesting the flux calibration at the $B$ band is correct, however, we cannot rule out the shape difference as being due to calibration issues because we lack photometric measurements at multiple wavelengths taken at the same time. The grey bands show the telluric bands. The darker grey bands absorb more strongly than the light grey band. The SALT spectra have a telluric correction applied.}
    \label{spectral_evolution}
\end{figure}

\begin{figure*}
\includegraphics[angle=0, scale=0.68]{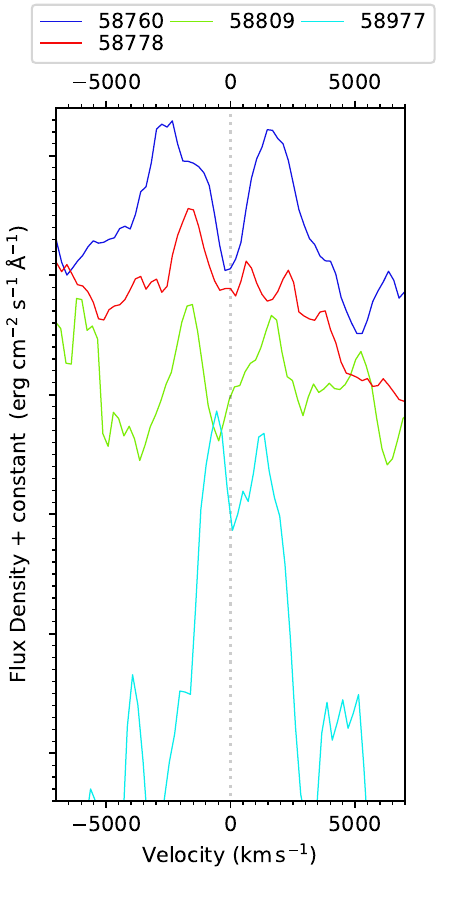}
\includegraphics[angle=0, scale=0.68]{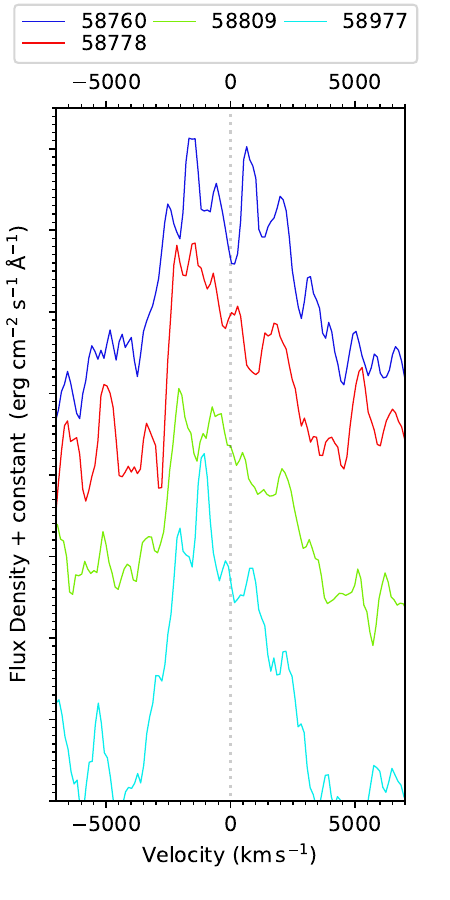}
\caption{Time evolution, in MJD, of the Mg II (left) and H$\beta$ (right) line profiles.}
\label{MgII+Hbeta}
\end{figure*}

We also see absorption lines of S~VI at 933.4, 944.5 \AA, N~V at 1238.8, 1242.8 \AA\ and O~VI at 1031.9, 1037.6 \AA. The red wings of the 1242.8 \AA\ and 1037.6 \AA\ lines match and we see extended absorption in the blue, consistent with a velocity of $-1800~{\rm km\,s^{-1}}$ in the absorption of O\,VI and $-1750~{\rm km\,s^{-1}}$ in the N\,V lines, with an accuracy of about $50~{\rm km\,s^{-1}}$. The absorption in these high ionisation lines extends to much larger velocities than we see in Ly-$\alpha$. Notably, the emission on the red part of the lines is small, suggesting no simple spherical geometry for the emitting region.

In addition to the Ly-$\alpha$ core P-Cygni profile which extends over only a $400~{\rm km\,s^{-1}}$/1.6 \AA~ range, a broader emission feature, which is likely also Ly-$\alpha$, surrounds that profile, best seen in Fig.~\ref{FUV_spec_comp}. The emission extends from $-2000$ to $2000~{\rm km\,s^{-1}}$ in the form of many peaks, or a broad emission that is cut through by many absorption lines, see Fig.~\ref{cosspectrum}. 

Apart from the previously mentioned lines, we find absorption lines of N\,I 964.1 \AA, C\,III 977.0 \AA. An emission feature at 1073 \AA~ might be a blend of Mg\,I 1073.5 \AA\ with Si\,IV 1072.96 \AA; increasing noise makes further identifications in the COS spectrum unreliable. Together these lines indicate various stages of ionisation. Emission features are seen, but cannot be identified as being either from host or Galaxy in origin. A prominent absorption line at a rest wavelength of 940.6 \AA\ cannot be identified.

The optical spectra, except for the X-shooter spectrum, are presented in Fig. \ref{spectral_evolution}. The continuum changes colour with time, in general becoming bluer. However, the Magellan spectrum observed on 2019 November 11 is considerably bluer than all other spectra but coincides with a minimum in the UV light curves. We checked if there were any issues with the acquisition or reduction of this spectrum by overlaying the blue band flux measurement derived from the $B$ band magnitude ($B=20.0\pm 0.1$\, mag) of the IMACS acquisition images, taken immediately beforehand. The $B$ band flux from the acquisition image matches well, suggesting the flux calibration at the $B$ band is correct, however, we cannot rule out the shape difference as being due to calibration issues because we lack photometric measurements at multiple wavelengths taken at the same time. 

The evolution of the Mg\,II 2800\,\AA\ and H$\beta$ line profiles is shown in Fig. \ref{MgII+Hbeta}. Mg\,II is observed to be initially double peaked, with the peaks narrowing over time and moving closer together, while the H$\beta$ shows a reduction in the emission of its red wing with time. A possible explanation for the double peaked Mg\,II is emission from a bipolar source or an accretion or debris disc. However, this does not explain why the H$\beta$ profile is different to that of Mg\,II. An alternative possibility is that it is a broad Mg \,II emission line with central Mg \,II absorption from the interstellar medium. If we examine Mg\,II and H$\beta$ at the time of the {\it HST} observation, we see the width is consistent with the width of the broad Ly$\alpha$ profile observed in the {\it HST} spectrum. This suggests that all three are broad emission lines with narrow absorption from the interstellar medium superimposed.

At the location of [O\,III] 5007 \AA~line, the spectra are affected by telluric absorption. Since this [O\,III] line is important in identifying the presence of an AGN, we used a SALT spectrum of Feige\,110, observed at a similar airmass as the SALT spectra of \srcname, to derive a telluric correction. In the telluric corrected spectra of \srcname, we do not see evidence for strong [O\,III] 5007 \AA~emission. Using the 58760 MJD spectrum, we measure a $3\sigma$ upper limit of $4.3\times 10^{-17} {\rm erg}\, {\rm cm^{-2}}\, {\rm s^{-1}}$ on the flux of the [O\,III] 5007 \AA~emission line. Dividing this by the continuum flux $3.1\times 10^{-17} {\rm erg}\, {\rm cm^{-2}}\, {\rm s^{-1}} {\rm \AA^{-1}}$ at 4861 \AA, following the method of \cite{bor92}, we find an equivalent width for the [O\,III] 5007 \AA~emission line of $<1.4$ \AA.

\begin{figure*}
    \centering
    \includegraphics[angle=0, scale=0.6]{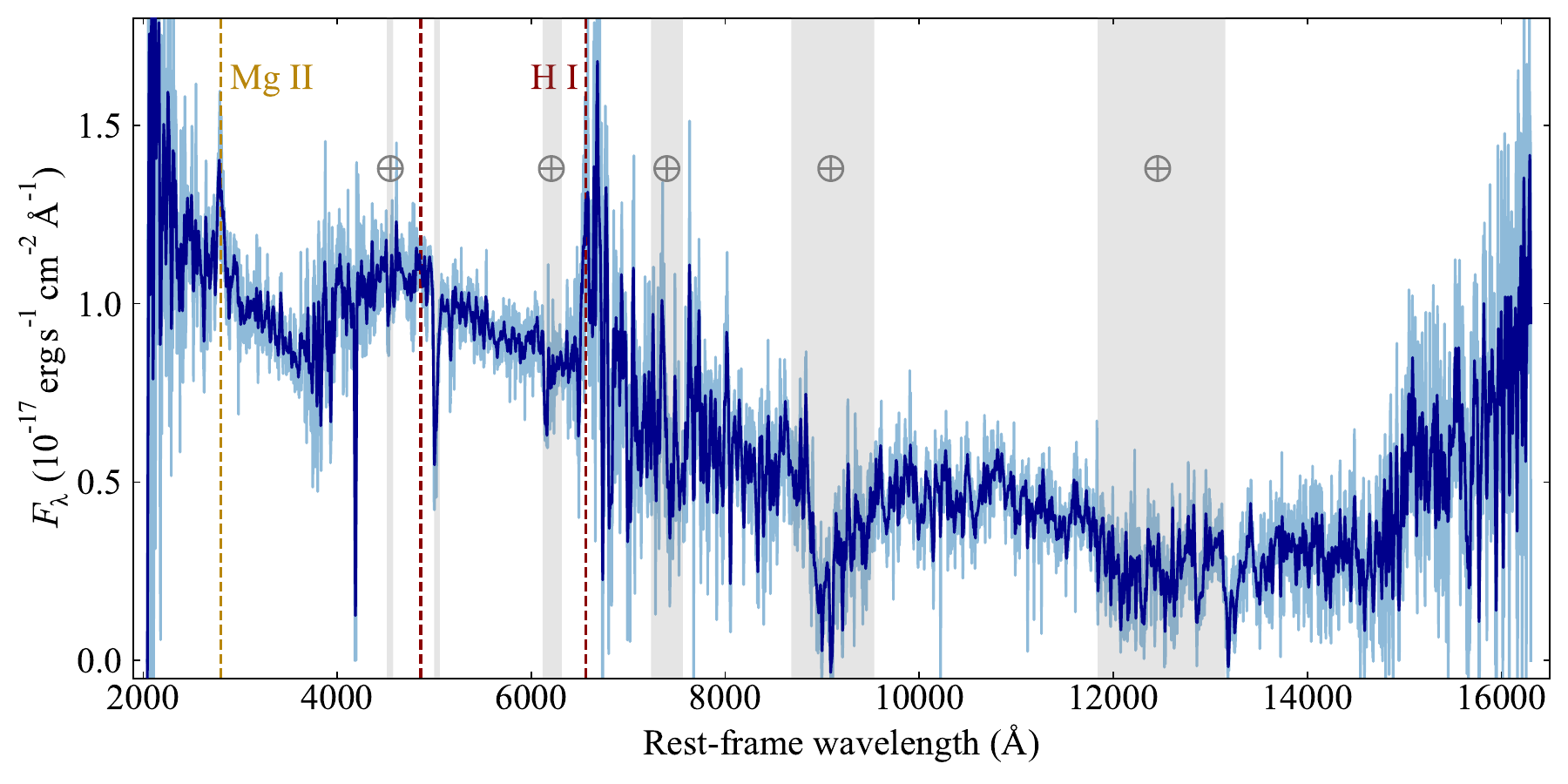}
    \caption{X-shooter spectrum taken on 59411 MJD. The spectrum has been binned using a $\rm 10\,\AA$ median filter (light blue) and then smoothed with a Savitsky-Golay filter (dark blue). ${\rm H}\alpha$ 6565 $\rm \AA$ and Mg II 2800 $\rm \AA$ emission lines have been labeled. The spectrum turns up at the IR end, consistent with a second thermal component observed in the SEDs, e.g Fig. \ref{fig:SEDs}.}
    \label{xshooter_spec}
\end{figure*}

\begin{figure}
    \centering
    \includegraphics[angle=0, scale=0.25, trim={0.5cm 0.5cm 0 0.7cm},clip]{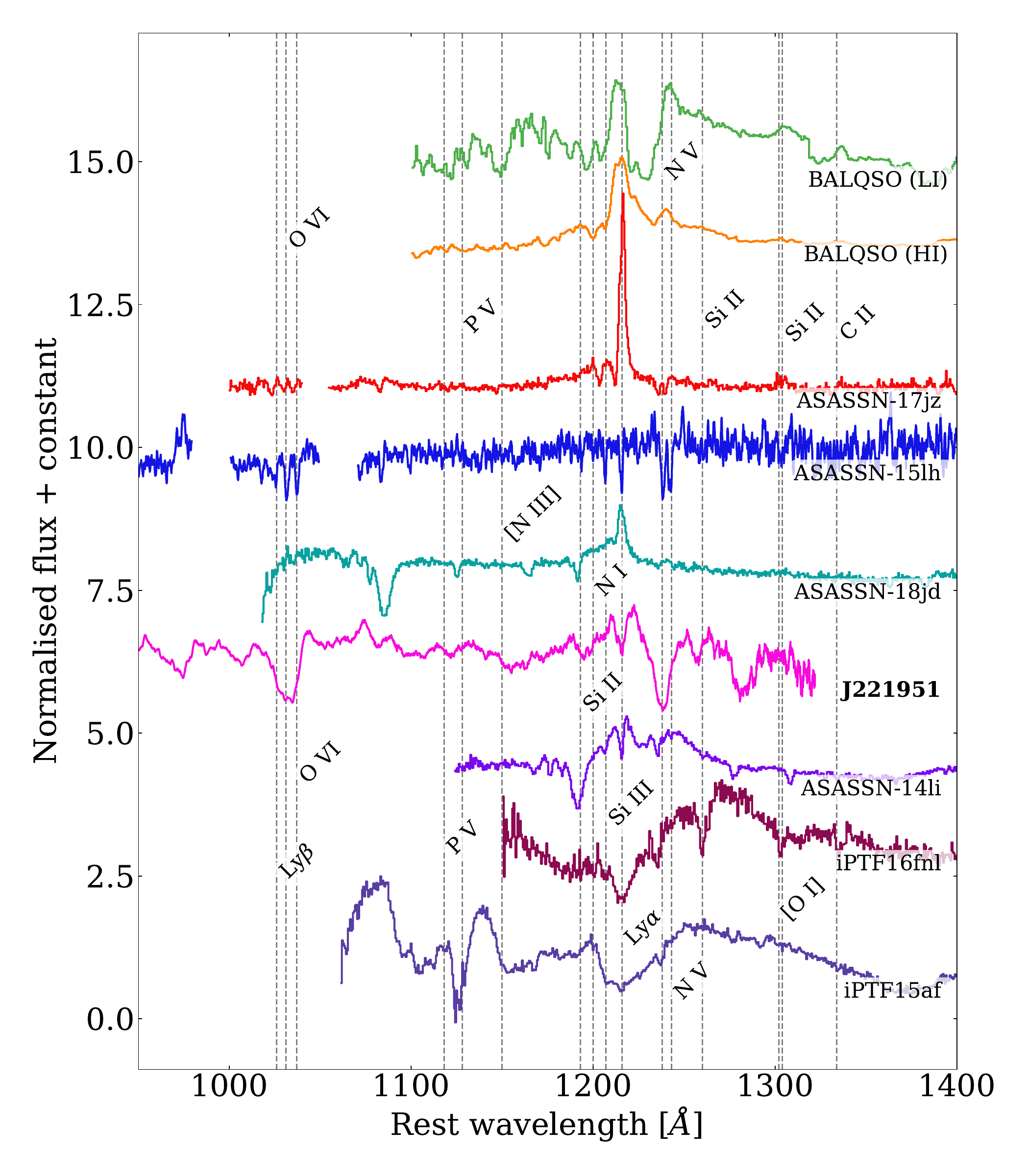}
    \caption{Comparison of the {\it HST} COS FUV spectrum of \srcname, taken on 58977 MJD, with FUV spectra of other transient objects: ASASSN-14li \citep{hol16b}, ASASSN-15lh \citep{bro16}, iPTF15af \citep{bla19}, iPTF16fnl \citep{bro18}, ASASSN-18jd \citep{neu20}, BALQSO low and high ionizaions \citep[LI and HI;][]{bro01}. \sourcename most closely resembles the UV spectra of the TDEs: ASASSN-14li, iPTF15af, iPTF16fnl and the low ionization 
    BALQSO, which all have broad absorption features.}
    \label{FUV_spec_comp}
\end{figure}

The X-shooter spectrum, taken at a much later time, is given in Fig. \ref{xshooter_spec}. The spectrum shows a broad H$\alpha$ emission line, though the S/N is low because it falls at the edge of the VIS arm of the spectrograph. We also detect emission from Mg~II 2800 \AA\, as shown also in Fig. \ref{MgII+Hbeta}. A red excess is observed in the X-shooter spectrum. This is consistent with a second blackbody component observed in the SEDs, see \S \ref{SED}.

\begin{figure}
    \centering
    \includegraphics[angle=0, scale=0.25, trim={0.5cm 0.5cm 0 0.7cm},clip]{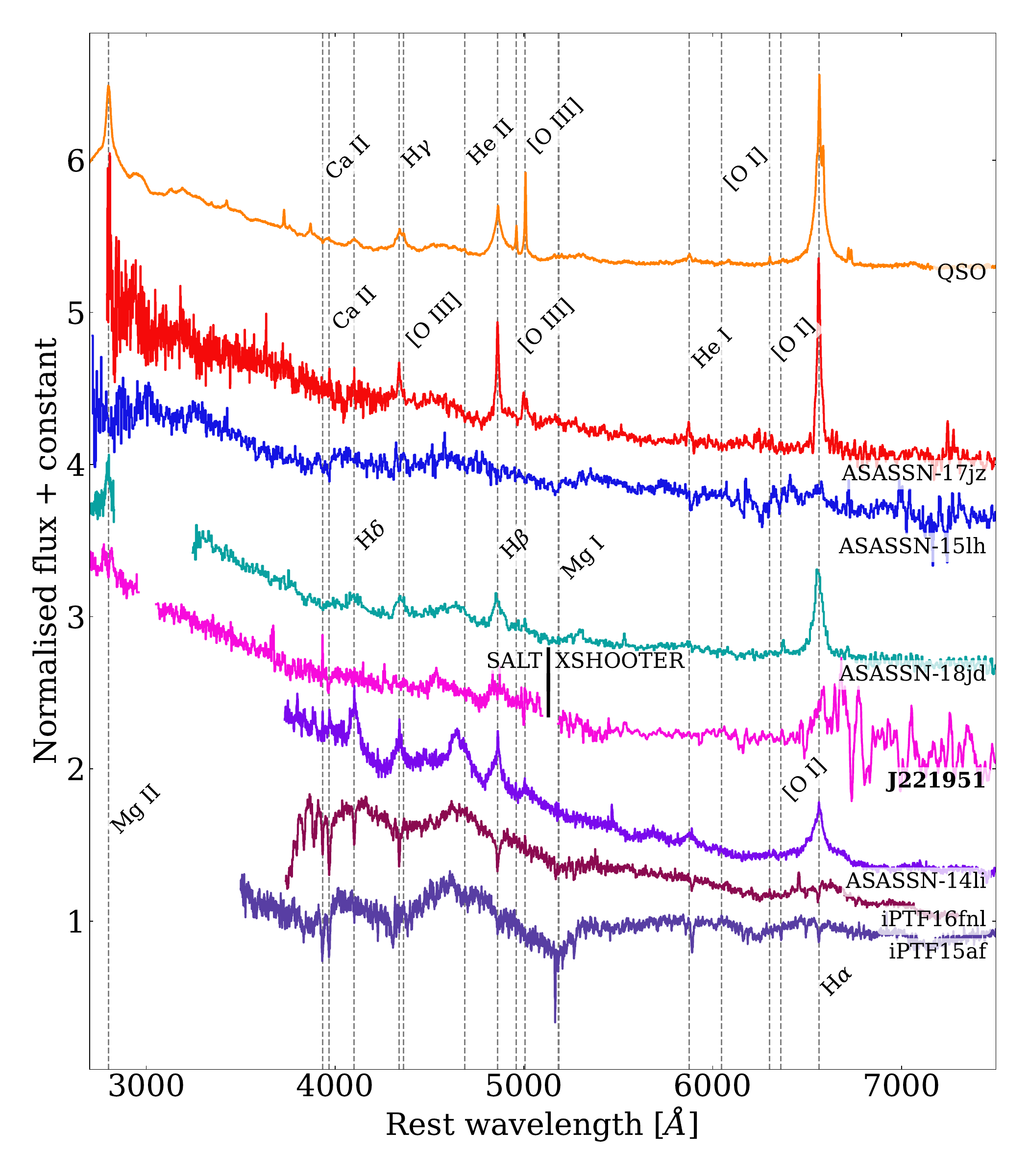}
    \caption{Comparison of the optical spectra of \sourcename (SALT spectrum taken on 58760 MJD and X-shooter taken on 59411 MJD)  with optical spectra of other transient objects: QSO \citep{van01}, ASASSN-14li \citep{hol16b}, iPTF15af \citep{bla19}, ASASSN-15lh \citep{bro16}, iPTF16fnl \citep{bla17b}, ASASSN-17jz \citep{hol21}, ASASSN-18jd \citep{neu20}. In terms of continuum shape and spectral lines, \sourcename most closely resembles the optical spectra of the ANT ASASSN-18jd.}
    \label{Opt_spec_comp}
\end{figure}

We compare the SALT spectrum of \sourcename taken at 58760 MJD and the X-shooter spectrum taken at 59411 MJD with a sample of spectra from other objects, including QSOs, known TDEs and ANTs in Fig. \ref{FUV_spec_comp} and Fig. \ref{Opt_spec_comp}. Examining the UV spectra of \srcname, the broad emission features resemble those found in the spectra of three TDEs, which were also shown to have broad absorption features: ASASSN-14li, iPTF15af, iPTF16fnl, and the low ionization broad absorption line QSO (BALQSO). The most significant absorptions in the \sourcename spectrum, N\,V and O\,VI, are observed in the optical spectrum of the ANTs ASASSN-15lh and ASASSN-17jz, however, the absorptions in the spectrum of \sourcename are much broader. When examining the optical spectra of \srcname, it most closely resembles the optical spectrum of the ANT, ASASSN-18jd.

\subsection{Host Galaxy Properties}
\label{host}
We built a host SED using archival observations from {\it GALEX}, DES, VISTA and {\it WISE}. 
We apply the stellar population synthesis models in {\sc Prospector} \citep{lej17,lej18} to the archival 
photometry, demonstrating that it is that of an underlying galaxy. We derive key physical 
parameters of the galaxy, which include the stellar mass, metallicity, current 
star-formation rate and the widths of five equal-mass bins for the star-formation history,
and three parameters controlling the dust fraction and reprocessing \citep[see][for details]{lej17}. 
\cite{lej17} identify important degeneracies between age–metallicity–dust, and the dust mass–dust
attenuation curve. {\sc Prospector} is specifically designed to account for such degeneracies in parameter 
estimation using Markov Chain Monte Carlo analysis to fully explore the posterior probability density.
The best-fitting model is shown compared to the archival photometry in Fig. \ref{Host_spectrum} and the 
two-dimensional posteriors are plotted in Supplementary Figure S.2. We find a stellar mass 
$\log({\rm M_\ast}/{\rm M_\odot}) = 10.8\pm0.1$ and a metallicity below or marginally consistent with solar, 
$\log\,Z/Z_\odot = -1.24^{+0.56}_{-0.47}$. {\sc Prospector} fits to TDE hosts have also favoured low 
metallicities \citep{ram22,ham23}. This may be expected since these galaxies tend to be below the mass of the Milky Way. 
We also find a low specific star-formation rate, log $sSFR = -12.0\pm1\,{\rm yr^{-1}}$ in
the last 50 Myr, where the reported values and uncertainties are the median and 16th/84th percentiles of 
the marginalized posterior distributions. 
The host galaxy SED is consistent with no AGN contribution, with the fraction of bolometric luminosity from an AGN, 
$f_{AGN}<0.06$. This is consistent with the {\it WISE} W1-W2 colour of 0.20 mag, which suggests the archival IR 
emission is galaxy-like and not dominated by an AGN \citep{ste12,yan13}. Overall, the {\sc Prospector} fit suggests that the host galaxy 
is consistent with a recently quenched galaxy with high star formation rate between 200-700 Myr ago and no strong AGN activity.

We compute the BH mass using the BH mass - bulge mass relation, $M_{BH} - M_{bulge}$ \citep{kor13,mcc13,ram22}.  We are unable to decompose the host galaxy light into bulge and disc components since the host is too faint. We estimate the bulge mass from the total mass of the galaxy using the average bulge to total light (B/T) ratio, for a $\log\,(M_{\ast}/M_\odot)\sim 10.8$ the ratio is $\sim 0.67$ \citep{sto18}, indicating that a large fraction of the mass of the galaxy is within the bulge. This gives $\log\,(M_{bulge}/M_\odot)\sim 10.6$. We first use the relationship derived from TDE host galaxies in \cite{ram22} and then compare this to the value produced by \cite{kor13}, which has been calibrated mainly at BH masses greater than those able to produce a TDE. Using the relationship derived using TDE host galaxies \citep{ram22}, we obtain an expected value of the SMBH mass of $\log\,(M_{BH}/M_\odot)=6.9$. This value is consistent with the mass derived from the TDE model in \S \ref{TDE_model}. This mass is at the top end of the SMBH masses associated with TDEs \citep{wev17,wev19,nic22} and within the theoretically expected mass range for TDEs \citep{koc16}. If we use the \cite{kor13} ${\rm M_{BH}} - {\rm M_{bulge}}$ relation the value of the SMBH mass we obtain a higher expected value of $\log\,(M_{BH}/M_\odot) =8.2$, such a value is larger than the Hills mass for a $1\,{\rm M_\odot}$ star \citep{hil75}\footnote{The Hills mass is the largest BH mass for a given stellar mass that will result in a TDE and not swallow the star whole without disruption \citep{hil75}. For a star of $0.1-1\,{\rm M_\odot}$, the Hills mass of a Schwarzschild BH is $10^7-10^8\,{\rm M_\odot}$.}, suggesting that if this is the correct BH mass, either the mass of the disrupted star is larger than 1\,M$_\odot$, contrary to the modelling in \S \ref{TDE_model}, or that \sourcename is not a TDE as the star should have been swallowed whole without disruption. 

\section{Discussion}
\label{discussion}
With an absolute magnitude of $M_{u,AB} = -23$ mag, peak bolometric luminosity $L_{max} = (1.1\pm0.7)\times10^{45}\,{\rm erg}\,{\rm s}^{-1}$ and total radiated energy of $E>2.6\times10^{52}\,{\rm erg}$, \sourcename is one of the brightest, most energetic and long-lived UV transients observed to date. Based on some of the basic properties of this event, we are immediately able to draw conclusions on the nature of \sourcename and exclude some classes of transient. The long-duration of the light curve and lack of an X-ray counterpart tends to rule out fast-evolving transients, including on-axis Gamma-ray bursts and most supernovae (SNe), except the long duration SN type IIn, which interact strongly with the circumstellar medium \citep[e.g.][]{sch90,smi17}. 

The broad absorption seen in N\,V and O\,VI, and the lack of undulations due to singly or doubly ionised metals are not expected in SNe \citep{bar00,fol13,yan17}. The total radiated energy of \sourcename of $>2.6\times 10^{52}$\,erg is also in tension with SN IIn theory \citep{suk16} and even the brightest confirmed interacting superluminous supernovae \citep[SLSNe; ][]{nic20}, although there is a population of nuclear transients with similar total radiated energies that have some properties consistent with supernovae \citep[e.g. PS1-10adi;][see \S\ref{AGN} for further discussion]{kan17}. \sourcename has a constant temperature at $\sim2.3\times 10^4\,{\rm K}$. Supernovae have been shown to have similar initial temperatures, however SNe cool below $10^4\, {\rm K}$ within a few weeks \citep{hol19}. The {\it WISE} W2 peak luminosity ($\sim 10^{44}\,{\rm erg s^{-1}}$) is also of order a factor 100 brighter than the brightest known SN in the IR \citep[see Fig. 8 of ][]{jia19}. The large bolometric luminosity suggests an association with a black hole, belonging to one of two main classes of events that also display broad absorption line features: TDEs or AGN. In the following, we investigate these two main classes of event and then discuss the origin of the second, lower-temperature blackbody component.

\begin{figure}
    \centering
\includegraphics[angle=0, scale=0.4]{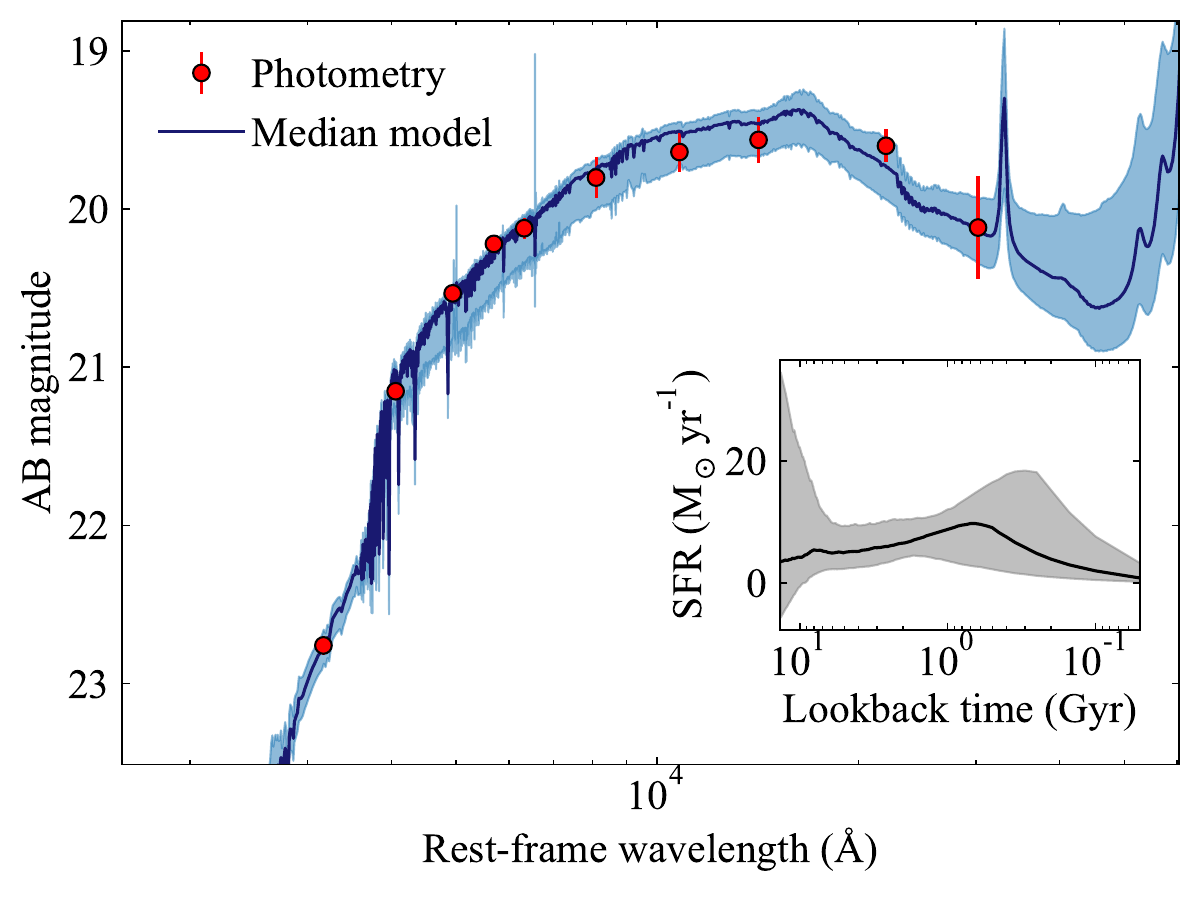}
    \caption{Archival photometry of the host galaxy of \srcname, and SED fit using {\sc Prospector}. The best-fitting model, as well as the 1$\sigma$ dispersion in model realizations, is shown. The inset shows the derived star-formation history, which peaks at 10 $M_\odot\,{\rm yr^{-1}}$ between a lookback time of 0.1 and 1 Gyr, and has a steep drop in the last $\sim 0.5$ Gyr. The shaded areas give the $1\sigma$ uncertainty on their respective parameter.}
    \label{Host_spectrum}
\end{figure}

\subsection{Tidal disruption events}
\sourcename is positionally coincident with the nucleus of its host galaxy and is therefore consistent with a TDE origin. In TDEs, approximately half of the disrupted material falls back onto the SMBH, likely forming an accretion disc, while the other half is unbound \citep{ree88, lac82}. The spectra of TDEs are typically blue and thermal in nature \citep[e.g.][]{van20}. \sourcename is consistent with this picture with the SEDs well fit by a blackbody with a typical temperature of $T\sim23000$ K. With H$\alpha$ and weak H$\beta$ being observed in the optical spectra, \sourcename would be classified as a H-only TDE in terms of the \cite{van21} spectral classification scheme for optical TDEs. In addition, the overall light curve evolution can be fit reasonably well with a TDE model. Comparing the model parameters of \sourcename to a sample of TDEs \citep{nic22} suggests that for \sourcename the BH mass is at the high end, the star mass is typical, and the low impact parameter is most similar to that of a H-only TDE rather than a He or Bowen TDE, consistent with the spectral classification obtained from the spectra. The radio luminosity of \sourcename is consistent with radio-quiet TDEs \citep{ale20}. If \sourcename does contain a jetted synchrotron source, we are unlikely to be viewing it on-axis.

We can use the SMBH mass estimates and the peak luminosity to determine the Eddington ratio. Using the lower mass estimate for the SMBH, based on the assumption that \sourcename is a TDE ($\log\,(M/M_\odot) \sim7.1$; see \S \ref{TDE_model} \& \ref{host}), we derive an Eddington luminosity of $L_{Edd} = 1.6\times10^{45}\,{\rm erg}\,{\rm s}^{-1}$, and an Eddington ratio of $L_{max}/L_{Edd} = 0.70$. This is consistent with the typical Eddington ratios measured for a sample of TDEs with well-constrained SMBH masses, for which the peak luminosities are $\sim L_{Edd}$ \citep{wev19}. The $L_{max}/L_{Edd}$ value computed for \sourcename is likely to be an underestimate of the true ratio since we may not have observed the event at peak brightness and therefore the peak luminosity was probably larger and closer to $L_{Edd}$. However, if we use the larger black hole estimate of $\log\,(M_{BH}/M_\odot)\sim 8.2$ determined from the \cite{kor13} BH mass -- bulge mass relation (see \S \ref{host}), which challenges the formation of a TDE in the first place, then this increases $L_{Edd} = 2\times10^{46}\,{\rm erg}\,{\rm s}^{-1}$ and reduces $L_{max}/L_{Edd} = 0.06$. 

For a (minimum) total energy of $E\sim3\times10^{52}$ erg, the accreted mass needed to radiate this energy is $\sim 0.14 {\rm M_\odot}$ for a typical efficiency $\epsilon = 0.1$. This is much larger than the accreted mass estimates of most TDEs e.g. $<0.01{\rm M_\odot}$ \citep[][]{hol16,hol19,hun21} and is consistent with radiatively efficient accretion of the bound stellar debris \citep{van19}. 

The peak luminosity of \sourcename of $L_{max}=1.1\times10^{45} {\rm erg\,s}^{-1}$ is larger than the bulk population of TDEs and is on par with the luminosity and energy of ASASSN-15lh, ASASSN-17jz and ASASSN-18jd \citep{don16, hol21, neu20}. The light curve shows bumps, which are not typical of TDE light curves \citep{neu20}, but are also seen in ASASSN-15lh \citep{bro16} and ASASSN-18jd. TDEs tend to show smooth monotonic declining behaviour, though some do show variability and moderate rebrightening episodes \citep[e.g. AT 2018fyk;][]{wev19b}. ASASSN-15lh, ASASSN-17jz and ASASSN-18jd are ANTs \citep{hol21} and so their nature is also uncertain and under debate. \cite{neu20} discuss ASASSN-18jd as being either due to a TDE or as a rapid turn-on AGN. \cite{hol21} suggest ASASSN-17jz was a SN IIn occurring in or near the disc of an existing AGN, and that the late-time emission is caused by the AGN transitioning to a more active state. For ASASSN-15lh, the literature is more extensive. Initially, ASASSN-15lh was deemed to be a hydrogen poor superluminous supernova \citep[SLSN-I;][]{don16}, but with an absolute peak magnitude more than 1 magnitude brighter than typical SLSNe-I. While some studies agreed with the SLSNe interpretation \citep{god17,bro16}, there is a larger consensus that the properties of the UV light curve, spectra and the host galaxy together implies ASASSN-15lh is more consistent with a TDE origin \citep[e.g.][]{bro16,lel16,mar17b,kru18}. Recently, an even more luminous ANT has been discovered, AT2021lwx \citep{sub23,wis23}. It is located at a redshift of 0.995, even higher than \srcname. As with the other events discussed in this paragraph, there is debate over whether AT2021lwx is a TDE or some other accretion event around a SMBH, particularly so in this case, because the light curve fit with MOSFIT requires the unlikely disruption of a $\sim 14\,{\rm M_\odot}$ star by a $10^8\,{\rm M_\odot}$ SMBH \citep{sub23,wis23}. This stellar mass is unusually large, whereas the MOSFIT parameters derived for \sourcename are more consistent with likely TDE configurations.

The host galaxies of \srcname, ASASSN-15lh ASASSN-17jz and ASASSN-18jd are also similar. For \sourcename we determine the host mass to be $\log\,(M/M_\odot) = 10.8\pm0.1$ and the specific star formation rate $\log\,sSFR = -12\pm1\,{\rm yr^{-1}}$. For ASASSN-15lj, \cite{lel16} showed the host to be a massive red galaxy with a small rate of ongoing star formation with host mass $\log\,(M/M_\odot) = 10.95^{+0.15}_{-0.11}$, a star formation rate $SFR < 0.02\,{\rm M_\odot\,yr^{-1}}$ and a specific star formation rate of $\log\,sSFR <\, -12.5\,{\rm yr^{-1}}$. For ASASSN-18jd the host mass is $\log\,(M/M_\odot) = 11.23^{+0.03}_{-0.33}$ and a star formation rate of $SFR = 0.6^{+0.1}_{-0.3}\,{\rm M_\odot\,yr^{-1}}$ \citep{neu20}. For ASASSN-17jz the host mass is $\log\,(M/M_\odot) = 10.74^{+0.11}_{-0.14}$, age $= 2.2^{+1.2}_{-1.0}\, {\rm Gyr}$, and a star formation rate of $SFR = 2.9^{+0.4}_{-0.5}\,{\rm M_\odot\,yr^{-1}}$ \citep{hol21}. The black hole mass at the centre of the host galaxy for ASASSN-15lh has been estimated from galactic scaling relationships to be $\sim10^9\,{\rm M_\odot}$ \citep{lel16,kru18} and is consistent with the value derived from TDE light curve model fits \citep{mum20}. For ASASSN-18jd, the black hole mass is $\log\,(M_{BH}/M_\odot)=7.6\pm0.4$ \citep{neu20} and for ASASSN-17jd the black hole mass is $\log\,(M_{BH}/M_\odot)\sim7.5$ \citep{hol21}. For \sourcename we estimate a lower value of $\log\,(M_{BH}/M_\odot)\sim7.1$ and an upper value of $\log\,(M_{BH}/M_\odot)\sim 8.2$ depending on the scaling relation. For ASASSN-15lh, the black hole mass is larger than the Hills mass and the same is true for the upper value of the black hole mass for \srcname. For a black hole, bigger than this Hills mass, the star should have been swallowed whole without disruption. For ASASSN-15lh, to overcome this in order to allow the tidal disruption to occur, it has been suggested that the black hole is a rapidly spinning Kerr SMBH \citep{lel16,kru18,mum20}. In this case, the Hills mass increases by approximately an order of magnitude for extreme Kerr spins \citep{kes12}. There is some evidence to suggest that TDEs fade more slowly as the SMBH mass increases \citep{bla17b,wev17,van19}. Considering \sourcename as a TDE, it would be consistent with this picture since the black hole mass of \sourcename is at the high end of the TDE black hole mass distribution \citep{nic22} and the decay rate is shallower than typical TDE light curves (see Fig. \ref{luminosiy_comparison}). 

X-ray-selected TDEs have harder spectra with $\alpha_{OX}\sim 1.5$ and optically selected TDEs are softer with $\alpha_{OX}\sim 2.4$.  In TDEs, high (soft) values of $\alpha_{OX}$, associated with high Eddington ratios, are thought to arise from disc dominated spectra, while low (hard) values of $\alpha_{OX}$, associated with low Eddington ratios, indicate power-law spectra \citep{wev20}. The latter may be more consistent with a jet rather than a disc. For \srcname, $\alpha_{OX}$ is $>1.6$ upon initial detection. This excludes the hardest spectra, allowing for either a thermal spectrum or a mixture of power-law and thermal components. 

Examining the light curve behaviour of \srcname, the light curve decays approximately as a broken power-law, with a change to a steeper decay after 200 days. Around this time, the bolometric light curve behaviour of \sourcename appears to have a similar decay rate to that of ASASSN-15lh after its second light curve peak. \citet{mum20} and \citet{lel16} showed that this part of the light curve of ASASSN-15lh, at T+100 days, is consistent with being disc-dominated. Usually, the transition from fallback-dominated to disc-dominated emission is expected as a flattening of the TDE light curves and has been observed for a number of optical/UV TDEs \citep[$>$few hundred days;][]{van19}. However, \cite{mum20} showed that disc-dominated light curves may not be flat and may actually decay. If ASASSN-15lh and \sourcename are similar in origin, this suggests that the late-time behaviour of \sourcename may also be disc dominated. However, \cite{mum20} note that their model is unable to reproduce the very latest UV emission of ASASSN-15lh, which is above their model prediction. Instead, they suggest that the late emission in ASASSN-15lh may be due to AGN activity \citep{kru18}, additional material from the initial disruption returning to the disc, or even a state transition within the disc at low Eddington ratio \citep{mum20}.

Comparing the UV spectrum of \sourcename with other objects in Fig. \ref{FUV_spec_comp}, we see that \sourcename is different to ASASSN-18jd and ASASSN-15lh. ASASSN-18jd shows weak N~V emission at 1238.8, 1242.8 \AA\, while we see it in absorption for \srcname. ASASSN-15lh does not show any broad absorption features \citep{bro16}, though it does show N~V and O\, VI absorptions, which are broad in the spectrum of \srcname, but narrow for ASASSN-15lh. In terms of the broad absorption lines, \sourcename most closely resembles TDEs ASASSN-14li \citep{cen16} and iPTFi6fnl \citep{bro18} and that of the low-ionisation BALQSO. In general, the UV spectra of the other objects do not typically extend bluer than 1000-1100 \AA, however, the UV spectrum of ASASSN-15lh \citep{bro16} does cover a similar wavelength range to that of \srcname. In both spectra, absorption from the Lyman series is observed although the features are more evident for \srcname. 

Comparing the optical spectra of \sourcename with other objects in Fig. \ref{Opt_spec_comp}, we see that \sourcename lacks the narrow line features that are present in the composite QSO spectrum. In contrast to the UV comparison \sourcename looks least like ASASSN-14li and iPTF16fnl. No absorption features are present in the optical spectrum of \srcname, while they are present in that of iPTF16fnl and the emission features in ASASSN-14li are much stronger than \srcname. Overall, \sourcename most closely resembles ASASSN-18jd in the optical, though has less prominent hydrogen Balmer lines. The emission lines of \sourcename with $FWHM \sim 2200\,{\rm km \, s^{-1}}$ are much narrower than typically observed for TDEs ($\sim 10^4\,{\rm km \, s^{-1}}$), but are consistent with that observed for ASASSN-18jd \citep{neu20}.

\cite{par20} investigated why some TDEs show broad absorption lines (BALs) while others display broad emission lines (BELs) in their UV spectra, using synthetic UV spectra for disc and wind-hosting TDEs, produced by a state-of-the-art Monte Carlo ionization and radiative transfer code. Using a variety of disc wind geometries and kinematics they naturally reproduce both BALs and BELs with winds. Sight lines looking into the wind cone, at low angles relative to the plane of the disc, preferentially produce BALs, while other orientations preferentially produce BELs. Clumpy winds may also be a factor as clumping increases both the emission measure and the abundance of the relevant ionic species. Clumpier winds tend to produce stronger UV emission and absorption lines. This model suggests that we are viewing this event at high inclinations, towards the plane of a disc, if \sourcename is indeed a TDE with a wind. In the scheme of \cite{cha22}, which assumes a reprocessing scenario whereby the optical emission of the TDE is produced by reprocessed X-rays, at high inclinations TDEs would only be H-only and lack X-ray emission, which is consistent with that observed for \srcname. This inclination would also suggest we may not observe the relativistic jet, if present, which is also consistent with the lack of strong X-ray and radio emission.

Using the properties of the spectrum we can form a picture of this event if it is indeed a TDE. The Ly-$\alpha$ profile shows two broad emission peaks on either side in addition to a saturated central absorption (see Fig. \ref{FUV_spec_comp}). The absorption is likely to be located in our line of sight far from the ionising source. The emission peaks would be produced through recombination of hydrogen much closer into the core and may be from a rotating disc-like object \citep{sanbui1994}. The wavelength separation of each of the broad peaks of H~Ly$\alpha$ from the centre suggests a velocity of $\approx 2000~{\rm km\,s^{-1}}$. We can determine the location of the emitting material that is causing the broad emission, specifically in the Hydrogen lines. Assuming it is orbiting the SMBH in a circular orbit, such that the kinetic energy equals the gravitational potential energy, then using the SMBH mass estimates from \S \ref{host} and the velocity of the broad H~Ly$\alpha$ emission, the radius would be $R\sim 5\times10^{16} - 1\times10^{18}$~cm (0.017 - 0.36 pc).

The Ly-$\beta$ absorption profile blends with the blue absorption trough of the O\,VI 1031.9 \AA~ line. The O\,VI 1037.6 \AA\ line shows a blue-shifted absorption profile. The extent of the absorption to the blue of Ly-$\beta$ is much smaller than the velocities derived from the blue wing of the absorption in O\,VI of $-1800~{\rm km\,s^{-1}}$, consistent with the lower ionisation lines being from a different component. The  N\,V lines have a velocity edge consistent with a $1750~{\rm km\,s^{-1}}$ outflow, so both are formed in the same component.

\begin{figure}
\centering      
\includegraphics[angle=0, scale=0.4]{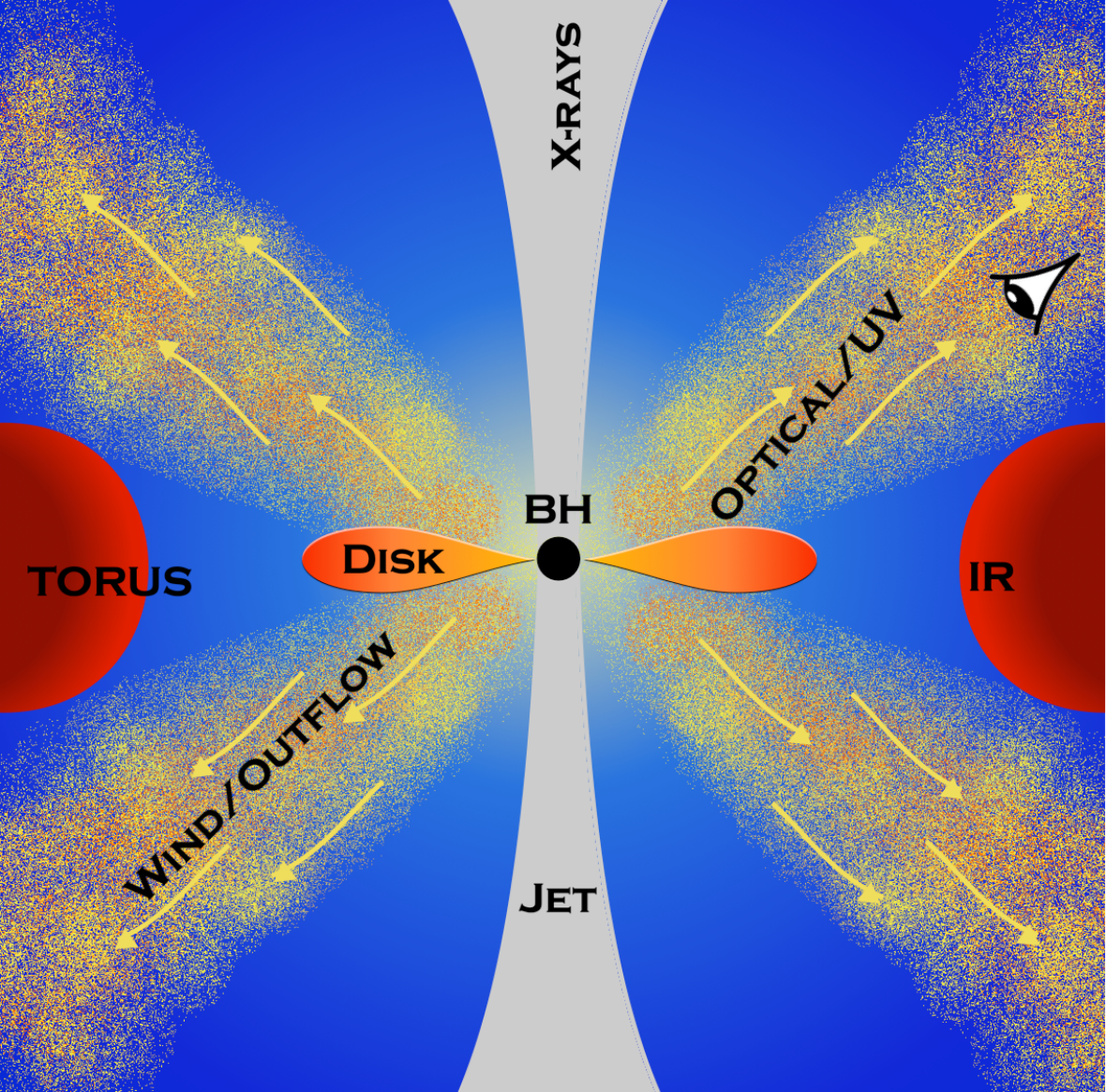}
\caption{Artist impression of the geometry of the source. We likely see closer to the equatorial plane since we do not see the X-rays from the jet. The centre is a SMBH, the disc surface has wind that causes the blue absorptions in resonance lines, and near the equatorial plane, the disc is cooler and provides the narrow absorption lines seen in the low ionisation lines. Over time the disc expands and accretion decreases.}
\label{schetch}
\end{figure}

Examination of Fig.~\ref{MgII+Hbeta} shows that over time the red wing of H$\beta$ becomes fainter and eventually the profile appears as an asymmetric blue-peaking hump. The He\,I 5017 \AA~ line is present but weak and no He\,II lines were found. The P-Cygni lines of O\,VI, N\,V, etc. in the hotter outflows reach projected outflow velocities of $1800\,{\rm km\,s^{-1}}$, but there is also a cooler outflow or turbulence of $\approx 90\,{\rm km\,s^{-1}}$ as seen in Hydrogen absorption lines. The hotter outflow is likely produced in the inner region, within a fast outflow, while the velocities in the low-ionisation lines are located further out. 
Ly-$\alpha$ is optically thick enough to show an outflow. There, the higher Lyman lines are optically thin, at least from the Ly-$\gamma$ on, which suggest a H column density of $\sim 10^{18}$ cm$^{-2}$. 
Entrained in the hot outflow is ionised H, which recombines, giving the broad emission seen in Ly$\alpha$. The slow decay of the emission may suggest on-going accretion, while the evolution of the width of the Mg\,II resonance line emission suggests decreasing densities reduce opacity in the line wings and a possible expansion of the disc-like structure. A possible model (Fig. \ref{schetch}) can be envisioned based on the data taking the central source to be a SMBH with a disc of material that is flared, a wind outflow and with the observer looking at a high inclination, close to the plane of the disc, consistent with the scheme of \cite{cha22}. Over time the disc expands outward, becoming cooler on the outside, whilst maintaining the inner disc radius. 

Overall, the peak luminosity of \sourcename is higher than the bulk TDE population, consistent with the luminosity of ANTs: ASASSN-15lh ASASSN-17jz and ASASSN-18jd, suggesting \sourcename is not a standard TDE. However, TDE light curve modelling suggests that \sourcename is a TDE, with a typical star mass but with a BH mass at the high end. The spectra and modelling suggest it is a H-only TDE. \sourcename has spectral properties consistent with ASASSN-18jd, in the optical, and BAL TDEs in the UV. The broad emission line features and lack of X-ray and radio emission may be because we are observing at a high inclination, close to the plane of the disc.

\subsection{Active Galactic Nucleus}
\label{AGN}
Given the potential nuclear nature of \srcname, we now look at whether an AGN may be the cause of this transient. The optical spectra of \sourcename show a double-peaked Mg II 2800 \AA\ profile. Mg II 2800 \AA\ is an emission line common to AGN, but not usually TDEs. ASASSN-18jd, suspected to be either a TDE or a rapid turn-on AGN, also has Mg\,II in emission \citep{neu20}. The narrow line widths, observed for \srcname, of order $2000\, {\rm km\,s^{-1}}$, are also commonly observed in AGN spectra. In addition, the broad absorption in the UV N\,V and O\,VI resonance lines, suggestive of outflows, also mimic the behaviour seen in C\,IV 1550 \AA\ in the rest-frame UV of BALQSOs. However, the lack of a strong [O\,III] 5007 \AA~emission line, with an EW of $<1$ \AA, is not typical of AGN \citep{she11}. Out of 105000 QSOs in the SDSS DR7 catalogue, 532 have a bolometric luminosity less than $L_{bol} = 10^{45}\,{\rm erg\,s^{-1}}$ and of these none has an EW $\lesssim 3$ \AA~\citep{she11}; their average EW is 57 \AA. The UV to X-ray spectral slope, for \sourcename is $\alpha_{OX}>1.6$, this is not consistent with QSOs but is consistent with BALQSOs; for QSOs $\alpha_{OX}\sim 1.4$ \citep[see e.g.][]{mar04}; for X-ray BALQSOs, $\alpha_{OX}\sim 1.90$; and for optical BALQSOs $\alpha_{OX}\sim 2.20$ \citep{blu08}. 

Based on an argument used for ASASSN-18jd, outlined by \cite{neu20} using the SDSS survey of quasars \citep{mac12}, it is unlikely that the variability of \sourcename is due to normal QSO variability. \sourcename is 3.4 magnitudes brighter than the archival $g$-band value (which is also likely a lower limit due to host contamination); the probability of achieving a $|\Delta m_g|> 3.5$ mag on a timescale of $< 5$ years is $P<2\times10^{-6}$ \citep{mac12}. \cite{gra17} use CRTS data to provide even tighter constraints on the likelihood of observing large-magnitude changes. For $\Delta m = 3.0$ mag, they find that the probability of achieving this change after 3200 days is $10^{-7}$, which would be even larger for shorter time lags. These probabilities make it unlikely that the variability of \sourcename is due to normal QSO variability.

However, just because such a large flare is unlikely, it is not impossible, and observations of AGN have discovered new and more extreme forms of variability, indicating that we are yet to discover the full range in AGN variability. A class of slow-blue transients, with $\Delta m > 1.5\,{\rm mag}$ over $\sim$years, were identified in \cite{law16} and a similar population was discovered by \cite{gra17}. \cite{law16} state that around 1 AGN in $10^4$ displays such behaviour at any given time. The origin of these large magnitude changes is unknown and may be due to rare eruptive events from accretion, or microlensing \citep{law16}. Although some instead may be attributed to stellar-related activity, such as TDEs, SLSNe and mergers of binary black holes \citep{gra17}. 

While \sourcename does share some characteristics of AGN, the lack of narrow line features, together with no AGN required in the host SED fitting and the {\it WISE} $W1-W2$ colour, we can conclude that the host galaxy of \sourcename did not host a strong AGN prior to the transient. This suggests that there has been no recent AGN activity prior to it turning on; the narrow line region (NLR) has not been ionized. We can estimate the distance from the SMBH to the NLR and thus estimate the minimum time that the AGN must have been inactive for us to observe no [O\,III]  5007\AA~emission. Using the same argument, we can also estimate when we would expect to see the [O\,III] 5007\AA~emission line in the spectrum if the AGN has newly turned on. \cite{bas05} state that for a single zone model, the distance to the $R_{NLR} = 40L_{44}^{0.45}\,{\rm pc}$ where $L_{44}$ is $\nu L_{\nu}/10^{44}\,{\rm erg\,s^{-1}}$ at 4861 \AA. From the first optical spectrum of \srcname, we derive $\nu\,L(4861\,{\rm \AA}) = 1.61\times10^{44}\, {\rm erg\,s^{-1}}$. This results in a distance to the NLR, $R_{NLR}$ of 49.5 pc. This implies that the AGN has not been active for at least 160 years. Therefore, we can conclude that if \sourcename is due to AGN activity, it is `turning on'. In this case, we may have to wait decades before the [O\,III] 5007 \AA~emission line is detectable. Historically, several AGN have been noted to `turn-on' \citep{fre19,yan19}, whereby galaxies transition from being LINERS to more active galaxies, such as narrow-line Seyfert 1s or radio-quiet QSOs \citep{gez17,fre19}. Some such objects show similar temporal behaviour to \sourcename \citep[e.g. SDSS1115+0544A;][]{yan19}. 

Overall, \sourcename did not host a strong AGN prior to the transient but has properties consistent with an AGN turning on. The clearest evidence of an AGN nature would be for \sourcename to deviate from its current steadily decaying behaviour and to show a sustained period of increased flux - not just a bump or flare. Continuous UV monitoring of this source will therefore be important for monitoring the late-time behaviour. Deep X-ray observations of this source, for instance with {\it Chandra} or {\it XMM-Netwon}, would provide tighter constraints on the X-ray brightness of the AGN.

\subsection{Origin of the low-temperature blackbody}
One interesting feature of \srcname, is that there is evidence of two blackbody components in at least two of the SEDs. Two blackbody components were also observed in PS1-10adi \citep{van16, kan17}, PS16dtm \citep{jia17,pet23} and most recently AT2021lwx \citep{wis23}. PS1-10adi is an AGN-associated transient that may be produced by a TDE, SNe or AGN activity \citep{kan17}. Spectroscopically, the transient has features similar to a narrow line Seyfert 1 galaxy and to certain types of supernovae. For PS1-10adi, the blackbody temperatures (11000 K and 8000 K evolving to 2500 K and 1200 K) are lower than those observed for \srcname. \cite{jia19} argue that the IR excess observed in PS1-10adi is a dust echo of a TDE in an AGN, with the UV emission from the TDE heating and sublimating the dust in the AGN torus. PS16dtm is a TDE in a narrow line Seyfert 1 galaxy \citep{bla17}. It similarly displayed a MIR flare that was also interpreted as a dust echo of a TDE in an AGN \citep{jia17}. MIR flares have also been found in other TDEs \citep{van16, ono22} and a systematic search of {\it WISE} observations of galaxies discovered over 100 with IR outbursts, thought to be the dust echoes of transient accretion events of SMBHs \citep{jia21}. \cite{jia17} noted that PS16dtm seemed to be detected a few days earlier in the MIR compared to the optical/UV. This may also be the case for \srcname, see Fig. \ref{J221951_archival}, however, it is difficult to confirm this given the large errors on the {\it WISE} data and the lack of optical data points in between the last DES visit and the UVOT detection. For PS16dtm, \cite{jia17} note that the blackbody temperature in the MIR decreases with time, to a value below the sublimation temperature.

For \srcname, a two blackbody component best fits only one of our SEDs which includes photometry redder than the $v$ band. While a model with two blackbodies can be fitted to the three subsequent SEDs, it does not provide a better fit for any, although the fits do suggest that the temperature of this second component is decreasing as was observed for PS16dtm \citep{jia17}. Assuming the IR excess observed in the 58787 MJD SED and the X-shooter spectrum of \sourcename is also due to UV heating of nearby dust, then with a temperature of $\sim 2800\pm400$\,K, the dust is consistent with the sublimation temperature. Using the formula for the sublimation radius given in \cite{nam16}, their Eqn. 2, we can calculate the distance of the inner edge of the dusty torus from \srcname. Taking the peak bolometric luminosity of $L_{bol}=8.91\times 10^{44}\, {\rm erg\, s^{-1}}$, assuming a grain radius of 0.1$\,\mu{\rm m}$ and a sublimation temperature of $\sim 2000$\,K, we compute a sublimation radius of $3\times 10^{17}\,{\rm cm}$ corresponding to 0.09pc or 110 light days. This distance is typical of the distances expected of the inner edge of an AGN torus \citep{sug06} and suggests that any preexisting dust within this radius will have been evaporated by UV emission from \srcname.

Another TDE candidate observed to have low temperature blackbody component is Arp 299-B AT1, which was discovered in the galaxy merger Arp 299 and is associated with an AGN \citep{mat18}. For Arp 299-B AT1, the temperature remains constant at $800\,$K beyond 2000 days after the transient was first observed to rise. At late times ($T_0>+800$), the flux of \sourcename is comparable to the host value in the reddest filters making it difficult to constrain the temperature of the second blackbody component. Observations with JWST would be important in enabling us to measure the IR flux and determine how the temperature of this second blackbody component evolves with time.

\section{Conclusion}
\label{conclusions}
\sourcename was discovered during the follow-up of a gravitational event: S190930t. It brightened by $>3$ magnitudes in the UV compared to archival data and coincides with the centre of an optical/IR archival source, previously observed by DES and VISTA, which we show to be an underlying galaxy. Our spectroscopic redshift of 0.5205 rules out its association with the gravitational wave event. However, \sourcename is a very unusual and long-lived UV-luminous nuclear transient. In this paper, we presented our follow-up of this transient and investigated its nature, whether it is a supernova, tidal disruption event or related to AGN activity. Below we summarise our key findings:

\begin{itemize}
  \item A {\it HST} UV spectrum determines a redshift 0.5205 and reveals broad absorption lines from ionised species such as NV and OVI, along with narrow, low-ionisation lines of H and N\,I.
  \item \sourcename has been observed at regular cadence for $\sim 1000$ days and continues to be detected in the UV, making it one of the longest observed UV transients with one of the best-sampled UV light curves.
  \item In the optical/UV the light curve decays from the start of observations. Several bumps that are more pronounced in the UV are present and appear to reset the brightness level, such that the light curve resumes its decay from close to the peak of the bump. 
  \item A supernova explosion is ruled out by a total radiated energy of $\gtrsim 3\times 10^{52}\, {\rm erg}$, as well as the lack of broad absorption lines in the optical spectrum
  \item Coincident {\it Swift}/XRT observations, do not detect X-ray emission from \srcname, providing an upper limit to the X-ray luminosity of $L_X<6\times 10^{42}\,{\rm erg\,s^{-1}}$ (0.3-10\,keV). Radio observations by ACTA also do not detect any radio emission with 3\,$\sigma$ upper limits of 117\,$\mu$Jy at 5.5 GHz and 90\,$\mu$Jy at 9 GHz, with a 5.5GHz luminosity of $<2\times10^{39}\, {\rm erg\,s^{-1}}$.
  \item The optical spectra are blue and relatively featureless, displaying only H$\beta$ and Mg II in emission.
  \item Spectral energy distributions, created from UVOT data only (with filters $uvw2$ through to $v$) for which we have the most epochs, are well fit by a power-law with a slope of $\beta=0.49\pm0.04$ or a blackbody with an average temperature of $23000\pm410\,{\rm K}$. In SEDs constructed using UVOT and ground-based photometry, a two blackbody model is preferred in one SED, with evidence for two blackbody components observed in at least one other SED. The temperature of the second component is $\sim2800\pm400$K, which potentially cools across later SEDs.
  \item Examining the archival photometry, we determine the host galaxy to be a massive red galaxy, with a host galaxy stellar mass $\log\,(M/M_\odot) = 10.8\pm0.1$ and a low specific star formation rate $\log\,sSFR = -12\pm1\,{\rm yr^{-1}}$ in the last 50 Myr. 
  \item From the host SED fitting and the {\it WISE} $W1-W2$ colour, we can conclude that the host galaxy of \sourcename did not host a strong AGN prior to the transient.
  \item Using the \cite{kor13} black hole mass -- bulge mass scaling relation we estimate that the mass of the BH is $\log\,(M_{BH}/M_\odot)\sim 8.2$, bigger than this Hills mass, which implies for a $\sim 0.6{\rm M_\odot}$ star it should have been swallowed whole without disruption and no emission should have been observed. One solution to this may be that the black hole is a rapidly spinning Kerr SMBH \citep{lel16,kru18,mum20}. However, using the black hole mass -- bulge mass scaling relation derived from TDE host galaxies in \cite{ram22}, we estimate a BH mass of $\log\,(M_{BH}/M_\odot)=6.9$, which is consistent with the value derived from the TDE light curve model fits, $\log\,(M_{BH}/M_\odot)\sim7.1$.
  \item The probability of seeing such a large flare from normal AGN activity is $P<2\times10^{-6}$, characterising this as one of the most extreme nuclear flares to date.
  \item If due to AGN activity, the lack of narrow emission lines together with the host fitting and the {\it WISE} colour, implies it is caused by the AGN `turning on'.
\end{itemize}

The progenitor of \sourcename is unclear. The optical and UV spectra show features resembling both TDEs and AGN. Overall its spectral, temporal and host properties and its energetics are closest in nature to ASASSN-15lh and ASASSN-18jd. ASASSN-15lh, ASASSN-18jd and \sourcename belong to an increasing population of luminous blue transients, dubbed ambiguous nuclear transients for which the progenitors are not well constrained, but may be TDEs or due to AGN activity. Observing the late time evolution of \sourcename will provide important clues as to its nature. For instance, if this source is associated with an AGN turning on we may expect it to deviate from its current steadily decaying behaviour. The clearest evidence of an AGN nature would be for \sourcename to show a sustained period where it increased in flux - not just a bump or flare. Deep X-ray observations of this source, for instance with {\it Chandra} or {\it XMM-Newton}, would provide tight constraints on the X-ray brightness, a late-time continued detection would be indicative of an AGN and disfavour a TDE origin. Observations with JWST would be important in enabling us to understand the nature of the second lower-temperature blackbody component, potentially due to a dusty torus, and how it evolves with time.

The increase in the number of ambiguous nuclear transients, such as \srcname, is blurring the boundary between what is considered TDE and AGN activity. Sources such as \sourcename are important to pinpoint SMBHs that are otherwise hidden and provide the means to study SMBHs across various degrees of activity.

\section{Acknowledgments}
This research has made use of data obtained from the High Energy Astrophysics Science Archive Research Center (HEASARC) and the Leicester Database and Archive Service (LEDAS), provided by NASA's Goddard Space Flight Center and the School of Physics and Astronomy, University of Leicester, UK, respectively. 
This publication makes use of data products from the Wide-field Infrared Survey Explorer, which is a joint project of the University of California, Los Angeles, and the Jet Propulsion Laboratory/California Institute of Technology, funded by the National Aeronautics and Space Administration. 
This research has made use of the VizieR catalogue access tool, CDS, Strasbourg, France (DOI : 10.26093/cds/vizier). The original description of the VizieR service was published in 2000, A\&AS 143, 23.
This research has made use of the NASA/IPAC Infrared Science Archive, which is funded by the National Aeronautics and Space Administration and operated by the California Institute of Technology.
This paper includes data gathered with the 6.5 meter Magellan Telescopes located at Las Campanas Observatory, Chile.
Some of the observations reported in this paper were obtained with the Southern African Large Telescope (SALT).
This research is based on observations made with the NASA/ESA Hubble Space Telescope obtained from the Space Telescope Science Institute, which is operated by the Association of Universities for Research in Astronomy, Inc., under NASA contract NAS 5–26555. These observations are associated with program \#16076.
The Australia Telescope Compact Array is part of the Australia Telescope National Facility (https://ror.org/05qajvd42) which is funded by the Australian Government for operation as a National Facility managed by CSIRO.We acknowledge the Gomeroi people as the traditional owners of the ATCA observatory site.
This research uses services or data provided by the Astro Data Lab at NSF's National Optical-Infrared Astronomy Research Laboratory. NOIRLab is operated by the Association of Universities for Research in Astronomy (AURA), Inc. under a cooperative agreement with the National Science Foundation.
This publication uses the data from the {\it AstroSat} mission of the Indian Space Research Organisation (ISRO), archived at the Indian Space Science Data Centre (ISSDC). Part of the funding for GROND (both hardware as well as personnel) was generously granted from the Leibniz-Prize to Prof. G. Hasinger (DFG grant HA 1850/28-1). 
This research made use of Astropy, a community-developed core Python package for Astronomy \citep{ast13}.
AAB, NPMK, MJP, KLP, PAE, APB and JPO acknowledge funding from the UK Space Agency. MDP acknowledges support for this work by the Scientific and Technological Research Council of Turkey (T\"UBITAK), Grant No: MFAG-119F073. RG and SBP acknowledge the financial support of ISRO under {\it AstroSat} archival Data utilization program (DS$\_$2B-13013(2)/1/2021-Sec.2). RG and SBP are also thankful to the {\it AstroSat} UVIT team for helping with UVIT data analysis.  EA, MGB, SC, GC, AD, PDA, AM and GT acknowledge funding from the Italian Space Agency, contract ASI/INAF n. I/004/11/4. This work is also partially supported by a grant from the Italian Ministry of Foreign Affairs and International Cooperation Nr. MAE0065741. PDA acknowledges support from PRIN-MIUR 2017 (grant 20179ZF5KS). DBM is supported by research grant 19054 from Villum Fonden. MN is supported by the European Research Council (ERC) under the European Union’s Horizon 2020 research and innovation programme (grant agreement No.~948381) and by a Fellowship from the Alan Turing Institute. MG is supported by the EU Horizon 2020 research and innovation programme under grant agreement No 101004719. ET acknowledges support from the European Research Council (ERC) under the European Union’s Horizon 2020 research and innovation programme (grant agreement 101002761). DBM is supported by the European Research Council (ERC) under the European Union’s Horizon 2020 research and innovation programme (grant agreement No.~725246). The Cosmic Dawn Center is supported by the Danish National Research Foundation.

\section{Data Availability}
The {\it Swift} data underlying this article are available in the {\it Swift} archives at https://www.swift.ac.uk/swift\_live/, https://heasarc.gsfc.nasa.gov/cgi-bin/W3Browse/swift.pl, https://www.ssdc.asi.it/mmia/index.php?mission=swiftmastr. The photometry of \sourcename is available in the online supplementary material. 
The ATCA data are available from the Australia Telescope Online Archive -- https://atoa.atnf.csiro.au/. The {\it HST} and {\it GALEX} observations are available from the MAST Portal -- https://mast.stsci.edu/portal/Mashup/Clients/Mast/Portal.html. The X-shooter spectrum and GROND data are available from the ESO main archive -- http://archive.eso.org/eso/eso\_archive\_main.html. 

\vspace{-1mm}

\bibliographystyle{mn2e}   
\bibliography{J221951-484240} 

\IfFileExists{\jobname.bbl}{}
 {\typeout{}
  \typeout{******************************************}
  \typeout{** Please run "bibtex \jobname" to optain}
  \typeout{** the bibliography and then re-run LaTeX}
  \typeout{** twice to fix the references!}
  \typeout{******************************************}
  \typeout{}
 }

 \section{Affiliations}
  $^{1}$ School of Physics and Astronomy \& Institute for Gravitational Wave Astronomy, University of Birmingham, B15 2TT, UK\\
  $^{2}$ University College London, Mullard Space Science Laboratory, Holmbury St. Mary, Dorking, RH5 6NT, UK\\
  $^{3}$ Astrophysics Research Centre, School of Mathematics and Physics, Queens University Belfast, Belfast BT7 1NN, UK\\
  $^{4}$ Astrophysics Science Division, NASA Goddard Space Flight Center, Greenbelt, MD 20771, USA\\
  $^{5}$ Las Campanas Observatory, Carnegie Observatories, Colina El Pino, Casilla 601, La Serena, Chile\\
  $^{6}$ South African Astronomical Observatory, PO Box 9, Observatory 7935, Cape Town, South Africa\\
  $^{7}$ Department of Astronomy, University of Cape Town, Private Bag X3, Rondebosch 7701, South Africa\\
  $^{8}$ Department of Physics, University of the Free State, PO Box 339, Bloemfontein 9300, South Africa\\
  $^{9}$ Joint Space-Science Institute, Computer and Space Sciences Building, University of Maryland, College Park, MD 20742, USA\\
  $^{10}$ University of Messina, MIFT Department, Polo Papardo, Viale F.S. D’Alcontres 31, 98166 Messina, Italy \\
  $^{11}$ Australia Telescope National Facility, CSIRO Space and Astronomy, PO Box 76, Epping, NSW 1710, Australia\\
  $^{12}$ Astronomical Observatory, University of Warsaw, Al. Ujazdowskie 4, 00-478 Warszawa, Poland\\
  $^{13}$ Aryabhatta Research Institute of Observational Sciences (ARIES), Manora Peak, Nainital-263002, India \\
  $^{14}$ Department of Physics, Deen Dayal Upadhyaya Gorakhpur University, Gorakhpur-273009, India \\
  $^{15}$ Center for Space Science and Technology, University of Maryland Baltimore County, 1000 Hilltop Circle, Baltimore, MD 21250, USA.\\
  $^{16}$ Center for Research and Exploration in Space Science and Technology, NASA/GSFC, Greenbelt, Maryland 20771, USA\\
  $^{17}$ Department of Astronomy, University of Wisconsin, 475 N. Charter Str., Madison, WI 53704, USA\\
  $^{18}$ INAF–Osservatorio di Padova, vicolo dell Osservatorio 5, I-35122 Padova, Italy\\
  $^{19}$ School of Physics and Astronomy, University of Leicester, LE1 7RH, UK\\
  $^{20}$ Max-Planck-Institut f{\"u}r Extraterrestrische Physik, Giessenbachstra{\ss}e 1, 85748 Garching, Germany\\
  $^{21}$ Department of Physics, University of Bath, Bath, BA2 7AY, UK\\
  $^{22}$ Australia Telescope National Facility, CSIRO Space and Astronomy, 1828 Yarrie Lake Road, Narrabri, NSW 2390, Australia\\
  $^{23}$ George P. and Cynthia Woods Mitchell Institute for Fundamental Physics and Astronomy, Mitchell Physics Building, \\
          Texas A.\&M. University, Department of Physics and Astronomy, College Station, TX 77843, USA\\
  $^{24}$ Department of Astronomy and Astrophysics, The Pennsylvania State University, University Park, PA 16802, USA\\
  $^{25}$ Institute for Gravitation and the Cosmos, The Pennsylvania State University, University Park, PA 16802, USA\\
  $^{26}$ Department of Astronomy and Astrophysics, University of Toronto, Toronto, ON, Canada\\
  $^{27}$ INAF -- IASF Palermo, Via Ugo La Malfa 153, I-90146, Palermo, Italy\\
  $^{28}$ INAF -- Osservatorio Astronomico di Brera, Via Bianchi 46, I-23807 Merate, Italy\\
  $^{29}$ INAF-Osservatorio Astronomico di Roma, Via Frascati 33, I-00040 Monteporzio Catone, Italy\\
  $^{30}$ University of Padova, Italy\\
  $^{31}$ Space Science Data Center (SSDC) - Agenzia Spaziale Italiana (ASI), I-00133 Roma, Italy\\
  $^{32}$ Department of Physics and Astronomy, Clemson University, Kinard Lab of Physics, Clemson, SC 29634-0978, USA\\
  $^{33}$ National Science Foundation, Alexandria, VA 22314, USA\\
  $^{34}$ Department of Astrophysics/IMAPP, Radboud University, 6525 AJ Nijmegen, The Netherlands.\\
  $^{35}$ Cosmic Dawn Center (DAWN), Denmark.\\
  $^{36}$ Niels Bohr Institute, University of Copenhagen, Jagtvej 128, 2200 Copenhagen N, Denmark.\\
  $^{37}$ Los Alamos National Laboratory, B244, Los Alamos, NM, 87545, USA\\
  $^{38}$ Department of Physics and Mathematics, Aoyama Gakuin University, Sagamihara, Kanagawa, 252-5258, Japan\\
  $^{39}$ Department of Physics, University of Rome Tor Vergata, via della Ricerca Scientifica 1, I-00100 Rome, Italy \\
  $^{40}$ INAF, Via del Fosso del Cavaliere, 100, 00133, Rome
  $^{41}$ Key Laboratory of Space Astronomy and Technology, National Astronomical Observatories, Chinese Academy of Sciences, Beijing, 100101, China\\

\end{document}


\maketitle
\renewcommand{\thesection}{S.\arabic{section}}
\section{Supplementary Section}
\renewcommand\thefigure{S.\arabic{figure}} 
 \begin{figure*}
    \centering
    \includegraphics[angle=0, scale=0.3]{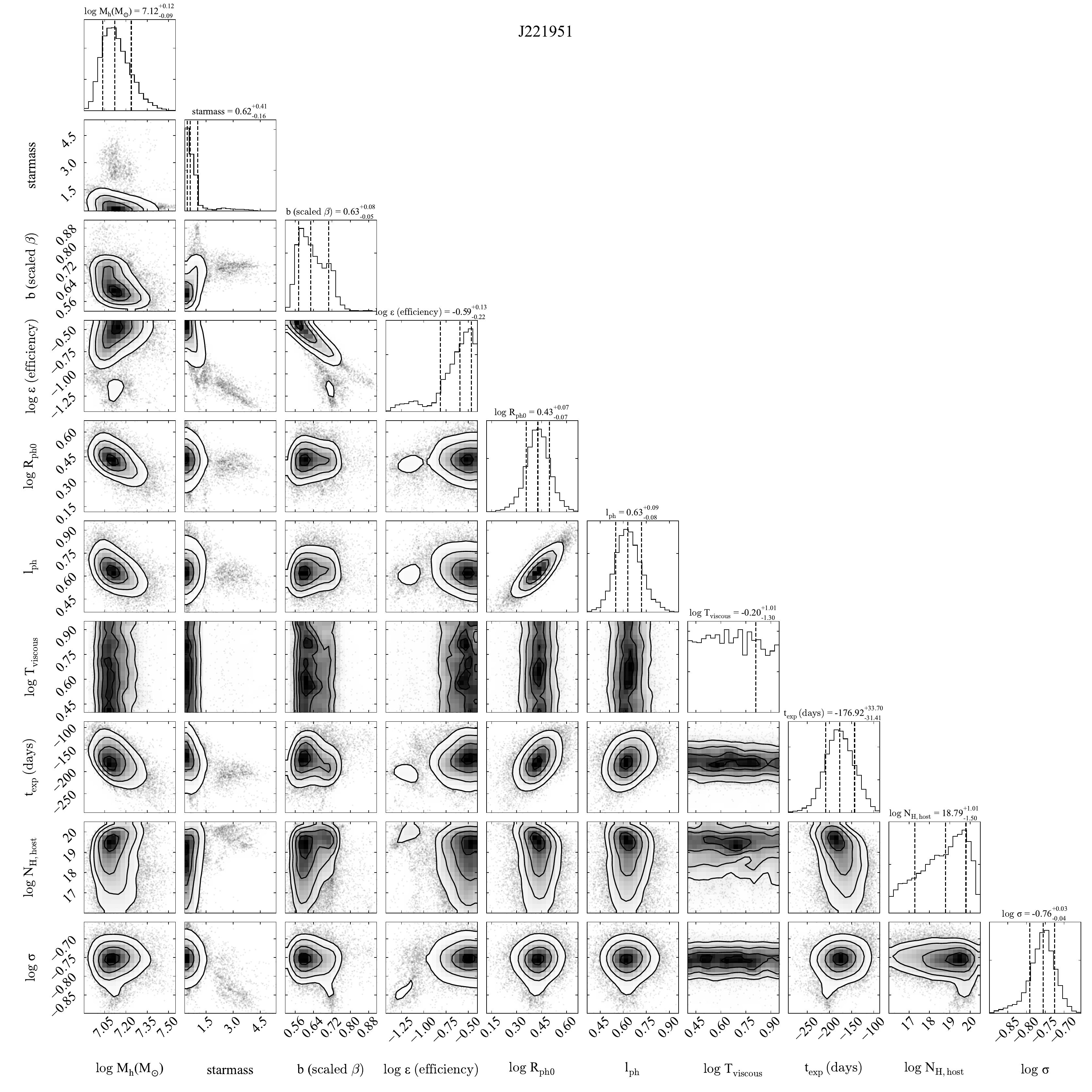}
    \caption{ Posterior probability density functions for the free parameters of the model light curves in Fig. 8 of the main paper.}
    \label{Mosfit_fit}
\end{figure*}

\begin{figure*}
    \centering
    \includegraphics[angle=0, scale=0.3]{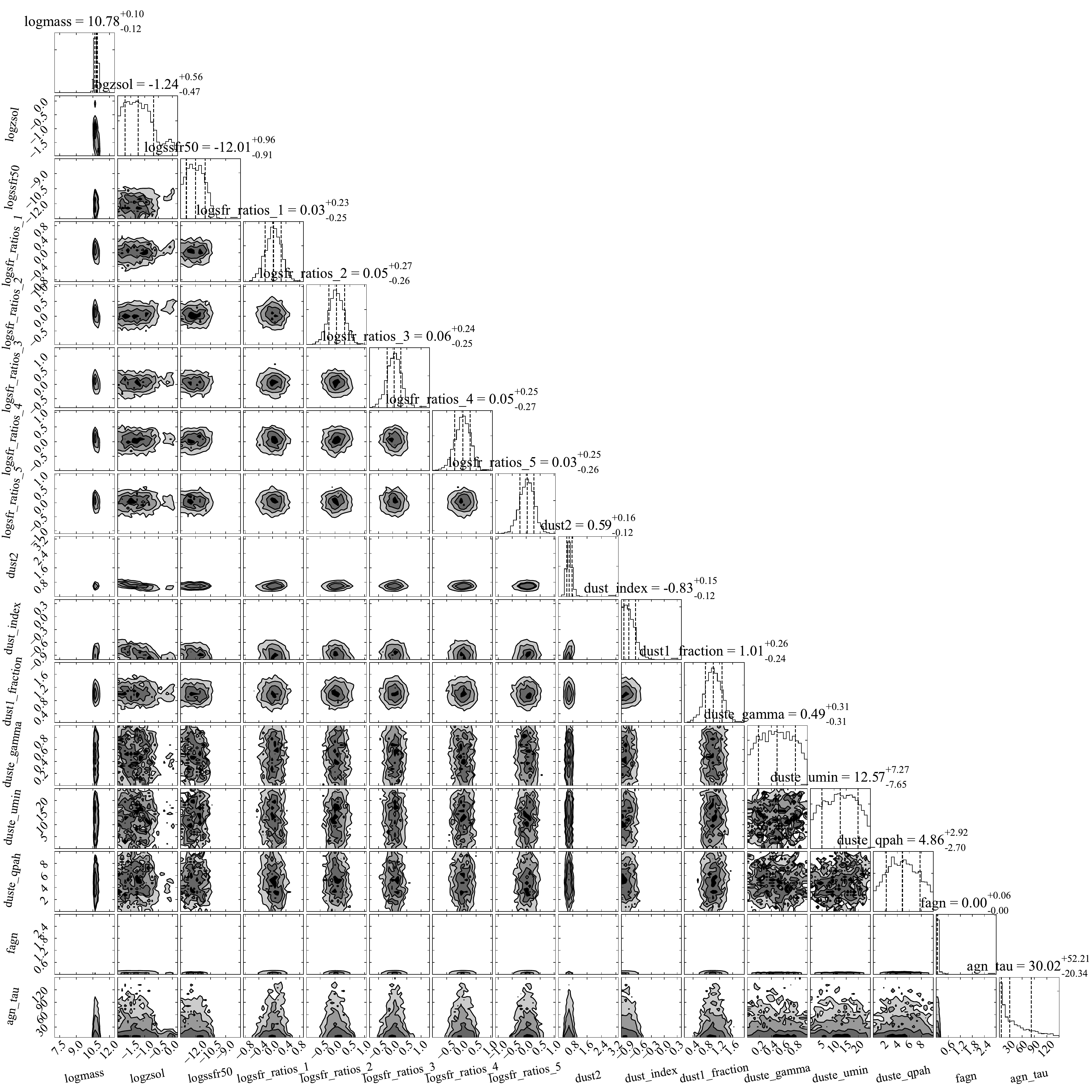}
    \caption{Posterior probability density functions for the free parameters of the {\sc Prospector} fit to the host photometry as shown in Fig. 15 of the main paper.}
    \label{Host_fit}
\end{figure*}

\onecolumn
\renewcommand\thetable{S.\arabic{table}}  
\begin{longtable}{lccccc} 
\caption{Log of the photometric observations. All magnitudes are given in AB, except the WISE magnitudes, which are Vega. No correction for extinction has been applied. Three sigma upper limits are given for data with signal-to-noise $<2$. \label{tab:obs}}
\\
\hline
Mid Time (MJD) & Bin Width (days) & Magnitude & Filter & Telescope & Reference \\
\hline
58758.68465 & 0.00046 & $19.04^{+0.62}_{-0.39}$ & $v$ & {\it Swift} & -- \\ 
58760.44175 & 0.16689 & $>18.82$ & $v$ & {\it Swift} & -- \\
58766.12753 & 0.06537 & $19.80^{+0.72}_{-0.43}$ & $v$ & {\it Swift} & -- \\ 
58773.26051 & 0.09620 & $>19.11$ & $v$ & {\it Swift} & -- \\
58780.19608 & 0.00160 & $19.52^{+0.50}_{-0.34}$ & $v$ & {\it Swift} & -- \\ 
58787.60099 & 0.09998 & $>18.43$ & $v$ & {\it Swift} & -- \\
58791.15503 & 0.79531 & $19.59^{+0.57}_{-0.37}$ & $v$ & {\it Swift} & -- \\ 
58801.08327 & 0.03526 & $>18.78$ & $v$ & {\it Swift} & -- \\
58808.94396 & 0.00029 & $>18.10$ & $v$ & {\it Swift} & -- \\
58815.55075 & 0.43356 & $19.60^{+0.68}_{-0.41}$ & $v$ & {\it Swift} & -- \\ 
58822.49320 & 0.47062 & $>18.72$ & $v$ & {\it Swift} & -- \\
58829.20136 & 0.13200 & $19.33^{+0.49}_{-0.34}$ & $v$ & {\it Swift} & -- \\ 
58836.50985 & 0.39919 & $19.20^{+0.44}_{-0.31}$ & $v$ & {\it Swift} & -- \\ 
58843.44692 & 0.10026 & $>18.95$ & $v$ & {\it Swift} & -- \\
58919.64104 & 0.33148 & $>18.26$ & $v$ & {\it Swift} & -- \\
58921.29634 & 0.06691 & $>18.28$ & $v$ & {\it Swift} & -- \\
58935.40477 & 0.29896 & $>18.22$ & $v$ & {\it Swift} & -- \\
58942.18132 & 0.16527 & $>18.48$ & $v$ & {\it Swift} & -- \\
58949.84458 & 0.00010 & $>17.89$ & $v$ & {\it Swift} & -- \\
58956.58525 & 0.35982 & $>18.65$ & $v$ & {\it Swift} & -- \\
58963.29886 & 0.09964 & $>19.62$ & $v$ & {\it Swift} & -- \\
58970.36505 & 0.19878 & $>18.70$ & $v$ & {\it Swift} & -- \\
58976.76503 & 0.03359 & $>18.86$ & $v$ & {\it Swift} & -- \\
58977.82978 & 0.69454 & $>18.85$ & $v$ & {\it Swift} & -- \\
58984.30065 & 0.06308 & $>18.54$ & $v$ & {\it Swift} & -- \\
58990.81583 & 0.99696 & $19.32^{+0.48}_{-0.33}$ & $v$ & {\it Swift} & -- \\ 
58998.54952 & 0.29946 & $19.56^{+0.55}_{-0.36}$ & $v$ & {\it Swift} & -- \\ 
59005.58852 & 0.36204 & $>19.14$ & $v$ & {\it Swift} & -- \\
59012.85190 & 0.13290 & $>19.20$ & $v$ & {\it Swift} & -- \\
59019.29892 & 0.19909 & $>19.39$ & $v$ & {\it Swift} & -- \\
59026.49603 & 0.36396 & $>19.11$ & $v$ & {\it Swift} & -- \\
59034.82707 & 1.59350 & $>19.30$ & $v$ & {\it Swift} & -- \\
59054.78952 & 0.23353 & $>19.86$ & $v$ & {\it Swift} & -- \\
59060.65864 & 0.33888 & $>19.18$ & $v$ & {\it Swift} & -- \\
59071.67341 & 3.32464 & $>19.50$ & $v$ & {\it Swift} & -- \\
59076.66080 & 0.52967 & $>19.08$ & $v$ & {\it Swift} & -- \\
59081.69581 & 1.85366 & $>18.78$ & $v$ & {\it Swift} & -- \\
59087.21008 & 1.59133 & $>19.10$ & $v$ & {\it Swift} & -- \\
59091.79808 & 1.73226 & $>19.37$ & $v$ & {\it Swift} & -- \\
59097.90413 & 0.59843 & $>19.07$ & $v$ & {\it Swift} & -- \\
59102.46257 & 0.43191 & $>19.40$ & $v$ & {\it Swift} & -- \\
59116.48845 & 0.40562 & $>19.13$ & $v$ & {\it Swift} & -- \\
59133.31357 & 0.30064 & $>19.37$ & $v$ & {\it Swift} & -- \\
59152.84635 & 0.03491 & $>18.68$ & $v$ & {\it Swift} & -- \\
59162.16786 & 0.67411 & $>19.05$ & $v$ & {\it Swift} & -- \\
59166.84168 & 0.62755 & $>18.81$ & $v$ & {\it Swift} & -- \\
59172.61692 & 0.36363 & $>19.64$ & $v$ & {\it Swift} & -- \\
59186.69754 & 0.23546 & $>19.07$ & $v$ & {\it Swift} & -- \\
59200.54414 & 0.27191 & $>19.43$ & $v$ & {\it Swift} & -- \\
59214.25326 & 0.10185 & $>18.93$ & $v$ & {\it Swift} & -- \\
59300.14463 & 2.22472 & $>18.79$ & $v$ & {\it Swift} & -- \\
59325.79693 & 0.13248 & $>19.32$ & $v$ & {\it Swift} & -- \\
59353.51931 & 0.43237 & $>19.59$ & $v$ & {\it Swift} & -- \\
59381.98648 & 0.96558 & $>19.37$ & $v$ & {\it Swift} & -- \\
59409.46649 & 0.36167 & $>19.48$ & $v$ & {\it Swift} & -- \\
59437.70631 & 0.27157 & $>19.51$ & $v$ & {\it Swift} & -- \\
58758.67991 & 0.00046 & $19.40^{+0.37}_{-0.28}$ & $b$ & {\it Swift} & -- \\ 
58760.43780 & 0.16744 & $19.55^{+0.31}_{-0.24}$ & $b$ & {\it Swift} & -- \\ 
58766.12436 & 0.06465 & $19.80^{+0.28}_{-0.22}$ & $b$ & {\it Swift} & -- \\ 
58773.25637 & 0.09751 & $19.45^{+0.19}_{-0.16}$ & $b$ & {\it Swift} & -- \\ 
58787.59860 & 0.10012 & $19.18^{+0.32}_{-0.25}$ & $b$ & {\it Swift} & -- \\ 
58791.15074 & 0.79620 & $20.10^{+0.36}_{-0.27}$ & $b$ & {\it Swift} & -- \\ 
58801.34234 & 0.29722 & $19.69^{+0.40}_{-0.29}$ & $b$ & {\it Swift} & -- \\ 
58808.94091 & 0.00029 & $>18.65$ & $b$ & {\it Swift} & -- \\
58815.54647 & 0.43277 & $>19.36$ & $b$ & {\it Swift} & -- \\
58822.48861 & 0.47004 & $20.09^{+0.57}_{-0.37}$ & $b$ & {\it Swift} & -- \\ 
58829.19654 & 0.13279 & $19.95^{+0.53}_{-0.36}$ & $b$ & {\it Swift} & -- \\ 
58836.50415 & 0.39892 & $19.27^{+0.28}_{-0.22}$ & $b$ & {\it Swift} & -- \\ 
58843.44253 & 0.10099 & $19.59^{+0.34}_{-0.26}$ & $b$ & {\it Swift} & -- \\ 
58919.63885 & 0.33105 & $19.77^{+0.72}_{-0.43}$ & $b$ & {\it Swift} & -- \\ 
58921.29220 & 0.06712 & $>19.10$ & $b$ & {\it Swift} & -- \\
58935.39978 & 0.29868 & $>19.11$ & $b$ & {\it Swift} & -- \\
58942.17691 & 0.16381 & $>19.03$ & $b$ & {\it Swift} & -- \\
58949.83996 & 0.00040 & $>18.98$ & $b$ & {\it Swift} & -- \\
58956.58184 & 0.36151 & $20.07^{+0.50}_{-0.34}$ & $b$ & {\it Swift} & -- \\ 
58963.29520 & 0.10129 & $>19.81$ & $b$ & {\it Swift} & -- \\
58970.36193 & 0.19847 & $20.22^{+0.75}_{-0.44}$ & $b$ & {\it Swift} & -- \\ 
58976.75721 & 0.03382 & $19.93^{+0.30}_{-0.23}$ & $b$ & {\it Swift} & -- \\ 
58977.82715 & 0.69416 & $20.11^{+0.59}_{-0.38}$ & $b$ & {\it Swift} & -- \\ 
58984.26557 & 0.09500 & $19.95^{+0.41}_{-0.30}$ & $b$ & {\it Swift} & -- \\ 
58990.80882 & 0.99638 & $>20.05$ & $b$ & {\it Swift} & -- \\
58998.54137 & 0.29942 & $20.68^{+0.63}_{-0.40}$ & $b$ & {\it Swift} & -- \\ 
59005.58619 & 0.36235 & $20.39^{+0.43}_{-0.31}$ & $b$ & {\it Swift} & -- \\ 
59012.84402 & 0.13290 & $20.18^{+0.45}_{-0.32}$ & $b$ & {\it Swift} & -- \\ 
59019.29437 & 0.19828 & $>19.72$ & $b$ & {\it Swift} & -- \\
59026.49056 & 0.36150 & $20.68^{+0.59}_{-0.38}$ & $b$ & {\it Swift} & -- \\ 
59034.82263 & 1.59611 & $20.42^{+0.50}_{-0.34}$ & $b$ & {\it Swift} & -- \\ 
59054.78463 & 0.23364 & $>20.18$ & $b$ & {\it Swift} & -- \\
59060.65477 & 0.33857 & $>19.83$ & $b$ & {\it Swift} & -- \\
59071.66996 & 3.32522 & $21.05^{+0.72}_{-0.43}$ & $b$ & {\it Swift} & -- \\ 
59076.65815 & 0.52984 & $>19.90$ & $b$ & {\it Swift} & -- \\
59081.69338 & 1.85378 & $>19.76$ & $b$ & {\it Swift} & -- \\
59087.20397 & 1.59347 & $20.64^{+0.68}_{-0.41}$ & $b$ & {\it Swift} & -- \\ 
59091.79305 & 1.73048 & $20.69^{+0.72}_{-0.43}$ & $b$ & {\it Swift} & -- \\ 
59098.13531 & 0.83517 & $20.61^{+0.48}_{-0.33}$ & $b$ & {\it Swift} & -- \\ 
59102.45759 & 0.43209 & $20.36^{+0.38}_{-0.28}$ & $b$ & {\it Swift} & -- \\ 
59108.71876 & 0.00118 & $>20.09$ & $b$ & {\it Swift} & -- \\
59116.48588 & 0.40599 & $20.77^{+0.62}_{-0.39}$ & $b$ & {\it Swift} & -- \\ 
59133.30920 & 0.30032 & $>19.86$ & $b$ & {\it Swift} & -- \\
59153.34541 & 0.53662 & $>19.77$ & $b$ & {\it Swift} & -- \\
59162.16531 & 0.67358 & $>19.42$ & $b$ & {\it Swift} & -- \\
59166.83896 & 0.62842 & $>19.61$ & $b$ & {\it Swift} & -- \\
59172.61206 & 0.36545 & $>19.69$ & $b$ & {\it Swift} & -- \\
59186.69305 & 0.23786 & $>19.62$ & $b$ & {\it Swift} & -- \\
59200.53793 & 0.27023 & $>19.97$ & $b$ & {\it Swift} & -- \\
59214.24674 & 0.10175 & $>20.10$ & $b$ & {\it Swift} & -- \\
59299.74921 & 2.61736 & $>19.66$ & $b$ & {\it Swift} & -- \\
59325.79085 & 0.13404 & $>20.41$ & $b$ & {\it Swift} & -- \\
59353.51215 & 0.43141 & $>20.14$ & $b$ & {\it Swift} & -- \\
59381.98167 & 0.96269 & $21.06^{+0.61}_{-0.39}$ & $b$ & {\it Swift} & -- \\ 
59409.46071 & 0.36403 & $>20.23$ & $b$ & {\it Swift} & -- \\
59437.70091 & 0.27164 & $>20.25$ & $b$ & {\it Swift} & -- \\
58756.77361 & 0.00043 & $19.40^{+0.21}_{-0.17}$ & $u$ & {\it Swift} & -- \\ 
58758.67894 & 0.00046 & $19.58^{+0.23}_{-0.19}$ & $u$ & {\it Swift} & -- \\ 
58760.43698 & 0.16755 & $19.57^{+0.18}_{-0.15}$ & $u$ & {\it Swift} & -- \\ 
58766.12374 & 0.06454 & $19.75^{+0.14}_{-0.13}$ & $u$ & {\it Swift} & -- \\ 
58773.25552 & 0.09777 & $19.70^{+0.12}_{-0.11}$ & $u$ & {\it Swift} & -- \\ 
58780.20476 & 0.00054 & $19.62^{+0.22}_{-0.18}$ & $u$ & {\it Swift} & -- \\ 
58787.59809 & 0.10015 & $19.50^{+0.22}_{-0.18}$ & $u$ & {\it Swift} & -- \\ 
58791.14985 & 0.79638 & $19.58^{+0.12}_{-0.11}$ & $u$ & {\it Swift} & -- \\ 
58801.34179 & 0.29728 & $20.23^{+0.32}_{-0.24}$ & $u$ & {\it Swift} & -- \\ 
58808.94026 & 0.00029 & $19.85^{+0.61}_{-0.39}$ & $u$ & {\it Swift} & -- \\ 
58815.54558 & 0.43261 & $19.88^{+0.23}_{-0.19}$ & $u$ & {\it Swift} & -- \\ 
58822.48767 & 0.46992 & $19.76^{+0.20}_{-0.17}$ & $u$ & {\it Swift} & -- \\ 
58829.19554 & 0.13295 & $19.96^{+0.25}_{-0.20}$ & $u$ & {\it Swift} & -- \\ 
58836.50298 & 0.39887 & $19.63^{+0.20}_{-0.17}$ & $u$ & {\it Swift} & -- \\ 
58843.44163 & 0.10114 & $19.76^{+0.19}_{-0.17}$ & $u$ & {\it Swift} & -- \\ 
58919.63845 & 0.33098 & $20.12^{+0.43}_{-0.31}$ & $u$ & {\it Swift} & -- \\ 
58921.29147 & 0.06715 & $20.47^{+0.47}_{-0.33}$ & $u$ & {\it Swift} & -- \\ 
58935.39892 & 0.29864 & $>19.78$ & $u$ & {\it Swift} & -- \\
58942.17614 & 0.16357 & $20.28^{+0.44}_{-0.31}$ & $u$ & {\it Swift} & -- \\ 
58949.81379 & 0.02572 & $20.12^{+0.41}_{-0.30}$ & $u$ & {\it Swift} & -- \\ 
58956.58124 & 0.36180 & $20.04^{+0.23}_{-0.19}$ & $u$ & {\it Swift} & -- \\ 
58963.29455 & 0.10156 & $20.56^{+0.41}_{-0.30}$ & $u$ & {\it Swift} & -- \\ 
58970.36138 & 0.19842 & $19.96^{+0.26}_{-0.21}$ & $u$ & {\it Swift} & -- \\ 
58976.75562 & 0.03386 & $20.54^{+0.25}_{-0.20}$ & $u$ & {\it Swift} & -- \\ 
58977.82668 & 0.69410 & $20.36^{+0.34}_{-0.26}$ & $u$ & {\it Swift} & -- \\ 
58984.26465 & 0.09527 & $20.55^{+0.35}_{-0.27}$ & $u$ & {\it Swift} & -- \\ 
58990.80739 & 0.99627 & $20.87^{+0.33}_{-0.25}$ & $u$ & {\it Swift} & -- \\ 
58998.53971 & 0.29942 & $20.58^{+0.25}_{-0.20}$ & $u$ & {\it Swift} & -- \\ 
59005.58569 & 0.36241 & $20.59^{+0.24}_{-0.20}$ & $u$ & {\it Swift} & -- \\ 
59012.84241 & 0.13290 & $20.52^{+0.24}_{-0.20}$ & $u$ & {\it Swift} & -- \\ 
59019.29342 & 0.19811 & $20.53^{+0.28}_{-0.22}$ & $u$ & {\it Swift} & -- \\ 
59026.48943 & 0.36102 & $20.79^{+0.29}_{-0.23}$ & $u$ & {\it Swift} & -- \\ 
59034.82171 & 1.59663 & $20.96^{+0.38}_{-0.28}$ & $u$ & {\it Swift} & -- \\ 
59054.78362 & 0.23365 & $21.27^{+0.40}_{-0.29}$ & $u$ & {\it Swift} & -- \\ 
59060.65397 & 0.33850 & $21.11^{+0.50}_{-0.34}$ & $u$ & {\it Swift} & -- \\ 
59071.66923 & 3.32533 & $>20.98$ & $u$ & {\it Swift} & -- \\
59076.65759 & 0.52987 & $20.95^{+0.37}_{-0.28}$ & $u$ & {\it Swift} & -- \\ 
59081.69286 & 1.85381 & $21.07^{+0.47}_{-0.33}$ & $u$ & {\it Swift} & -- \\ 
59087.20272 & 1.59389 & $>20.57$ & $u$ & {\it Swift} & -- \\
59091.79201 & 1.73012 & $21.17^{+0.49}_{-0.34}$ & $u$ & {\it Swift} & -- \\ 
59098.13429 & 0.83530 & $21.02^{+0.32}_{-0.25}$ & $u$ & {\it Swift} & -- \\ 
59102.45657 & 0.43212 & $21.23^{+0.41}_{-0.29}$ & $u$ & {\it Swift} & -- \\ 
59108.71635 & 0.00118 & $21.35^{+0.66}_{-0.41}$ & $u$ & {\it Swift} & -- \\ 
59116.48534 & 0.40606 & $21.27^{+0.45}_{-0.32}$ & $u$ & {\it Swift} & -- \\ 
59133.30829 & 0.30025 & $20.89^{+0.37}_{-0.28}$ & $u$ & {\it Swift} & -- \\ 
59153.34460 & 0.53636 & $21.37^{+0.68}_{-0.42}$ & $u$ & {\it Swift} & -- \\ 
59162.16477 & 0.67348 & $20.87^{+0.62}_{-0.39}$ & $u$ & {\it Swift} & -- \\ 
59166.83839 & 0.62859 & $>20.29$ & $u$ & {\it Swift} & -- \\
59172.61106 & 0.36581 & $20.76^{+0.38}_{-0.28}$ & $u$ & {\it Swift} & -- \\ 
59186.69212 & 0.23834 & $20.70^{+0.43}_{-0.31}$ & $u$ & {\it Swift} & -- \\ 
59200.53604 & 0.26927 & $20.72^{+0.32}_{-0.25}$ & $u$ & {\it Swift} & -- \\ 
59214.24541 & 0.10173 & $>20.70$ & $u$ & {\it Swift} & -- \\
59299.74830 & 2.61769 & $21.14^{+0.58}_{-0.38}$ & $u$ & {\it Swift} & -- \\ 
59325.78959 & 0.13435 & $21.39^{+0.37}_{-0.28}$ & $u$ & {\it Swift} & -- \\ 
59353.51069 & 0.43122 & $>20.76$ & $u$ & {\it Swift} & -- \\
59381.98068 & 0.96211 & $21.51^{+0.40}_{-0.29}$ & $u$ & {\it Swift} & -- \\ 
59409.45952 & 0.36451 & $21.75^{+0.53}_{-0.36}$ & $u$ & {\it Swift} & -- \\ 
59437.69981 & 0.27165 & $21.57^{+0.45}_{-0.32}$ & $u$ & {\it Swift} & -- \\ 
59470.18666 & 1.55946 & $22.02^{+0.38}_{-0.28}$ & $u$ & {\it Swift} & -- \\ 
59497.00830 & 0.96442 & $21.88^{+0.32}_{-0.25}$ & $u$ & {\it Swift} & -- \\ 
59528.82470 & 0.73400 & $>21.19$ & $u$ & {\it Swift} & -- \\
59558.39374 & 0.36575 & $>21.57$ & $u$ & {\it Swift} & -- \\
59582.28177 & 0.16837 & $>21.06$ & $u$ & {\it Swift} & -- \\
59652.35462 & 2.48827 & $>21.09$ & $u$ & {\it Swift} & -- \\
58758.67750 & 0.00092 & $19.73^{+0.18}_{-0.15}$ & $uvw1$ & {\it Swift} & -- \\ 
58760.43573 & 0.16809 & $19.86^{+0.15}_{-0.13}$ & $uvw1$ & {\it Swift} & -- \\ 
58766.12285 & 0.06463 & $19.53^{+0.09}_{-0.08}$ & $uvw1$ & {\it Swift} & -- \\ 
58773.25413 & 0.09856 & $19.70^{+0.09}_{-0.08}$ & $uvw1$ & {\it Swift} & -- \\ 
58780.16931 & 0.03486 & $19.79^{+0.10}_{-0.09}$ & $uvw1$ & {\it Swift} & -- \\ 
58787.59735 & 0.10042 & $19.59^{+0.17}_{-0.14}$ & $uvw1$ & {\it Swift} & -- \\ 
58791.14846 & 0.79706 & $19.79^{+0.10}_{-0.09}$ & $uvw1$ & {\it Swift} & -- \\ 
58801.34097 & 0.29761 & $20.09^{+0.16}_{-0.14}$ & $uvw1$ & {\it Swift} & -- \\ 
58808.93934 & 0.00058 & $19.73^{+0.31}_{-0.24}$ & $uvw1$ & {\it Swift} & -- \\ 
58813.98795 & 0.00206 & $20.29^{+0.19}_{-0.16}$ & $uvw1$ & {\it Swift} & -- \\ 
58814.26177 & 0.00254 & $20.39^{+0.17}_{-0.14}$ & $uvw1$ & {\it Swift} & -- \\ 
58815.44657 & 0.00247 & $20.17^{+0.16}_{-0.14}$ & $uvw1$ & {\it Swift} & -- \\ 
58815.57782 & 0.39932 & $20.38^{+0.20}_{-0.17}$ & $uvw1$ & {\it Swift} & -- \\ 
58816.65269 & 0.00260 & $20.22^{+0.14}_{-0.13}$ & $uvw1$ & {\it Swift} & -- \\ 
58817.31819 & 0.00238 & $20.45^{+0.18}_{-0.15}$ & $uvw1$ & {\it Swift} & -- \\ 
58818.24210 & 0.00347 & $20.09^{+0.12}_{-0.11}$ & $uvw1$ & {\it Swift} & -- \\ 
58819.29672 & 0.00301 & $20.30^{+0.16}_{-0.14}$ & $uvw1$ & {\it Swift} & -- \\ 
58820.02573 & 0.00039 & $20.20^{+0.52}_{-0.35}$ & $uvw1$ & {\it Swift} & -- \\ 
58820.10494 & 0.00173 & $19.94^{+0.15}_{-0.13}$ & $uvw1$ & {\it Swift} & -- \\ 
58820.16116 & 0.00246 & $20.29^{+0.18}_{-0.15}$ & $uvw1$ & {\it Swift} & -- \\ 
58820.23554 & 0.00308 & $20.27^{+0.14}_{-0.12}$ & $uvw1$ & {\it Swift} & -- \\ 
58820.36324 & 0.00246 & $20.16^{+0.15}_{-0.13}$ & $uvw1$ & {\it Swift} & -- \\ 
58820.43264 & 0.00321 & $20.28^{+0.14}_{-0.13}$ & $uvw1$ & {\it Swift} & -- \\ 
58820.50219 & 0.00310 & $20.25^{+0.14}_{-0.12}$ & $uvw1$ & {\it Swift} & -- \\ 
58820.56871 & 0.00325 & $20.31^{+0.14}_{-0.12}$ & $uvw1$ & {\it Swift} & -- \\ 
58820.63438 & 0.00286 & $20.07^{+0.13}_{-0.11}$ & $uvw1$ & {\it Swift} & -- \\ 
58820.70349 & 0.00319 & $20.37^{+0.14}_{-0.13}$ & $uvw1$ & {\it Swift} & -- \\ 
58820.77275 & 0.00337 & $20.11^{+0.12}_{-0.11}$ & $uvw1$ & {\it Swift} & -- \\ 
58820.82374 & 0.00238 & $20.23^{+0.19}_{-0.16}$ & $uvw1$ & {\it Swift} & -- \\ 
58820.89275 & 0.00281 & $20.32^{+0.17}_{-0.15}$ & $uvw1$ & {\it Swift} & -- \\ 
58820.97266 & 0.00276 & $20.35^{+0.15}_{-0.13}$ & $uvw1$ & {\it Swift} & -- \\ 
58822.48634 & 0.47019 & $20.25^{+0.18}_{-0.16}$ & $uvw1$ & {\it Swift} & -- \\ 
58829.19400 & 0.13366 & $19.88^{+0.11}_{-0.10}$ & $uvw1$ & {\it Swift} & -- \\ 
58836.50128 & 0.39934 & $19.98^{+0.13}_{-0.12}$ & $uvw1$ & {\it Swift} & -- \\ 
58843.44022 & 0.10179 & $19.78^{+0.10}_{-0.09}$ & $uvw1$ & {\it Swift} & -- \\ 
58919.63779 & 0.33118 & $20.40^{+0.18}_{-0.15}$ & $uvw1$ & {\it Swift} & -- \\ 
58921.29004 & 0.06790 & $20.26^{+0.14}_{-0.13}$ & $uvw1$ & {\it Swift} & -- \\ 
58935.39729 & 0.29935 & $20.39^{+0.18}_{-0.16}$ & $uvw1$ & {\it Swift} & -- \\ 
58942.17490 & 0.16380 & $20.53^{+0.18}_{-0.15}$ & $uvw1$ & {\it Swift} & -- \\ 
58949.81187 & 0.02679 & $20.45^{+0.18}_{-0.15}$ & $uvw1$ & {\it Swift} & -- \\ 
58956.57981 & 0.36292 & $20.50^{+0.15}_{-0.14}$ & $uvw1$ & {\it Swift} & -- \\ 
58963.29305 & 0.10271 & $20.48^{+0.15}_{-0.14}$ & $uvw1$ & {\it Swift} & -- \\ 
58970.36037 & 0.19883 & $20.44^{+0.16}_{-0.14}$ & $uvw1$ & {\it Swift} & -- \\ 
58976.75324 & 0.03470 & $20.76^{+0.18}_{-0.15}$ & $uvw1$ & {\it Swift} & -- \\ 
58977.82585 & 0.69438 & $20.80^{+0.19}_{-0.16}$ & $uvw1$ & {\it Swift} & -- \\ 
58984.26318 & 0.09610 & $20.59^{+0.18}_{-0.15}$ & $uvw1$ & {\it Swift} & -- \\ 
58990.80533 & 0.99678 & $20.55^{+0.13}_{-0.11}$ & $uvw1$ & {\it Swift} & -- \\ 
58998.53726 & 0.30021 & $21.10^{+0.23}_{-0.19}$ & $uvw1$ & {\it Swift} & -- \\ 
59005.58495 & 0.36272 & $20.73^{+0.17}_{-0.15}$ & $uvw1$ & {\it Swift} & -- \\ 
59012.84003 & 0.13368 & $20.90^{+0.20}_{-0.17}$ & $uvw1$ & {\it Swift} & -- \\ 
59019.29211 & 0.19832 & $20.66^{+0.19}_{-0.16}$ & $uvw1$ & {\it Swift} & -- \\ 
59026.48801 & 0.36082 & $20.74^{+0.16}_{-0.14}$ & $uvw1$ & {\it Swift} & -- \\ 
59034.82011 & 1.59784 & $21.30^{+0.27}_{-0.21}$ & $uvw1$ & {\it Swift} & -- \\ 
59054.78212 & 0.23416 & $21.11^{+0.18}_{-0.16}$ & $uvw1$ & {\it Swift} & -- \\ 
59060.65282 & 0.33879 & $20.85^{+0.22}_{-0.18}$ & $uvw1$ & {\it Swift} & -- \\ 
59068.77368 & 0.43139 & $20.94^{+0.16}_{-0.14}$ & $uvw1$ & {\it Swift} & -- \\ 
59074.99342 & 0.00055 & $21.21^{+0.74}_{-0.44}$ & $uvw1$ & {\it Swift} & -- \\ 
59076.65677 & 0.53017 & $21.16^{+0.24}_{-0.20}$ & $uvw1$ & {\it Swift} & -- \\ 
59081.69210 & 1.85407 & $21.37^{+0.32}_{-0.25}$ & $uvw1$ & {\it Swift} & -- \\ 
59087.20066 & 1.59513 & $21.53^{+0.30}_{-0.24}$ & $uvw1$ & {\it Swift} & -- \\ 
59091.79067 & 1.73007 & $21.78^{+0.43}_{-0.30}$ & $uvw1$ & {\it Swift} & -- \\ 
59098.13274 & 0.83597 & $21.85^{+0.34}_{-0.26}$ & $uvw1$ & {\it Swift} & -- \\ 
59102.45505 & 0.43266 & $21.39^{+0.24}_{-0.20}$ & $uvw1$ & {\it Swift} & -- \\ 
59108.71275 & 0.00236 & $21.91^{+0.55}_{-0.36}$ & $uvw1$ & {\it Swift} & -- \\ 
59116.48452 & 0.40641 & $21.60^{+0.31}_{-0.24}$ & $uvw1$ & {\it Swift} & -- \\ 
59133.30700 & 0.30058 & $21.47^{+0.29}_{-0.23}$ & $uvw1$ & {\it Swift} & -- \\ 
59153.34354 & 0.53637 & $21.78^{+0.48}_{-0.33}$ & $uvw1$ & {\it Swift} & -- \\ 
59162.16404 & 0.67357 & $21.59^{+0.49}_{-0.34}$ & $uvw1$ & {\it Swift} & -- \\ 
59166.83748 & 0.62911 & $21.14^{+0.37}_{-0.27}$ & $uvw1$ & {\it Swift} & -- \\ 
59172.60940 & 0.36683 & $20.79^{+0.17}_{-0.15}$ & $uvw1$ & {\it Swift} & -- \\ 
59186.69052 & 0.23950 & $21.13^{+0.26}_{-0.21}$ & $uvw1$ & {\it Swift} & -- \\ 
59200.53435 & 0.26937 & $21.07^{+0.21}_{-0.17}$ & $uvw1$ & {\it Swift} & -- \\ 
59214.24345 & 0.10235 & $21.98^{+0.45}_{-0.32}$ & $uvw1$ & {\it Swift} & -- \\ 
59299.74679 & 2.61862 & $22.07^{+0.50}_{-0.34}$ & $uvw1$ & {\it Swift} & -- \\ 
59325.78760 & 0.13541 & $21.88^{+0.34}_{-0.26}$ & $uvw1$ & {\it Swift} & -- \\ 
59353.47348 & 0.46678 & $21.65^{+0.29}_{-0.23}$ & $uvw1$ & {\it Swift} & -- \\ 
59381.97950 & 0.96170 & $22.04^{+0.33}_{-0.25}$ & $uvw1$ & {\it Swift} & -- \\ 
59409.45753 & 0.36578 & $21.70^{+0.25}_{-0.20}$ & $uvw1$ & {\it Swift} & -- \\ 
59437.69817 & 0.27220 & $22.15^{+0.38}_{-0.28}$ & $uvw1$ & {\it Swift} & -- \\ 
59470.19564 & 1.55752 & $22.63^{+0.47}_{-0.33}$ & $uvw1$ & {\it Swift} & -- \\ 
59497.01559 & 0.96466 & $22.32^{+0.31}_{-0.24}$ & $uvw1$ & {\it Swift} & -- \\ 
59528.82910 & 0.73239 & $22.28^{+0.37}_{-0.27}$ & $uvw1$ & {\it Swift} & -- \\ 
59558.40521 & 0.36809 & $22.55^{+0.44}_{-0.31}$ & $uvw1$ & {\it Swift} & -- \\ 
59582.29516 & 0.16910 & $>21.67$ & $uvw1$ & {\it Swift} & -- \\
59652.36026 & 2.49039 & $22.21^{+0.58}_{-0.38}$ & $uvw1$ & {\it Swift} & -- \\ 
59687.97614 & 2.89604 & $22.85^{+0.47}_{-0.33}$ & $uvw1$ & {\it Swift} & -- \\ 
58758.68662 & 0.00146 & $19.78^{+0.16}_{-0.14}$ & $uvm2$ & {\it Swift} & -- \\ 
58760.44331 & 0.16753 & $19.98^{+0.14}_{-0.12}$ & $uvm2$ & {\it Swift} & -- \\ 
58766.12912 & 0.06645 & $19.85^{+0.09}_{-0.09}$ & $uvm2$ & {\it Swift} & -- \\ 
58773.26209 & 0.09667 & $19.86^{+0.09}_{-0.08}$ & $uvm2$ & {\it Swift} & -- \\ 
58780.16607 & 0.03485 & $19.81^{+0.11}_{-0.10}$ & $uvm2$ & {\it Swift} & -- \\ 
58787.60203 & 0.10049 & $20.13^{+0.18}_{-0.16}$ & $uvm2$ & {\it Swift} & -- \\ 
58791.15664 & 0.79587 & $20.00^{+0.10}_{-0.10}$ & $uvm2$ & {\it Swift} & -- \\ 
58801.08494 & 0.03633 & $19.87^{+0.14}_{-0.12}$ & $uvm2$ & {\it Swift} & -- \\ 
58808.94528 & 0.00098 & $21.03^{+0.47}_{-0.33}$ & $uvm2$ & {\it Swift} & -- \\ 
58815.55292 & 0.43501 & $20.41^{+0.12}_{-0.11}$ & $uvm2$ & {\it Swift} & -- \\ 
58822.49525 & 0.47185 & $20.47^{+0.12}_{-0.11}$ & $uvm2$ & {\it Swift} & -- \\ 
58829.20317 & 0.13267 & $20.17^{+0.10}_{-0.09}$ & $uvm2$ & {\it Swift} & -- \\ 
58836.51200 & 0.40023 & $20.17^{+0.11}_{-0.10}$ & $uvm2$ & {\it Swift} & -- \\ 
58843.44837 & 0.10067 & $19.78^{+0.09}_{-0.09}$ & $uvm2$ & {\it Swift} & -- \\ 
58919.64300 & 0.33311 & $20.43^{+0.10}_{-0.09}$ & $uvm2$ & {\it Swift} & -- \\ 
58921.29933 & 0.06915 & $20.44^{+0.10}_{-0.09}$ & $uvm2$ & {\it Swift} & -- \\ 
58935.40872 & 0.30211 & $20.33^{+0.09}_{-0.09}$ & $uvm2$ & {\it Swift} & -- \\ 
58942.15225 & 0.13567 & $20.45^{+0.12}_{-0.10}$ & $uvm2$ & {\it Swift} & -- \\ 
58956.58698 & 0.36066 & $20.39^{+0.09}_{-0.08}$ & $uvm2$ & {\it Swift} & -- \\ 
58963.30063 & 0.10048 & $20.48^{+0.11}_{-0.10}$ & $uvm2$ & {\it Swift} & -- \\ 
58970.36763 & 0.20086 & $20.57^{+0.11}_{-0.10}$ & $uvm2$ & {\it Swift} & -- \\ 
58976.76814 & 0.03506 & $20.56^{+0.15}_{-0.13}$ & $uvm2$ & {\it Swift} & -- \\ 
58977.83157 & 0.69593 & $20.54^{+0.10}_{-0.09}$ & $uvm2$ & {\it Swift} & -- \\ 
58984.30219 & 0.06350 & $20.79^{+0.23}_{-0.19}$ & $uvm2$ & {\it Swift} & -- \\ 
58990.81898 & 0.99880 & $20.61^{+0.12}_{-0.10}$ & $uvm2$ & {\it Swift} & -- \\ 
58998.55279 & 0.30110 & $21.01^{+0.19}_{-0.16}$ & $uvm2$ & {\it Swift} & -- \\ 
59005.58942 & 0.36238 & $20.74^{+0.16}_{-0.14}$ & $uvm2$ & {\it Swift} & -- \\ 
59012.85565 & 0.13504 & $20.95^{+0.17}_{-0.14}$ & $uvm2$ & {\it Swift} & -- \\ 
59019.30086 & 0.20025 & $21.38^{+0.28}_{-0.22}$ & $uvm2$ & {\it Swift} & -- \\ 
59026.49850 & 0.36580 & $20.90^{+0.16}_{-0.14}$ & $uvm2$ & {\it Swift} & -- \\ 
59034.82847 & 1.59347 & $20.92^{+0.23}_{-0.19}$ & $uvm2$ & {\it Swift} & -- \\ 
59054.79156 & 0.23455 & $20.92^{+0.14}_{-0.12}$ & $uvm2$ & {\it Swift} & -- \\ 
59060.66010 & 0.33961 & $21.50^{+0.30}_{-0.24}$ & $uvm2$ & {\it Swift} & -- \\ 
59071.67466 & 3.32505 & $21.33^{+0.16}_{-0.14}$ & $uvm2$ & {\it Swift} & -- \\ 
59076.66190 & 0.53018 & $21.58^{+0.26}_{-0.21}$ & $uvm2$ & {\it Swift} & -- \\ 
59081.69689 & 1.85421 & $21.64^{+0.32}_{-0.25}$ & $uvm2$ & {\it Swift} & -- \\ 
59087.21215 & 1.59173 & $22.09^{+0.40}_{-0.29}$ & $uvm2$ & {\it Swift} & -- \\ 
59091.80046 & 1.73398 & $22.05^{+0.42}_{-0.30}$ & $uvm2$ & {\it Swift} & -- \\ 
59097.90605 & 0.59922 & $21.79^{+0.28}_{-0.22}$ & $uvm2$ & {\it Swift} & -- \\ 
59102.46432 & 0.43261 & $22.13^{+0.41}_{-0.29}$ & $uvm2$ & {\it Swift} & -- \\ 
59116.48950 & 0.40605 & $21.53^{+0.21}_{-0.18}$ & $uvm2$ & {\it Swift} & -- \\ 
59133.31542 & 0.30166 & $22.33^{+0.37}_{-0.28}$ & $uvm2$ & {\it Swift} & -- \\ 
59152.84822 & 0.03621 & $21.57^{+0.39}_{-0.29}$ & $uvm2$ & {\it Swift} & -- \\ 
59162.16882 & 0.67464 & $21.61^{+0.30}_{-0.23}$ & $uvm2$ & {\it Swift} & -- \\ 
59166.84271 & 0.62784 & $21.23^{+0.29}_{-0.23}$ & $uvm2$ & {\it Swift} & -- \\ 
59172.61851 & 0.36384 & $21.00^{+0.14}_{-0.12}$ & $uvm2$ & {\it Swift} & -- \\ 
59186.69896 & 0.23547 & $21.35^{+0.18}_{-0.16}$ & $uvm2$ & {\it Swift} & -- \\ 
59200.54615 & 0.27300 & $21.35^{+0.19}_{-0.16}$ & $uvm2$ & {\it Swift} & -- \\ 
59214.25573 & 0.10301 & $21.17^{+0.15}_{-0.13}$ & $uvm2$ & {\it Swift} & -- \\ 
59300.14598 & 2.22525 & $21.83^{+0.25}_{-0.20}$ & $uvm2$ & {\it Swift} & -- \\ 
59325.79917 & 0.13317 & $>22.15$ & $uvm2$ & {\it Swift} & -- \\
59353.48095 & 0.47361 & $22.24^{+0.37}_{-0.28}$ & $uvm2$ & {\it Swift} & -- \\ 
59381.98898 & 0.96766 & $21.71^{+0.22}_{-0.18}$ & $uvm2$ & {\it Swift} & -- \\ 
59409.46837 & 0.36189 & $22.17^{+0.28}_{-0.22}$ & $uvm2$ & {\it Swift} & -- \\ 
59437.70840 & 0.27254 & $21.99^{+0.23}_{-0.19}$ & $uvm2$ & {\it Swift} & -- \\ 
59470.19263 & 1.55814 & $22.34^{+0.32}_{-0.25}$ & $uvm2$ & {\it Swift} & -- \\ 
59497.01313 & 0.96454 & $22.01^{+0.22}_{-0.18}$ & $uvm2$ & {\it Swift} & -- \\ 
59528.82758 & 0.73287 & $22.55^{+0.44}_{-0.31}$ & $uvm2$ & {\it Swift} & -- \\ 
59558.40141 & 0.36732 & $22.32^{+0.29}_{-0.23}$ & $uvm2$ & {\it Swift} & -- \\ 
59582.29084 & 0.16900 & $22.25^{+0.38}_{-0.28}$ & $uvm2$ & {\it Swift} & -- \\ 
59652.35844 & 2.48974 & $22.76^{+0.53}_{-0.36}$ & $uvm2$ & {\it Swift} & -- \\ 
59687.97400 & 2.89505 & $22.64^{+0.25}_{-0.20}$ & $uvm2$ & {\it Swift} & -- \\ 
58758.68229 & 0.00185 & $20.03^{+0.13}_{-0.12}$ & $uvw2$ & {\it Swift} & -- \\ 
58760.43962 & 0.16831 & $19.92^{+0.09}_{-0.09}$ & $uvw2$ & {\it Swift} & -- \\ 
58766.12616 & 0.06593 & $19.78^{+0.07}_{-0.06}$ & $uvw2$ & {\it Swift} & -- \\ 
58773.25805 & 0.09806 & $19.82^{+0.06}_{-0.06}$ & $uvw2$ & {\it Swift} & -- \\ 
58780.12636 & 0.00158 & $19.87^{+0.13}_{-0.12}$ & $uvw2$ & {\it Swift} & -- \\ 
58787.59976 & 0.10073 & $20.12^{+0.14}_{-0.12}$ & $uvw2$ & {\it Swift} & -- \\ 
58791.15262 & 0.79701 & $20.16^{+0.08}_{-0.07}$ & $uvw2$ & {\it Swift} & -- \\ 
58801.34273 & 0.29700 & $20.24^{+0.12}_{-0.11}$ & $uvw2$ & {\it Swift} & -- \\ 
58808.94244 & 0.00117 & $20.42^{+0.23}_{-0.19}$ & $uvw2$ & {\it Swift} & -- \\ 
58815.54885 & 0.43442 & $20.59^{+0.10}_{-0.09}$ & $uvw2$ & {\it Swift} & -- \\ 
58822.49109 & 0.47167 & $20.53^{+0.10}_{-0.09}$ & $uvw2$ & {\it Swift} & -- \\ 
58829.19872 & 0.13380 & $20.29^{+0.08}_{-0.07}$ & $uvw2$ & {\it Swift} & -- \\ 
58836.50709 & 0.40073 & $20.18^{+0.08}_{-0.07}$ & $uvw2$ & {\it Swift} & -- \\ 
58843.44451 & 0.10191 & $20.01^{+0.07}_{-0.06}$ & $uvw2$ & {\it Swift} & -- \\ 
58919.64010 & 0.33196 & $20.51^{+0.12}_{-0.11}$ & $uvw2$ & {\it Swift} & -- \\ 
58921.29420 & 0.06836 & $20.40^{+0.10}_{-0.09}$ & $uvw2$ & {\it Swift} & -- \\ 
58935.40237 & 0.30045 & $20.37^{+0.11}_{-0.10}$ & $uvw2$ & {\it Swift} & -- \\ 
58942.17961 & 0.16597 & $20.75^{+0.12}_{-0.11}$ & $uvw2$ & {\it Swift} & -- \\ 
58949.84242 & 0.00200 & $20.66^{+0.20}_{-0.17}$ & $uvw2$ & {\it Swift} & -- \\ 
58956.58299 & 0.36177 & $20.62^{+0.11}_{-0.10}$ & $uvw2$ & {\it Swift} & -- \\ 
58963.29649 & 0.10165 & $20.37^{+0.10}_{-0.09}$ & $uvw2$ & {\it Swift} & -- \\ 
58970.36360 & 0.19962 & $20.56^{+0.12}_{-0.10}$ & $uvw2$ & {\it Swift} & -- \\ 
58976.76106 & 0.03602 & $20.75^{+0.11}_{-0.10}$ & $uvw2$ & {\it Swift} & -- \\ 
58977.82860 & 0.69519 & $20.70^{+0.12}_{-0.11}$ & $uvw2$ & {\it Swift} & -- \\ 
58984.26742 & 0.09564 & $21.01^{+0.15}_{-0.13}$ & $uvw2$ & {\it Swift} & -- \\ 
58990.81250 & 0.99873 & $20.76^{+0.09}_{-0.09}$ & $uvw2$ & {\it Swift} & -- \\ 
58998.54546 & 0.30185 & $21.01^{+0.12}_{-0.11}$ & $uvw2$ & {\it Swift} & -- \\ 
59005.58727 & 0.36286 & $20.87^{+0.12}_{-0.10}$ & $uvw2$ & {\it Swift} & -- \\ 
59012.84796 & 0.13523 & $21.12^{+0.13}_{-0.12}$ & $uvw2$ & {\it Swift} & -- \\ 
59019.29689 & 0.20001 & $20.85^{+0.13}_{-0.12}$ & $uvw2$ & {\it Swift} & -- \\ 
59026.49403 & 0.36433 & $20.91^{+0.11}_{-0.10}$ & $uvw2$ & {\it Swift} & -- \\ 
59034.82408 & 1.59610 & $21.01^{+0.17}_{-0.14}$ & $uvw2$ & {\it Swift} & -- \\ 
59054.78705 & 0.23502 & $21.33^{+0.13}_{-0.11}$ & $uvw2$ & {\it Swift} & -- \\ 
59060.65681 & 0.33985 & $21.28^{+0.16}_{-0.14}$ & $uvw2$ & {\it Swift} & -- \\ 
59071.67152 & 3.32593 & $21.50^{+0.13}_{-0.11}$ & $uvw2$ & {\it Swift} & -- \\ 
59076.65943 & 0.53051 & $21.37^{+0.15}_{-0.13}$ & $uvw2$ & {\it Swift} & -- \\ 
59081.69456 & 1.85442 & $21.50^{+0.19}_{-0.16}$ & $uvw2$ & {\it Swift} & -- \\ 
59087.20638 & 1.59420 & $21.55^{+0.18}_{-0.15}$ & $uvw2$ & {\it Swift} & -- \\ 
59091.79610 & 1.73284 & $21.63^{+0.19}_{-0.16}$ & $uvw2$ & {\it Swift} & -- \\ 
59098.13731 & 0.83602 & $21.71^{+0.16}_{-0.14}$ & $uvw2$ & {\it Swift} & -- \\ 
59102.46004 & 0.43346 & $21.79^{+0.17}_{-0.15}$ & $uvw2$ & {\it Swift} & -- \\ 
59108.72391 & 0.00390 & $22.14^{+0.35}_{-0.26}$ & $uvw2$ & {\it Swift} & -- \\ 
59116.48706 & 0.40654 & $21.73^{+0.17}_{-0.15}$ & $uvw2$ & {\it Swift} & -- \\ 
59133.31149 & 0.30176 & $21.85^{+0.17}_{-0.15}$ & $uvw2$ & {\it Swift} & -- \\ 
59153.34655 & 0.53719 & $21.61^{+0.23}_{-0.19}$ & $uvw2$ & {\it Swift} & -- \\ 
59162.16675 & 0.67457 & $21.50^{+0.20}_{-0.17}$ & $uvw2$ & {\it Swift} & -- \\ 
59166.84007 & 0.62876 & $21.36^{+0.22}_{-0.18}$ & $uvw2$ & {\it Swift} & -- \\ 
59172.61395 & 0.36597 & $21.14^{+0.11}_{-0.10}$ & $uvw2$ & {\it Swift} & -- \\ 
59186.69458 & 0.23797 & $21.19^{+0.12}_{-0.11}$ & $uvw2$ & {\it Swift} & -- \\ 
59200.54155 & 0.27290 & $21.25^{+0.11}_{-0.10}$ & $uvw2$ & {\it Swift} & -- \\ 
59214.25004 & 0.10372 & $21.37^{+0.12}_{-0.11}$ & $uvw2$ & {\it Swift} & -- \\ 
59299.75095 & 2.61782 & $22.15^{+0.26}_{-0.21}$ & $uvw2$ & {\it Swift} & -- \\ 
59325.79343 & 0.13505 & $22.08^{+0.21}_{-0.17}$ & $uvw2$ & {\it Swift} & -- \\ 
59353.47849 & 0.47154 & $22.06^{+0.23}_{-0.19}$ & $uvw2$ & {\it Swift} & -- \\ 
59381.98494 & 0.96554 & $22.37^{+0.26}_{-0.21}$ & $uvw2$ & {\it Swift} & -- \\ 
59409.46290 & 0.36455 & $22.27^{+0.20}_{-0.17}$ & $uvw2$ & {\it Swift} & -- \\ 
59437.70360 & 0.27319 & $22.24^{+0.19}_{-0.16}$ & $uvw2$ & {\it Swift} & -- \\ 
59470.18966 & 1.55880 & $22.51^{+0.29}_{-0.23}$ & $uvw2$ & {\it Swift} & -- \\ 
59497.01072 & 0.96448 & $22.68^{+0.30}_{-0.24}$ & $uvw2$ & {\it Swift} & -- \\ 
59528.82616 & 0.73344 & $22.55^{+0.35}_{-0.27}$ & $uvw2$ & {\it Swift} & -- \\ 
59558.39759 & 0.36654 & $22.29^{+0.24}_{-0.20}$ & $uvw2$ & {\it Swift} & -- \\ 
59582.28632 & 0.16869 & $22.32^{+0.34}_{-0.26}$ & $uvw2$ & {\it Swift} & -- \\ 
59652.35654 & 2.48900 & $22.85^{+0.58}_{-0.38}$ & $uvw2$ & {\it Swift} & -- \\ 
59687.97175 & 2.89395 & $22.60^{+0.20}_{-0.17}$ & $uvw2$ & {\it Swift} & -- \\ 

59169.23517 & 0.0136 & $21.86\pm0.24$ & F169M & {\it AstroSat} & -- \\ 
59176.60971  & 0.0111 & $21.72\pm0.25$ & F169M & {\it AstroSat} & -- \\ 
59192.51029  & 0.0114 & $22.15\pm0.30$ & F169M & {\it AstroSat} & -- \\ 
59176.67736  & 0.0132 & $21.84\pm0.40$ & F172M & {\it AstroSat} & -- \\ 
59192.64563 & 0.0113 & $22.07\pm0.49$ & F172M & {\it AstroSat} & -- \\ 

55324.99868 & 0.62841 & $16.83\pm0.15$ & W1 & {\it WISE} & -- \\ 
55504.33386 & 0.56214 & $16.87\pm0.14$ & W1 & {\it WISE} & -- \\ 
56787.28494 & 0.55940 & $16.90\pm0.21$ & W1 & {\it WISE} & -- \\ 
57966.78907 & 280.142 & $16.70\pm0.29$ & W1 & {\it WISE} & -- \\ 
58409.98672 & 1.93012 & $16.69\pm0.16$ & W1 & {\it WISE} & -- \\ 
58611.61711 & 0.62147 & $16.39\pm0.26$ & W1 & {\it WISE} & -- \\ 
58775.56522 & 0.49063 & $15.72\pm0.18$ & W1 & {\it WISE} & -- \\ 
58975.75518 & 0.55583 & $16.00\pm0.20$ & W1 & {\it WISE} & -- \\ 
59143.18162 & 4.08684 & $16.20\pm0.29$ & W1 & {\it WISE} & -- \\ 
59506.69516 & 0.62078 & $16.54\pm0.25$ & W1 & {\it WISE} & -- \\ 
56055.67957 & 731.3093 & $16.61\pm0.30$ & W2 & {\it WISE} & -- \\ 
58775.56522 & 0.49063  & $15.16\pm0.21$ & W2 & {\it WISE} & -- \\ 
56915.14 & -- & $22.82\pm0.07$ & g & DES & -- \\ 
57656.10 & -- & $22.68\pm0.05$ & g & DES & -- \\ 
56915.14 & -- & $22.80\pm0.07$ & g & DES & -- \\ 
56924.12 & -- & $22.70\pm0.05$ & g & DES & -- \\ 
57283.16 & -- & $22.87\pm0.08$ & g & DES & -- \\ 
56953.18 & -- & $22.97\pm0.09$ & g & DES & -- \\ 
56953.18 & -- & $22.95\pm0.09$ & g & DES & -- \\ 
57251.23 & -- & $23.06\pm0.12$ & g & DES & -- \\ 
56924.12 & -- & $22.71\pm0.05$ & g & DES & -- \\ 
58758.9588 &  -- & $19.60$ & g & {\it J-GEM} & GCN 25941\\ 
58787.06513 & 0.00510 & $19.38\pm0.09$ & g' & {\it GROND} & -- \\ 
59428.17707 & 0.00898 & $21.07\pm0.08$ & g' & {\it GROND} & -- \\ 
59430.16758 & 0.00898 & $21.05\pm0.07$ & g' & {\it GROND} & -- \\ 
59438.34115 & 0.00409 & $21.18\pm0.06$ & g' & {\it GROND} & -- \\ 
59470.14427 & 0.00394 & $21.24\pm0.06$ & g' & {\it GROND} & -- \\ 
59498.01320 & 0.00890 & $21.14\pm0.04$ & g' & {\it GROND} & -- \\ 
59529.02932 & 0.00902 & $21.22\pm0.05$ & g' & {\it GROND} & -- \\ 
59558.03155 & 0.00406 & $21.08\pm0.10$ & g' & {\it GROND} & -- \\ 
59725.33557 & 0.00895 & $21.27\pm0.08$ & g' & {\it GROND} & -- \\ 
57283.15 & -- & $21.20\pm0.09$ & r & DES & -- \\ 
57656.10 & -- & $21.19\pm0.05$ & r & DES & -- \\ 
56926.11 & -- & $21.22\pm0.05$ & r & DES & -- \\ 
56888.22 & -- & $21.25\pm0.05$ & r & DES & -- \\ 
56926.11 & -- & $21.22\pm0.05$ & r & DES & -- \\ 
56888.22 & -- & $21.25\pm0.09$ & r & DES & -- \\ 
56924.12 & -- & $21.21\pm0.02$ & r & DES & -- \\ 
56898.18 & -- & $20.94\pm0.09$ & r & DES & -- \\ 
56924.12 & -- & $21.22\pm0.02$ & r & DES & -- \\ 
56898.18 & -- & $20.95\pm0.09$ & r & DES & -- \\ 
58758.9588 & -- & $19.30$ & r & {\it J-GEM} & GCN 25941\\ 
58761.19808 & 0.031 & $19.35\pm0.06$ & r & {\it Chilescope} & GCN 25963\\ 
58787.06436 & 0.00433 & $19.58\pm0.03$ & r' & {\it GROND} & -- \\ 
59428.17707 & 0.00898 & $20.79\pm0.01$ & r' & {\it GROND} & -- \\ 
59430.16758 & 0.00898 & $20.76\pm0.03$ & r' & {\it GROND} & -- \\ 
59438.34115 & 0.00409 & $20.83\pm0.04$ & r' & {\it GROND} & -- \\ 
59470.14427 & 0.00394 & $20.88\pm0.04$ & r' & {\it GROND} & -- \\ 
59498.01320 & 0.00890 & $20.78\pm0.03$ & r' & {\it GROND} & -- \\ 
59529.02932 & 0.00902 & $20.82\pm0.03$ & r' & {\it GROND} & -- \\ 
59558.03155 & 0.00406 & $20.86\pm0.06$ & r' & {\it GROND} & -- \\ 
59725.33557 & 0.00895 & $20.91\pm0.03$ & r' & {\it GROND} & -- \\ 
58429.09 & -- & $20.59\pm0.03$ & i & DES & -- \\ 
56927.12 & -- & $20.61\pm0.03$ & i & DES & -- \\ 
56927.12 & -- & $20.61\pm0.03$ & i & DES & -- \\ 
56916.13 & -- & $20.61\pm0.04$ & i & DES & -- \\ 
56924.12 & -- & $20.63\pm0.04$ & i & DES & -- \\ 
57656.10 & -- & $20.61\pm0.03$ & i & DES & -- \\ 
56924.12 & -- & $20.63\pm0.03$ & i & DES & -- \\ 
56916.13 & -- & $20.61\pm0.04$ & i & DES & -- \\ 
57299.10 & -- & $20.63\pm0.05$ & i & DES & -- \\ 
58758.9588 & -- & $19.30$ & i & {\it J-GEM} & GCN 25941\\ 
58787.06436 & 0.00433 & $19.37\pm0.03$ & i' & {\it GROND} & -- \\ 
59428.17707 & 0.00898 & $20.26\pm0.01$ & i' & {\it GROND} & -- \\ 
59430.16758 & 0.00898 & $20.25\pm0.03$ & i' & {\it GROND} & -- \\ 
59438.34115 & 0.00409 & $20.25\pm0.05$ & i' & {\it GROND} & -- \\ 
59470.14427 & 0.00394 & $20.35\pm0.04$ & i' & {\it GROND} & -- \\ 
59498.01320 & 0.00890 & $20.26\pm0.03$ & i' & {\it GROND} & -- \\ 
59529.02932 & 0.00902 & $20.28\pm0.04$ & i' & {\it GROND} & -- \\ 
59558.03155 & 0.00406 & $20.16\pm0.06$ & i' & {\it GROND} & -- \\ 
59725.33557 & 0.00895 & $20.35\pm0.04$ & i' & {\it GROND} & -- \\ 
56886.23 & -- & $20.34\pm0.04$ & z & DES & -- \\ 
57643.17 & -- & $20.34\pm0.04$ & z & DES & -- \\ 
57294.11 & -- & $20.36\pm0.04$ & z & DES & -- \\ 
56886.23 & -- & $20.34\pm0.04$ & z & DES & -- \\ 
56904.16 & -- & $20.27\pm0.03$ & z & DES & -- \\ 
56932.15 & -- & $20.32\pm0.04$ & z & DES & -- \\ 
56904.16 & -- & $20.27\pm0.03$ & z & DES & -- \\ 
56932.15 & -- & $20.32\pm0.04$ & z & DES & -- \\ 
58409.03 & -- & $20.33\pm0.03$ & z & DES & -- \\ 
58787.06436 & 0.00433 & $19.29\pm0.04$ & z' & {\it GROND} & -- \\ 
59428.17707 & 0.00898 & $20.09\pm0.02$ & z' & {\it GROND} & -- \\ 
59430.16758 & 0.00898 & $20.04\pm0.02$ & z' & {\it GROND} & -- \\ 
59438.34115 & 0.00409 & $20.07\pm0.06$ & z' & {\it GROND} & -- \\ 
59470.14427 & 0.00394 & $20.08\pm0.04$ & z' & {\it GROND} & -- \\ 
59498.01320 & 0.00890 & $20.00\pm0.07$ & z' & {\it GROND} & -- \\ 
59529.02932 & 0.00902 & $20.12\pm0.03$ & z' & {\it GROND} & -- \\ 
59725.33557 & 0.00895 & $20.18\pm0.03$ & z' & {\it GROND} & -- \\ 
56931.12 & -- & $20.18\pm0.07$ & Y & DES & -- \\ 
57291.09 & -- & $20.26\pm0.09$ & Y & DES & -- \\ 
57999.12 & -- & $20.40\pm0.14$ & Y & DES & -- \\ 
57286.11 & -- & $20.28\pm0.21$ & Y & DES & -- \\ 
56886.23 & -- & $20.28\pm0.10$ & Y & DES & -- \\ 
57288.11 & -- & $20.39\pm0.40$ & Y & DES & -- \\ 
57293.10 & -- & $20.09\pm0.07$ & Y & DES & -- \\ 
58787.06541 & 0.00537 & $19.19\pm0.08$ & J & {\it GROND} & -- \\ 
59428.17707 & 0.00898 & $19.80\pm0.07$ & J & {\it GROND} & -- \\ 
59430.16758 & 0.00898 & $19.79\pm0.09$ & J & {\it GROND} & -- \\ 
59438.34137 & 0.00430 & $19.71\pm0.12$ & J & {\it GROND} & -- \\ 
59470.14448 & 0.00414 & $20.13\pm0.15$ & J & {\it GROND} & -- \\ 
59498.01340 & 0.00907 & $19.84\pm0.10$ & J & {\it GROND} & -- \\ 
59529.02953 & 0.00921 & $20.00\pm0.12$ & J & {\it GROND} & -- \\ 
59558.03179 & 0.00425 & $20.40\pm0.30$ & J & {\it GROND} & -- \\ 
59725.33575 & 0.00914 & $19.74\pm0.10$ & J & {\it GROND} & -- \\ 
58787.06541 & 0.00537 & $18.86\pm0.10$ & H & {\it GROND} & -- \\ 
59430.16758 & 0.00898 & $19.27\pm0.09$ & H & {\it GROND} & -- \\ 
59438.34137 & 0.00430 & $19.21\pm0.13$ & H & {\it GROND} & -- \\ 
59470.14448 & 0.00414 & $19.34\pm0.12$ & H & {\it GROND} & -- \\ 
59529.02953 & 0.00921 & $19.55\pm0.13$ & H & {\it GROND} & -- \\ 
59558.03179 & 0.00425 & $19.42\pm0.18$ & H & {\it GROND} & -- \\ 
59725.33575 & 0.00914 & $19.48\pm0.12$ & H & {\it GROND} & -- \\ 
58787.06464 & 0.00460 & $18.69\pm0.19$ & K & {\it GROND} & -- \\ 
59428.17707 & 0.00898 & $19.56\pm0.34$ & K & {\it GROND} & -- \\ 
59430.16778 & 0.00917 & $19.82\pm0.47$ & K & {\it GROND} & -- \\ 
59438.34137 & 0.00430 & $19.42\pm0.34$ & K & {\it GROND} & -- \\ 
59470.14448 & 0.00414 & $18.95\pm0.18$ & K & {\it GROND} & -- \\ 
59498.01340 & 0.00907 & $19.17\pm0.21$ & K & {\it GROND} & -- \\ 
59529.02953 & 0.00921 & $19.45\pm0.31$ & K & {\it GROND} & -- \\ 
59725.33575 & 0.00914 & $19.30\pm0.21$ & K & {\it GROND} & -- \\
\hline
\hline 
\end{longtable} 

\twocolumn